\documentclass[aps,prb,superscriptaddress,twocolumn,preprintnumbers,amsmath,amssymb]{revtex4-1}
\usepackage{color}
\usepackage{array}
\usepackage{amsmath}
\usepackage{amssymb}
\usepackage[pdftex]{graphicx}
\usepackage{epstopdf}
\usepackage{bbm}
\usepackage{hyperref}
\usepackage{multirow}
\usepackage{float}

\newcommand{\dd}{\; \mathrm{d}}

\begin{document}
\title{Chiral spin density wave, spin-charge-Chern liquid and d+id superconductivity in 1/4-doped correlated electronic systems on the honeycomb lattice}
\author{Shenghan Jiang, Andrej Mesaros and Ying Ran}
\affiliation{Department of Physics, Boston College, Chestnut Hill, MA 02467}
\date{\today}

\begin{abstract}
Recently two interesting candidate quantum phases --- the chiral spin density wave state featuring anomalous quantum Hall effect and the d+id superconductor --- were proposed for the Hubbard model on the honeycomb lattice at 1/4 doping. Using a combination of exact diagonalization, density matrix renormalization group, the variational Monte Carlo method and quantum field theories, we study the quantum phase diagrams of both the Hubbard model and t-J model on the honeycomb lattice at 1/4-doping. The main advantage of our approach is the use of symmetry quantum numbers of ground state wavefunctions on finite size systems (up to 32 sites) to sharply distinguish different quantum phases. Our results show that for $1\lesssim U/t< 40$ in the Hubbard model and for $0.1< J/t<0.80(2)$ in the t-J model, the quantum ground state is either a chiral spin density wave state or a spin-charge-Chern liquid, but not a d+id superconductor. However, in the t-J model, upon increasing $J$ the system goes through a first-order phase transition at $J/t=0.80(2)$ into the d+id superconductor. Here the spin-charge-Chern liquid state is a new type of topologically ordered quantum phase with Abelian anyons and fractionalized excitations. Experimental signatures of these quantum phases, such as tunneling conductance, are calculated. These results are discussed in the context of 1/4-doped graphene systems and other correlated electronic materials on the honeycomb lattice.
\end{abstract}

\maketitle
\tableofcontents

\section{Introduction}
The reliable determination of quantum phase diagrams of correlated electronic systems has been one of the central issues in quantum condensed matter physics. In the past decades, different analytic and numeric methods have been developed to attack this problem, including renormalization group methods\cite{Shankar:1994p7862,Polchinski:1984p7873,Metzner:2012p7861,Platt:2013p7896}, quantum Monte Carlo methods (for a review see Ref.\onlinecite{Foulkes:2001p7937}), variational Monte Carlo methods\cite{Foulkes:2001p7937,Gros:1989p7551}, the density matrix renormalization group (DMRG) method\cite{Schollwock:2011p7856,White:1992p7946} and the recently developed tensor-network methods\cite{Vidal:2008p7925,Vidal:2007p7926,Corboz:2009p7934,Corboz:2010p7933,Verstraete:2004p7928,Kraus:2010p7935}. Although each method has its advantages and disadvantages, this growing list of theoretical techniques has enabled careful investigations, and sometimes reliable determinations of quantum phase diagrams of correlated systems. In particular, in 
presence of strong correlations, a reliable understanding of the quantum phase diagrams of realistic model Hamiltonians usually strongly relies on unbiased numerical techniques. For instance, the DMRG method has successfully determined quantum phase diagrams of various quantum spin systems, and exotic quantum spin liquid phases were revealed\cite{Yan:2011p7804,Jiang:2012p7939,Gong:2013p7942,Gong:2013p7940,Depenbrock:2012p7944}.

However, in the presence of doping, due to the larger dimension of Hilbert space and stronger quantum entanglement, a reliable determination of quantum phases remains challenging. The challenge is partially due to the fact that competing quantum phases cannot be sharply distinguished on finite-size systems in an obvious fashion, while most cutting-edge numerical simulations can only be performed on finite-size systems.

In this work, we show that a combination of different quantum many-body techniques allows, to a certain level, a sharp determination of quantum phases of correlated electronic systems at some commensurate dopings\footnote{The reason why we focus on commensurately doped systems is mainly due to technical considerations: at certain commensurate fillings, there can be very reasonable guesses for the candidate quantum phases, and explicitly constructing their wavefunctions is not too difficult within the currently available theoretical frameworks.}. Particularly, we demonstrate our approach in the 1/4-doped correlated systems on the honeycomb lattice. Our approach is based on our ability of analytically writing down symmetric quantum wavefunctions of different candidate quantum phases on finite size systems, studying their characteristic quantum numbers and other properties, and comparing with results from unbiased numerical simulations such as exact diagonalization and DMRG. When different quantum phases can be analytically shown to have different lattice quantum numbers, this approach has the power to sharply distinguish them even on small lattices.

\begin{figure}
 \includegraphics[width=0.48\textwidth]{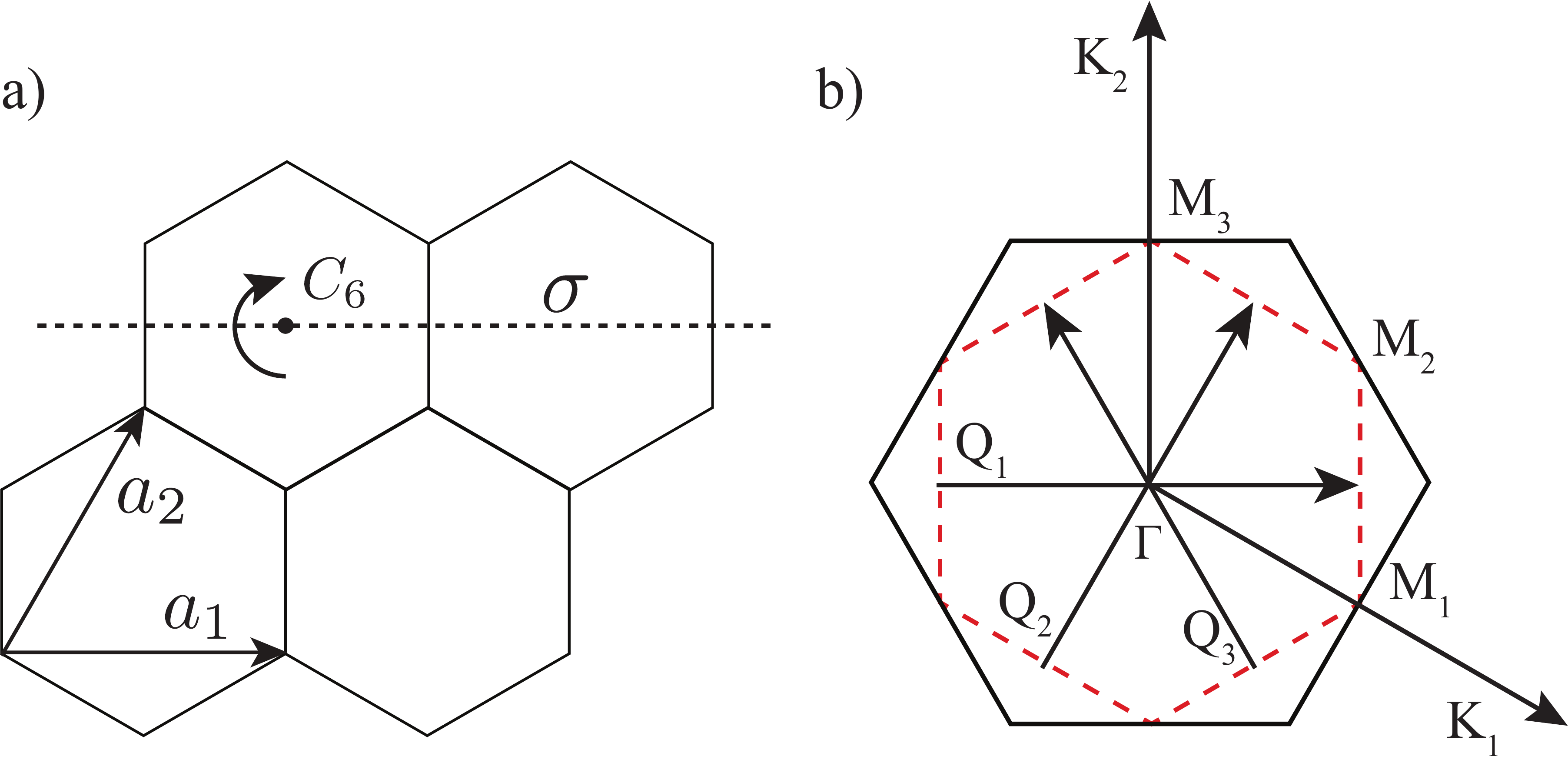}
 \caption{(a) Symmetries of the honeycomb lattice. (b): Nested Fermi surface at 1/4 doping.}
 \label{fig:fermi_surface}
\end{figure}

Recently, interesting quantum phases were proposed for the 1/4-doped Hubbard model on the honeycomb lattice. Considering the nearest neighbor single-band tight-binding model on the honeycomb lattice, both the 1/4 electron-doped and hole-doped systems feature a Fermi surface of hexagonal shape (see Fig.\ref{fig:fermi_surface}), which is unstable even in the presence of weak interactions. There are two important features of the hexagonal Fermi surface: The opposite sides of the Fermi surface are nested by three wavevectors $\mathbf{Q}_{1,2,3}$, and three Van Hove singularities are located at the mid-points of the Brillouin Zone boundary $M_{1,2,3}$. Previous studies have revealed two interesting candidate quantum phases: The chiral spin density wave state\cite{Li:2012p7905} and the d+id superconductor\cite{Raghu:2010p7897,Nandkishore:2012p7877}, both of which can be understood starting from the two features of the hexagonal Fermi surface.

\begin{figure}
 \includegraphics[width=0.48\textwidth]{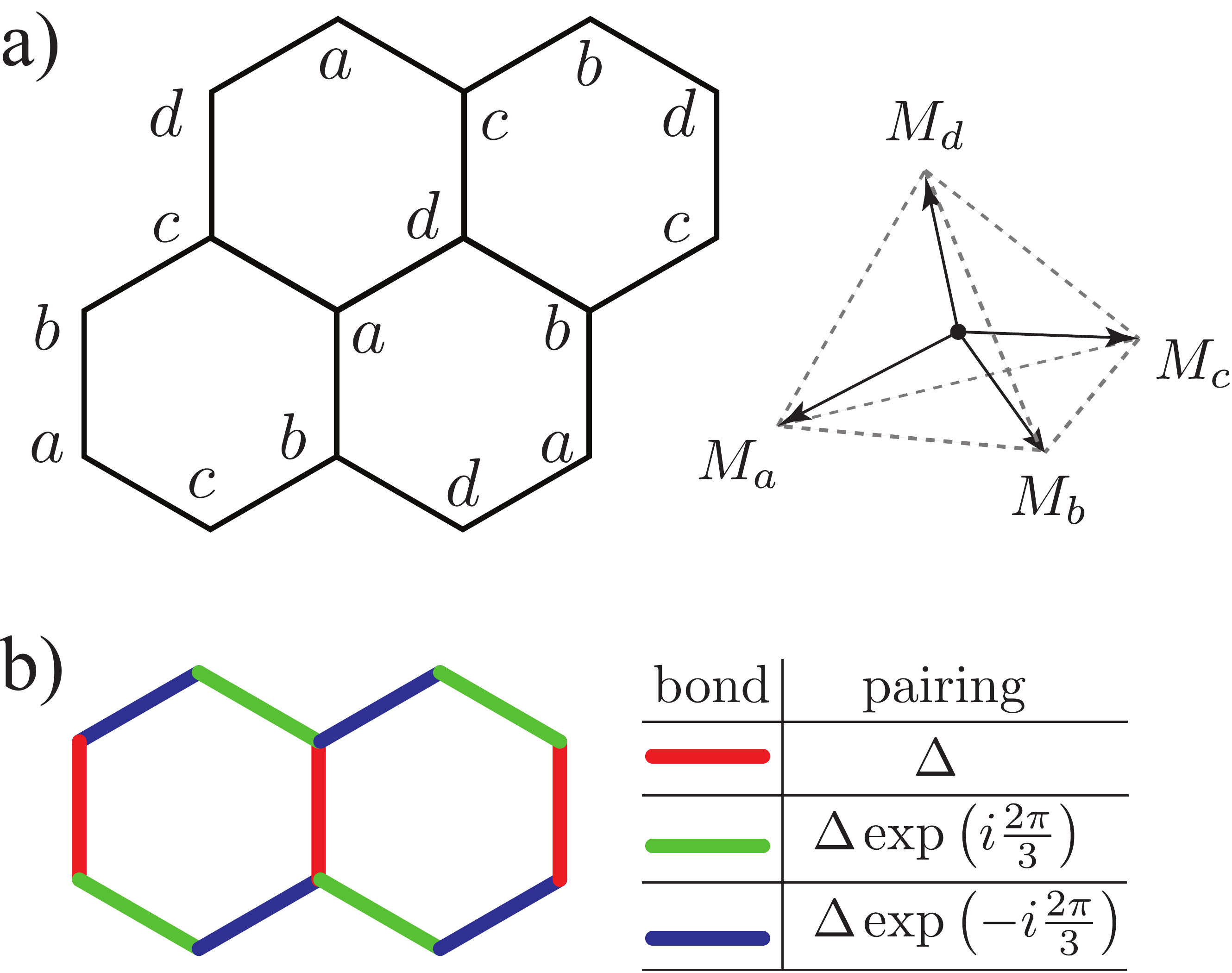}
 \caption{(a): The chiral spin density wave and (b) d+id pairing order parameters in real space.}
 \label{fig:cSDW_dpid_real_space}
\end{figure}

It is well-known that nested Fermi surfaces can cause magnetism. Based on Hartree-Fock mean-field calculations\cite{Li:2012p7905} and functional renormalization group calculations\cite{Wang:2012p7894,Kiesel:2012p7895}, it has been shown that the three nested wavevectors together could give rise to a rather exotic type of magnetic ordering at intermediate coupling strengths: The tetrahedral magnetic order which quadruples the unit cell (see Fig.\ref{fig:cSDW_dpid_real_space}a). Due to the non-coplanar magnetic ordering pattern, electrons pick up Berry's phase when hopping around the lattice, similarly to the effect of a non-uniform magnetic field. Consequently the electronic band structure is found to carry a non-zero Chern number. This magnetically ordered phase, termed the chiral spin density wave (c-SDW) state, is a topological phase featuring gapless electronic edge states and anomalous quantum Hall effect $\sigma_{xy}=e^2/h$.\cite{Li:2012p7905,Wang:2012p7894}

In addition, analytical renormalization group calculations, focusing on scattering involving electronic states at the Van Hove singularities, show a d+id superconductor as the ground state of the system, which in principle could be a high temperature phenomenon\cite{Nandkishore:2012p7877} (see Fig.\ref{fig:cSDW_dpid_real_space}b for the pairing order parameter in real space). The same d+id superconductor (SC) has also been proposed for the Hubbard model and t-J model on the honeycomb lattice over a large range of doping levels based on renormalization group studies\cite{Raghu:2010p7897,Kiesel:2012p7895,Nandkishore:2012p7877,Wang:2012p7894,Wu:2013p7952}, variational Monte Carlo approaches\cite{Pathak:2010p7901}, and tensor-network numerical simulations\cite{Gu:2013p7902}. The d+id SC phase turns out to be a topological superconductor hosting a spin-quantum-Hall effect.\cite{Senthil:1999p7954}

Apart from these two phases, in this work we propose yet another candidate quantum phase, denoted as spin-charge-Chern liquid (SCCL), which could also be realized in correlated electronic systems on the honeycomb lattice at 1/4 doping. SCCL can be viewed as the resulting phase after the long-range magnetic order in the c-SDW phase is quantum melted. In the past, quantum melting of long-range magnetic order was discussed in the context of undoped quantum spin systems, and the resulting exotic phases, quantum spin liquids, have attracted considerable interest (see, e.g., \onlinecite{2010Natur.464..199B,Han:2012p7973,Watanabe:2012p7963,Jiang:2012p7939,Cheng:2011p7959,Yan:2011p7804,Pratt:2011p7961,Yamashita:2010p7955}). It is known that strong quantum fluctuations are necessary to stabilize such liquid phases. In fact, most candidate quantum spin liquid materials are spin-1/2 systems, where quantum fluctuations are strong. Intuitively, quantum fluctuations of spin degrees of freedom are likely to be even stronger in doped spin-1/2 systems, which can be justified by slave-fermion mean-field arguments (see Sec.\ref{sec:sym_wavefunctions}). This suggests that liquid phases such as SCCL may have a better chance to be stabilized in doped correlated electronic systems.

Unlike c-SDW, the SCCL phase respects spin-rotation and lattice translation and rotation symmetries, while breaking the time-reversal symmetry; nevertheless, both the charge and spin excitations are gapped in the bulk. This violation of Luttinger's theorem is due to the fact that SCCL is a fractionalized phase with topological order. For example, in the bulk SCCL features charge-$1/2$, spin-neutral anyon excitations with $\theta=\pi/4$ exchange statistics. On the boundary, SCCL hosts chiral gapless edge states of charge-$1$, spin-neutral fermions. We show that although the electromagnetic response in the bulk of the SCCL is described by an anomalous quantum Hall response similar to the c-SDW phase: $j_x=\sigma_{xy} E_y$, where $\sigma_{xy}=e^2/h$, the SCCL and c-SDW have very different signatures in transport experiments, which can be used to identify them in candidate materials. One important result of the current work is that \emph{the conductance through a weakly coupled tunneling junction with a metallic lead (namely, $G\ll e^2/h$) in a SCCL phase should vanish as $G(T)\propto T^4$ at low temperatures, while in the c-SDW phase this should obey $G(T)\propto \mbox{constant}$}.

Experimentally, single-band correlated electronic models on the honeycomb lattice are relevant for many materials. For instance, doped graphene may be described by the Hubbard model in the intermediate correlated regime $U/t=2\sim 3$.\cite{CastroNeto:2009p3917} More candidate materials, including certain transition metal oxide heterstructures, will be discussed later in this paper. Although experimental realization of 1/4-doping on these materials has not yet been reported, with the fast developing material science techniques on thin film synthesis, this doping level may be achievable within foreseeable future. This motivates us to carefully investigate the phase diagrams of the correlated electronic systems on the honeycomb lattice at 1/4 doping, especially over the intermediate to strong correlation strengths. The previous studies are either based on mean-field theories\cite{Li:2012p7905} which is biased, or renormalization group techniques \cite{Wang:2012p7894,Kiesel:2012p7895,Raghu:2010p7897,
Nandkishore:2012p7877} which presumably are under control only for the weak coupling regime.

We study both the 1/4-doped Hubbard model and t-J model on the honeycomb lattice:
\begin{align} H_{H}&=-t\sum_{<ij>,\alpha}(c_{i\alpha}^{\dagger}c_{j\alpha}+h.c.)+U\sum_{i}n_{i\uparrow}n_{i\downarrow},\notag\\
 H_{tJ}&=P_G \sum_{<ij>,\alpha} -t(c_{i\alpha}^{\dagger}c_{j\alpha}+h.c.)P_G\notag\\
 &+P_G \sum_{<ij>}J(\mathbf{S}_i\cdot\mathbf{S}_j-\frac{1}{4} n_i \cdot n_j) P_G .
\end{align}
Here $P_G$ is the usual Gutzwiller projection operator removing the double occupancies in the t-J model. Due to the particle-hole symmetry of these nearest neighbor ($t$ only) models, our study applies for both the electron-doped and hole-doped systems.

\begin{figure}
 \includegraphics[width=0.48\textwidth]{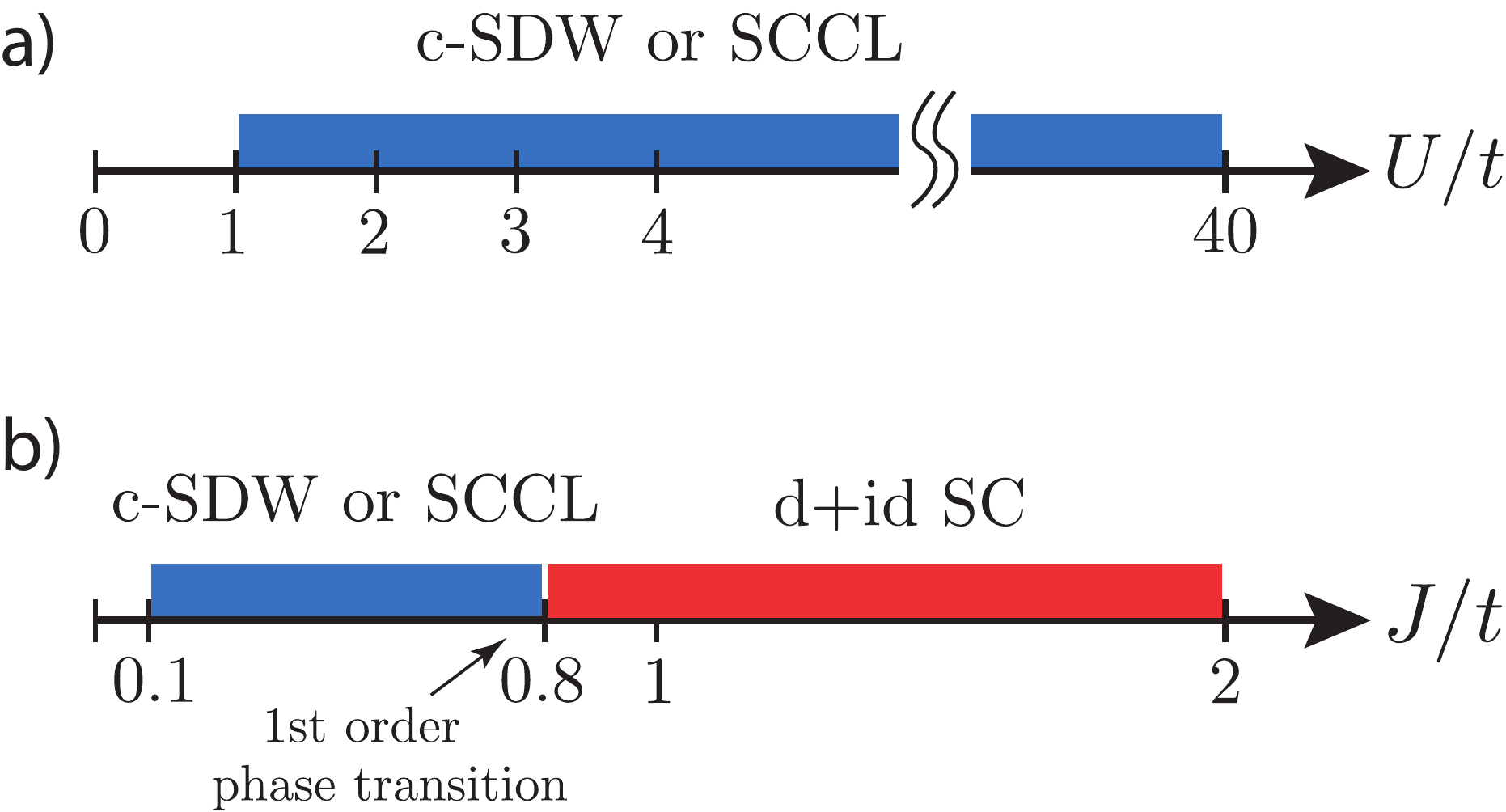}
 \caption{The phase diagrams of the correlated electronic systems on the honeycomb lattice at 1/4 doping. (a) In the Hubbard model, the ground state is found to be either in a c-SDW phase or in a SCCL phase over the majority of the parameter range $1\lesssim U/t< 40$. (b) In the t-J model, the c-SDW phase or the SCCL phase is identified in the regime $0.1<J/t<0.80(2)$, seperated from the d+id superconductor phase at larger $J/t$ by a first order phase transition.}
 \label{fig:phase_diagrams}
\end{figure}

We analytically constructed quantum wavefunctions, performed exact diagonalization on the 8-site sample, DMRG simulations on the 24-site and 32-site samples, and variational Monte Carlo simulations. These allow us to construct reliably, at least to a certain extent, the quantum phase diagrams from intermediate to strong coupling regimes. Our main results are summarized in Fig.\ref{fig:phase_diagrams}. The limitation of our calculations are two-fold. First, we cannot address the phase diagram reliably for the weak coupling regime in the Hubbard model: $U/t\lesssim 1$, because correlation lengths of the competing phases can be much larger than the investigated system sizes. Second, we cannot sharply distinguish the c-SDW phase from the SCCL phase, because they are distinguished only by long-range physics which requires careful finite size scaling and larger system sizes.

Despite these limitations, we find that c-SDW or SCCL is stablized in the majority of the physically realistic parameter regime: $1\lesssim U/t < 40$ in the Hubbard model and $0.1<J/t<0.80(2)$ in the t-J model. Between the c-SDW and SCCL phases, the measurements of correlation functions suggest that SCCL is more likely to be realized in the small $U/t$ and large $J/t$ regimes within these parameter windows. The d+id SC phase is found in the t-J model at $0.80(2)<J/t$. The sharp distinction between the c-SDW/SCCL phase and the d+id SC phase becomes possible because they have different lattice quantum numbers on the 32-site sample.

The remaining parameter regimes will be briefly discussed but will \emph{not} be the focus of the present paper, since it is unclear whether these regimes are relevant for correlated materials. For instance, in the t-J model with $J/t<0.1$ and in the Hubbard model with $U/t\gtrsim 50$, we find some inconclusive evidence for a different homogeneous phase. This parameter regime is adjacent to the infinite-$U$ Hubbard problem,\cite{Nagaoka:1966p8011,Kanamori:1963p8010} and will be left for future study.
% which is separated from the c-SDW/SCCL phase by a first-order phase transition.
In addition, when $J/t\gtrsim 3$ in the t-J model, evidence of charge inhomogeneity is observed, which is likely due to phase separation and consistent with physical intuition.

There is a useful by-product of our investigation. It was proposed that, on finite size lattices, the rotational symmetry eigenvalues in the ground state manifold of a topologically ordered phase can be determined by the modular transformation matrices.\cite{Zhang:2012p7534,Zhang:2013p8017,He:2013p8022,Wen:2012p8018,Cincio:2013p7854} However, the SCCL phase here serves as a counterexample of this claim, because the rotational eigenvalues are system-size dependent (see Section~\ref{sec:SCCL}).

The paper is organized as follows. In Section \ref{sec:sym_wavefunctions} we analytically construct symmetric quantum wavefunctions of the three phases: c-SDW, SCCL and d+id SC. We show that they have characteristic signatures in quantum numbers on finite size lattices. In Section \ref{sec:numerical_simulation} we present the results from a combination of different numerical simulations, which convincingly justify the phase diagrams in Fig.\ref{fig:phase_diagrams}. Because SCCL is a new topologically ordered quantum phase, its low energy effective theory, fundamental properties, and experimental signatures are studied in Section \ref{sec:SCCL}. Finally, we discuss our methodology and results, in particular in the context of a few candidate materials in Section \ref{sec:conclusion}.

\section{Symmetric wavefunctions of competing quantum phases}\label{sec:sym_wavefunctions}

To identify fingerprints of these candidate quantum phases in unbiased numerical simulations, we explicitly write down the symmetric quantum wavefunctions of the competing c-SDW, SCCL and d+id SC phases on finite-size lattices. Note that there is no sense of spontaneous symmetry breaking on finite-size lattices and these quantum wavefunctions are symmetric (i.e, forming irreducible representations) of the full symmetry group involving both the $SU(2)$ spin-rotation group and the lattice space group. This is also why the c-SDW and SCCL cannot be sharply distinguished on finite-size lattices because they share the same quantum numbers.

We construct the c-SDW/SCCL quantum wavefunctions using the slave-fermion approach,\cite{Wang:2006p6704,Sachdev:1992p12377,Sachdev:1991p8025,Read:1991p5262,Arovas:1988p7948} and construct the d+id SC wavefunctions by the slave-boson approach.\cite{Wen:2002p6309,Wen:2002p8024} Note that these wavefunctions are constructed in the Hilbert space of the t-J model; however, their quantum numbers on finite size lattices are unchanged in the Hubbard model. This is because the small $J/t$ regime in the t-J model and large $U/t$ regime of the Hubbard model are smoothly connected, and if two quantum states have different quantum numbers on finite size lattices, they cannot represent the same quantum phase. In addition, although these quantum wavefunctions are not the exact ground states of simple model Hamiltonians, they have the same \emph{universal} properties of the quantum phases that they belong to, including symmetry quantum numbers.

\subsection{d+id SC}
\label{sec:did}
Here we briefly describe these symmetric quantum wavefunctions. The details can be found in Appendix \ref{app:wavefunction_construction}). A d+id SC can be constructed using the slave-boson approach,\cite{Wen:2002p6309,Wen:2002p8024} in which the electrons are split into fermionic spinons and bosonic holons:
\begin{align}
 c_{i\alpha}=f_{i\alpha} b_i^{\dagger}.
\end{align}
This parton construction enlarges the Hilbert space and has a $U(1)$ gauge redundancy. This gauge redundancy is broken by boson condensation $|\langle b_i\rangle|=\sqrt{x}$ at zero temperature ($x$ is the doping fraction), which is required to accomodate the doped charge at the mean-field level. A d+id SC can be represented if $f_{i\alpha}$ fermions form a d+id SC band structure, and the bosons are condensed at the $\Gamma$-point. The associated  physical wavefunction is obtained after projecting out the unphysical states of the t-J model, i.e., it is a simple Gutzwiller projected d+id SC wavefunction:
\begin{align}
  |\Psi_{d+id}(\chi,\Delta)\rangle=P_GP_N|\Psi^{MF}_{d+id}(\chi, \Delta)\rangle,
  \label{eq:psi_did}
\end{align}
where $P_N$ is the projector into a fixed fermion number sector, enforcing that the total number of fermions equals $3/4$ of the total number of sites. $|\Psi^{MF}_{d+id}(\chi,\Delta)\rangle$ is the ground state of the d+id SC mean-field Hamiltonian:
\begin{align}
 H_{d+id}^{MF}(f)=&\sum_{<ij>}\big(-\chi f^{\dagger}_{i\alpha}f_{j\alpha}+\Delta_{ij}f_{i\alpha}f_{j\beta}\epsilon_{\alpha\beta} +h.c.\big)\notag\\
 &-\mu_f\sum_{i}f_{i\alpha}^{\dagger}f_{i\alpha}.
 \label{eq:Hmf_did}
\end{align}
Here $\chi$ is the real hopping, while singlet pairing $\Delta_{ij}$ has the real space pattern shown in Fig.\ref{fig:cSDW_dpid_real_space}b. Namely, $\Delta_{ij}=\Delta,\Delta e^{2\pi i/3}, \Delta e^{-2\pi i/3}$ depending on the orientations of the bonds ($\Delta$ is chosen to be real). For simplicity we include the nearest neighbor amplitudes only. The $\mu_f$ is tuned to satisfy $\langle f_{i\alpha}^{\dagger}f_{i\alpha}\rangle=3/4$ at the mean-field level, so it is not a variational parameter. Note that after the sign of $\chi$ is fixed, only the ratio $\Delta/\chi$ is a variational parameter of the constructed wavefunction $|\Psi_{d+id}(\chi,\Delta)\rangle$. This single-parameter variational wavefunction will be used in the variational Monte Carlo study in Section \ref{sec:VMC}, which reproduces $\sim 97 -99\%$ of the ground state energy in the $d+id$ phase shown in Fig.\ref{fig:phase_diagrams}b.

\begin{figure}
 \includegraphics[width=0.49\textwidth]{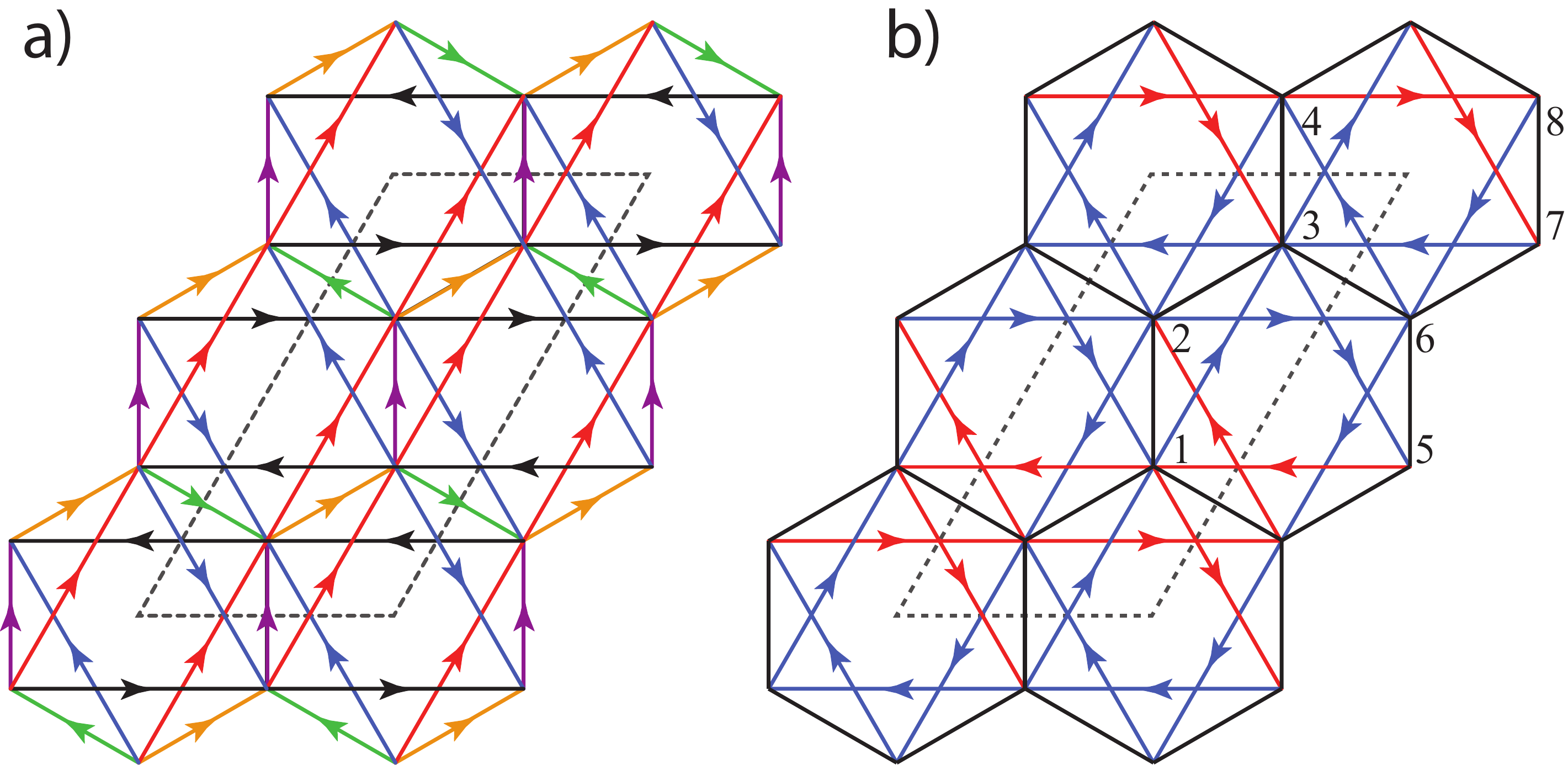}
\caption{(color online) The real space pattern of the slave-fermion amplitudes describing the c-SDW/SCCL phases. The dashed line encircles the doubled unit-cell. (a): The nearest neighbor(NN) and next nearest neighbor(NNN) boson pairing amplitudes $A_{ij}$ are directional (labeled by arrows) since $A_{ij}=-A_{ji}$. $A_{ij}$ on the NN(NNN) bonds have the same magnitude respectively. Their different phases are represented by different colors. Black: $1$; Violet: $e^{i\pi/2}$; Green: $e^{i5\pi/6}$; Orange: $e^{i\pi/6}$; Red: $e^{i\pi/3}$; Blue: $e^{i2\pi/3}$. (b): The NN(NNN) boson/fermion hopping amplitudes $B_{ij}$/$\chi_{ij}$ also have uniform magnitudes respectively. When they are complex, the amplitudes are directional $B_{ij}=B_{ji}^*,\chi_{ij}=\chi_{ji}^*$ (labeled by arrows). The phases are illustrated by colors. Black: $\pm 1$; Blue: $e^{i\phi}$; Red: $-e^{i\phi}$. Here the real number $\phi=\phi_b$ for bosons and $\phi=\phi_f$ for fermions. $\phi_b$ and $\phi_f$ can be viewed as two variational 
parameters. The above pattern is for one of the two degenerate ground states while the other one is its time-reversal image, which can be obtained by sending these amplitudes to their complex conjugates: $A_{ij}/B_{ij}/\chi_{ij}\rightarrow A^*_{ij}/B^*_{ij}/\chi^*_{ij}$. Sites numbered 1 to 8 label the quadrupled unit-cell used in Appendix~\ref{app:wf_quantum_number}.}
\label{fig:SCCL_pattern}
 \end{figure}

\subsection{c-SDW/SCCL}
To construct c-SDW/SCCL wavefunctions, we use the slave-fermion approach,\cite{Wang:2006p6704,Sachdev:1992p12377,Sachdev:1991p8025,Read:1991p5262,Arovas:1988p7948} in which electrons are split into bosonic spinons and fermionic spinless holons:
\begin{align}
 c_{i\alpha}=b_{i\alpha} f_i^{\dagger},
 \label{slave_fermion}
\end{align}
which also enlarges the Hilbert space and has a $U(1)$ gauge redundancy. At the mean-field level, the spin dynamics is described by a bosonic superconductor, and the charge dynamics is described by a spinless fermion band structure:
\begin{align}
 H_{c-SDW/SCCL}^{MF}(b)=&\sum_{ij}\big(B_{ij} b^{\dagger}_{i\alpha}b_{j\alpha}+A_{ij}b_{i\alpha}b_{j\beta}\epsilon_{\alpha\beta} +h.c.\big)\notag\\
 &-\mu_b\sum_{i}b_{i\alpha}^{\dagger}b_{i\alpha},\notag\\ H_{c-SDW/SCCL}^{MF}(f)=&\sum_{ij}\big(\chi_{ij}f_{i}^{\dagger}f_j+h.c.\big)-\mu_f\sum_{i}f_{i}^{\dagger}f_{i},\label{eq:sf_MF}
\end{align}
where $B_{ij}$ and $A_{ij}$ are boson singlet hopping and pairing on bond $ij$, and $\chi_{ij}$ is the spinless fermion hopping. Nonzero $B_{ij}$ and $A_{ij}$, which are required to describe c-SDW/SCCL, break the $U(1)$ gauge redundancy down to $Z_2$. The mean-field boson(fermion) wavefunction $|\Psi^{MF}_{b}\rangle$($|\Psi^{MF}_{f}\rangle$) is the ground state of the corresponding Hamiltonian in Eq.\ref{eq:sf_MF}, which can be mathematically represented as a permanent(determinant). The associated physical wavefunction $|\Psi_{c-SDW/SCCL}\rangle$ is obtained by gluing two parts together and going back to the Hilbert space of the t-J model.

More precisely, note that any physical state in the t-J model can be expanded in the spin-occupation basis $\{|s_1,s_2,s_3...s_N\rangle\}$, where $N$ is the number of sites, and $s_i=\uparrow,\downarrow,0$ depending on whether the site-$i$ is spin-up, spin-down or empty:
\begin{align}
 |s_1,s_2,...s_N\rangle\equiv \prod_{s_{i_a} =\uparrow } b_{i_a,\uparrow}^{\dagger}\prod_{s_{i_b}=\downarrow} b_{i_b, \downarrow}^{\dagger}\prod_{s_{i_c}=0}f_{i_c}^{\dagger}|0\rangle,
\end{align}
where a certain ordering of sites is required in the last product to take care of the fermion sign. The physical wavefunction $|\Psi_{c-SDW/SCCL}\rangle$ is defined as:
\begin{align}
 &\langle s_1,s_2,...s_N|\Psi_{c-SDW/SCCL}\rangle\notag\\
 =&\langle 0|\big[\prod_{s_{i_a} =\uparrow } b_{i_a,\uparrow}^{\dagger}\prod_{s_{i_b}=\downarrow} b_{i_b, \downarrow}^{\dagger}\big]^{\dagger}|\Psi^{MF}_{b}\rangle\cdot\notag\\
 &\cdot \langle 0|\big[\prod_{s_{i_c}=0}f_{i_c}^{\dagger}\big]^{\dagger}|\Psi^{MF}_{f}\rangle;
 \label{c-SDW_projection}
\end{align}
i.e., $|\Psi_{c-SDW/SCCL}\rangle$ is a product of a permanent(the second line) and a determinant(the third line).

It turns out that the real space pattern of $A_{ij},B_{ij},\chi_{ij}$ as shown in Fig.\ref{fig:SCCL_pattern} is describing the c-SDW/SCCL phases (see Appendix \ref{app:wavefunction_construction}). For simplicity we plot these amplitudes only on the nearest neighbor (NN) and next nearest neighbor (NNN) bonds. This complicated pattern ensures that the wavefunction is symmetric under lattice space group while capturing the tetrahedral spin correlation.

One can see that the unit-cell of the amplitudes doubles the original unit-cell of the honeycomb lattice, which indicates that the mean-field states $|\Psi^{MF}_{b}\rangle$($|\Psi^{MF}_{f}\rangle$) break translational symmetry. However, the physical state $|\Psi_{c-SDW/SCCL}\rangle$ is fully translationally symmetric, as shown in Appendix \ref{app:wavefunction_construction}. Similar states having doubled unit-cell of the mean-field amplitudes are often called $\pi$-flux states in the context of quantum spin liquids.

In addition, this doubling of unit-cell is physically important. This is why the spinless fermion filling $\langle f_i^{\dagger}f_i\rangle=1/4$, required by the 1/4-doping, actually corresponds to a fully filled lowest $f$-fermion band, which is separated from higher bands by an energy gap generated by the imaginary part of the NNN hopping $e^{i\phi_f}$. Similarly to the Haldane model of spinless fermions,\cite{Haldane:1988p5265} which preserves the original unit-cell of the honeycomb lattice, the lowest energy band of $f$-fermion here is found to carry non-zero Chern-number $C=1$. Because $f$-fermion describes the charge dynamics, the electromagnetic response of c-SDW/SCCL features an anomalous quantum Hall response, $\sigma_{xy}=e^2/h$.

Now we describe the difference between the c-SDW phase and the SCCL phase in the above slave-fermion formulation. At the mean-field level, $\mu_b$ is chosen so that $\langle b_{i\alpha}^{\dagger}b_{i\alpha}\rangle=3/4$ to be consistent with the doping level. On a finite size lattice, this is always achieved by tuning $\mu_b$ so that the boson band minima are close enough to, but not touching, zero. Note that when the bosonic band minima touch zero, boson condensation occurs and long-range tetrahedral magnetic order is established (see Appendix \ref{app:wavefunction_construction}). This is the c-SDW phase in the slave-fermion formulation. However, because boson condensation never occurs on finite size lattice due to the presence of boson pairing, the difference between the two phases appears only in the thermodynamic limit ($L\rightarrow \infty$). In this limit, if the boson band minima separate from zero by a finite gap, the resulting phase is a SCCL; however, if the gap closes the resulting phase is a c-SDW.

The SCCL phase is thus a fully gapped phase in the bulk, which will be studied in detail in Section \ref{sec:SCCL}. Nevertheless it is helpful to mention some of its basic properties here. Because the bosons do not condense, there is a remaining $Z_2$ gauge dynamics which dictates the existence of a topological order. However, the topological order in SCCL is fundamentally different from a usual $Z_2$ topological order such as the one in Kitaev's toric code.\cite{Kitaev:2003p6185} In a usual $Z_2$ topological order, there are three types of nontrivial quasiparticles: bosonic $Z_2$-gauge-charge $e$, bosonic vison ($\pi$-gauge-flux) $m$, and the fermionic bound state $em$. But in SCCL, the three nontrivial quasiparticles are: spin-1/2 and charge-neutral bosonic $Z_2$-gauge-charge $b_{\alpha}$ (which can be identified with the spinons), spin-neutral and charge-1/2 vison $v$ with statistical angle $\theta=\pm\pi/4$, and their bound states: spin-1/2-charge-1/2 anyon $b_{\alpha}v$ with statistical angle $\pm5\pi/4$. Here the two signs of the statistical angles correspond to the two degenerate ground states which are time-reversal images of each other. The charge-1/2 vison $v$ is simply due to the fact that fermion-$f$ fills a Chern band. Thus the vison, a $\pi$-gauge flux, will be bound with 1/2-charge.

The charge-1, spin-neutral fermionic holon-$f$ differs from a spinon only by an electron. Therefore mathematically it is not a new type of quasiparticle. However, there are gapless chiral edge states formed by $f$ on the boundary, which is clear at the mean-field level. This means that although the spin-gap is opened everywhere in the SCCL phase, the charge gap is closed on the boundary. Because of the spin gap, single electron tunneling into the edge states is forbidden at low energy. However, singlet-pairs of electrons can still tunnel into the edge, which is the origin of the $T^4$ power-law tunneling conductance at low energy.
 
\emph{Finally, the slave-fermion formulation of long-range magnetic order allows us to argue that the spin liquid phases, such as SCCL, may be easier to be stablized in the doped systems compared with the undoped spin-1/2 systems}. In the past, a great number of spin-1/2 models were investigated in a search for quantum spin liquids. Only few of these models can host spin liquid phases.\cite{Morita:2002p8014,Koretsune:2007p8016,Kyung:2006p8015,Tay:2011p7822} In the slave-fermion formulation (which in the undoped case, is the same as the Schwinger-boson formulation), this can be understood as follows. 

For a given value of mean-field parameters $A_{ij}/B_{ij}$, $\langle b_{i\alpha}^{\dagger}b_{i\alpha}\rangle$ increases as $\mu_b$ increases and the boson quasiparticle gap decreases. In most cases, boson condensation is required to accommodate the boson density $\langle b_{i\alpha}^{\dagger}b_{i\alpha}\rangle=1$ in the undoped systems. For example, in the $Q_1=Q_2$ state on the Kagome lattice, only for a rather small parameter window of $A_{ij}/B_{ij}$, a spin liquid state is stabilized.\cite{Wang:2006p6704,Wang:2010p6724} Interestingly, this small window appears to be energetically favored in a variational Monte Carlo study of the $J_1$-$J_2$ Heisenberg model,\cite{Tay:2011p7822} which could explain the quantum spin liquid phase discovered in DMRG simulation.\cite{Yan:2011p7804} However, in the doped case, $\langle b_{i\alpha}^{\dagger}b_{i\alpha}\rangle=1-x$ where $x$ is the doping level, suggesting a larger parameter range in which the liquid phase is stabilized. This is also consistent with physical intuition. In the slave-fermion mean-field description, in terms of spin dynamics, doping only means replacing $S=1/2$ by $S=1/2(1-x)$. Therefore doping effectively reduces the spin and increases the effects of quantum fluctuations.

\subsection{Quantum numbers}
After the symmetric wavefunctions are constructed on finite-size lattices, their symmetry quantum numbers can be analytically computed. In Table \ref{tab:irreps} we summarize the quantum numbers of the three competing phases on symmetric samples (see Appendix \ref{app:wf_quantum_number} for details). All wavefunctions are $SU(2)$ spin singlets. We find that on $2N\times2N\times2$ lattices\footnote{For $X\times Y\times2$ lattices, $X$=$Y$ is required to respect point-group symmetry of the honeycomb lattice, and $X$ needs to be an even integer so that $1/4$ doping can be accommodated.}, the ground state wavefunctions of all the three competing phases always form two-fold irreducible representations(irreps) of symmetry group.
\begin{table}
 \caption{ \label{tab:irreps} Two-fold symmetry irreps of the many-body ground state wavefunctions on $2N\times2N\times2$ lattices in the 60$^{\circ}$-rotation eigenbasis. (see Fig.\ref{fig:fermi_surface} for definitions of the symmetry operations.) Table b) also holds for 24-site sample in Fig.~\ref{fig:samples}b.}
 \center{(a): on $4N\times4N\times2$ lattices}
 \begin{tabular}{|c|c|c|}
  \hline
  Sym. & c-SDW or SCCL & d+id SC\\\hline
  Lattice mom. & $\Gamma$ & $\Gamma$\\\hline
  60$^{\circ}$-rot. $C_6$ & $\begin{pmatrix} e^{-\pi i/3} & 0\\ 0 & e^{\pi i/3}\end{pmatrix}$ & $\begin{pmatrix} e^{2\pi i/3} & 0\\ 0 & e^{-2\pi i/3}\end{pmatrix}$\\\hline
  Mirror $\sigma$ & $\begin{pmatrix} 0 & 1\\ 1 & 0\end{pmatrix}$ & $\begin{pmatrix} 0 & 1\\ 1 & 0\end{pmatrix}$\\\hline
  Time-Reveral & $\begin{pmatrix} 0 & 1\\ 1 & 0\end{pmatrix}$ & $\begin{pmatrix} 0 & 1\\ 1 & 0\end{pmatrix}$\\\hline
  Inversion($C_6^3$) & $-1$ & $1$\\\hline
 \end{tabular}\\
 \center{(b): on $(4N+2)\times(4N+2)\times2$ lattices and Fig.~\ref{fig:samples}b}
 \begin{tabular}{|c|c|}
  \hline
  Sym. & c-SDW or SCCL or d+id SC\\\hline
  Lattice mom. & $\Gamma$\\\hline
  60$^{\circ}$-rot. $C_6$ & $\begin{pmatrix} e^{2\pi i/3} & 0\\ 0 & e^{-2\pi i/3}\end{pmatrix}$\\\hline
  Mirror $\sigma$ & $\begin{pmatrix} 0 & 1\\ 1 & 0\end{pmatrix}$\\\hline
  Time-Reveral & $\begin{pmatrix} 0 & 1\\ 1 & 0\end{pmatrix}$\\\hline
  Inversion($C_6^3$) & $1$\\\hline 
 \end{tabular}
\end{table}

In particular, the two degenerate states in the angular momentum basis (rotational symmetry eigenbasis) exactly form time-reversal images of each other. This is a rather special case of time-reversal symmetry breaking phenomena. Although all three competing phases break time-reversal symmetry in the thermodynamic limit, without the analysis of lattice symmetries, naively one may expect that the time-reversal-related two-fold ground state sector is nondegenerate on finite size lattices due to tunneling. Here the quantum tunneling between the two ground states is forbidden by the lattice rotational symmetry.

One may wonder that in the thermodynamic limit, apart from two-fold degeneracy induced by time-reversal symmetry breaking, there should also be a topological order induced degeneracy in the SCCL phase. In fact we will show in Section \ref{sec:SCCL_parton_K_matrix} that there will be four-fold degeneracy induced by topological order and totally we have eight-fold degeneracy. The ground states of SCCL shown in Table \ref{tab:irreps} correspond to a particular one of the four topologically degenerate sectors. The center of mass momentum of the other three sectors is at the three $M_{1,2,3}$ points. We believe that on the  $2N\times2N\times2$ finite lattices the energies of these three sectors are higher than the one shown in Table \ref{tab:irreps}, because only in the sector with the center of mass momentum at $\Gamma$ the minima of spin-1/2 boson dispersion coincide with available momenta in Brillouin Zone; other three sectors are obtained by insertion of $\pi$-fluxes, which moves the momenta away from the position of boson minima and should lead to higher energy (see Appendix \ref{app:wavefunction_construction}).

From Table \ref{tab:irreps} one learns that the 32-site sample is the smallest system allowing a sharp distinction\footnote{On the honeycomb lattice, apart from the $2N\times2N\times2$ samples, there exists a second sequence of finite-size samples respecting the full point group symmetry, and accommodating the 1/4 doping. By tripling the unit cell, namely treating each hexagon in the honeycomb lattice as one unit cell, one can obtain this second sequence as $2N\times2N\times6$ lattices. Among this sequence, although the 6-site sample is very small, the 24-site sample considered here has a reasonable size to investigate the bulk physics.
  % However, we can show that the 24-site sample do NOT sharply distinguish the c-SDW/SCCL or d+id SC phases based on lattice quantum numbers. Please see Appendix \ref{app:24_site_sample} for details.
} between the c-SDW/SCCL phase and the d+id SC phase. However, it is likely that exact diagonalization on the 32-site sample is beyond the currently available computing power. This motivated us to perform the 32-site DMRG calculations in Section \ref{sec:32site_DMRG}.

\section{Numerical simulations}\label{sec:numerical_simulation}

\begin{figure}
 \includegraphics[width=0.48\textwidth]{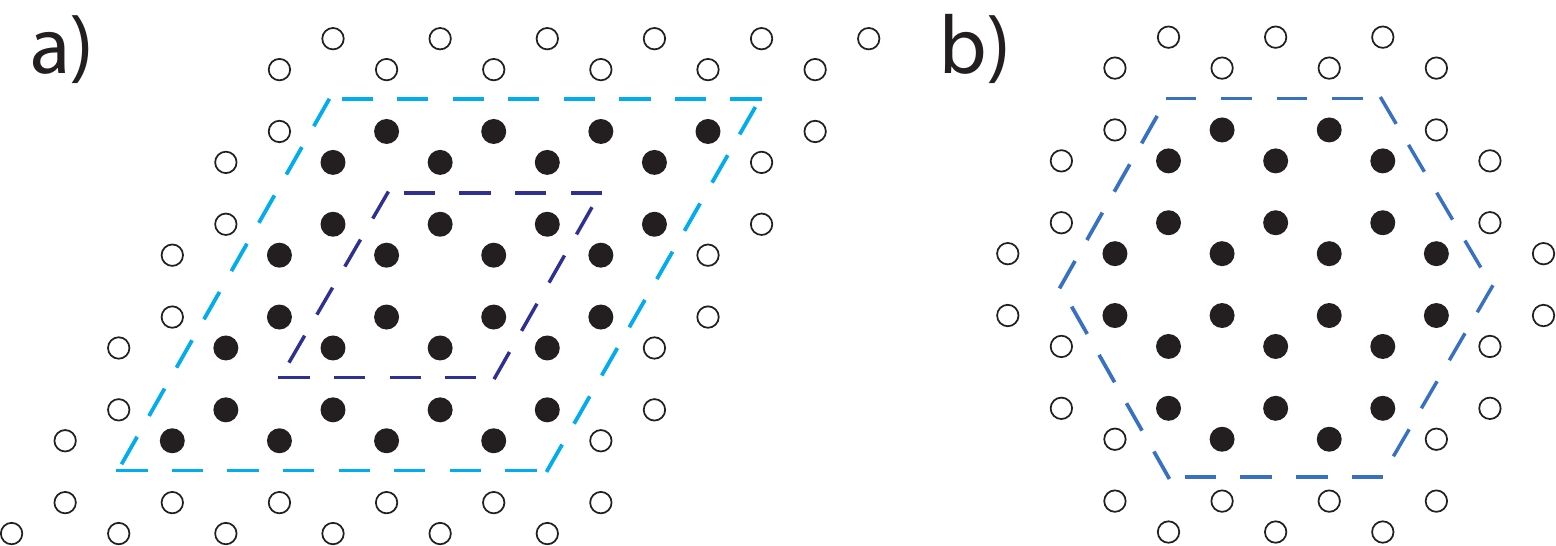}
 \caption{Three samples marked by dashed lines (8-site, 24-site and 32-site) are used in numerical calculations. Periodic boundary conditions are applied for each of them.}
 \label{fig:samples}
\end{figure}

Numerical calculations were performed on three samples shown in Fig.~\ref{fig:samples}, each defined with periodic boundary conditions. In all numerics, $t=1$ is fixed while $J$ or $U$ are varied.

\subsection{Exact diagonalization on the 8-site sample}

\begin{figure}
 \includegraphics[width=0.5\textwidth]{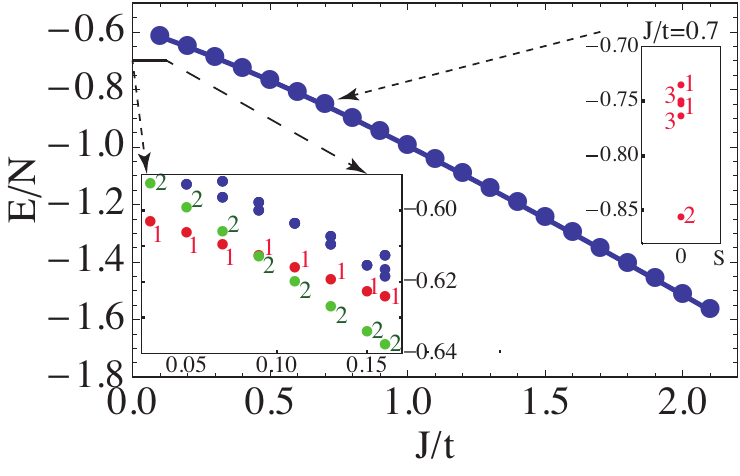}
 \caption{Energy per site of t-J model on 8-site sample obtained by exact diagonalization. Bottom inset: Crossover between two different ground states occurs at $J/t=0.089(1)$. Numbers show the state degeneracy. Top inset: Lowest states with spin zero at fixed $J/t=0.7$.}
 \label{fig:2x2x2_ED}
\end{figure}

\begin{figure}
 \includegraphics[width=0.48\textwidth]{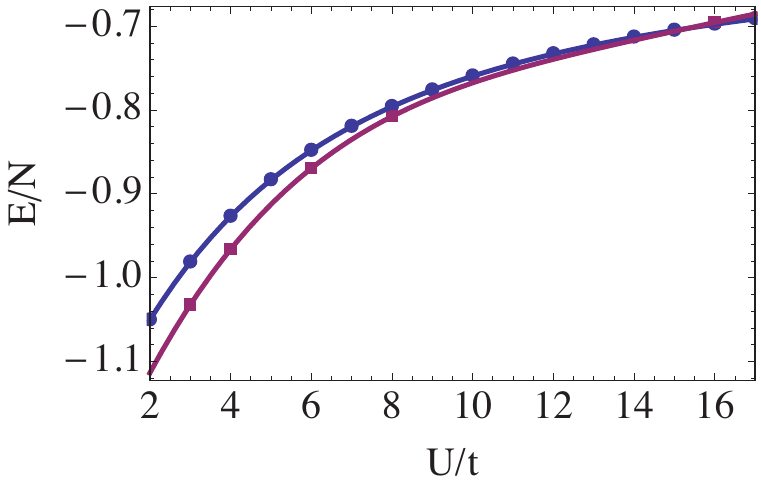}
 \caption{Ground state energy per site of the Hubbard model. Red line: 32-site sample, obtained by DMRG. Blue: 8-site sample, obtained by ED. For 8-site sample ED shows a twofold degenerate ground state.}
 \label{fig:Hubbard_energy_per_site}
\end{figure}

We first describe the results for the t-J model on the 8-site sample (Fig.~\ref{fig:samples}) with 6 fermions.
%Exact diagonalization (ED) was done separately in each (total-)$S_z$ quantum number sector.

Fig.~\ref{fig:2x2x2_ED} shows the ground state (GS) energy throughout the physically interesting parameter regime $0.1\leq J/t\leq 2$.\footnote{This energy is between values for two larger samples, inset of Fig.~\ref{fig:did_relative_energy}.} We find it is twofold degenerate and a spin singlet $S=0$.
% , even up to $J/t=3$.
Evaluating directly the matrix elements $\langle\psi_i|O_{sym}|\psi_j\rangle$ of symmetry operators in the GS doublet $i,j\in\{1,2\}$, we find that the translation, rotation, inversion, mirror and time inversion properties of the GS doublet match the ones shown in Table~\ref{tab:irreps}b to all available digits. The inset shows a more detailed scan revealing a level crossing to a singly degenerate GS below $J/t=0.089(1)$ which forms a trivial irrep of the symmetry group, but the relevance and nature of this very-low-$J$ state will be studied in future work.
%We have also found a (possibly related) transition at around $U/t=65$ in the 32-site sample DMRG GS, at which the GS inversion quantum number changes.

%For such a small sample it is expected that the topologically degenerate states are at relatively high energies. According to the prediction from parton wavefunction (see Section~\ref{sec:modular}) the rest of the topologically degenerate manifold should consist of three states (not counting time reversal degeneracy), each having momentum at an $M$-point, and all having inversion eigenvalue opposite to the case of the identified GS doublet at momentum $\Gamma$ (this is true for any sample).\footnote{Of course, in this 8-site sample the only available momenta are at $\Gamma$ and $M$ points.} Top inset of Fig.~\ref{fig:2x2x2_ED} shows a representative low energy spectrum in $S=0$ sector, for $J/t=0.7$. The first excited triplet state consists of states with $M$ momenta in a representation with inversion equal to minus identity. These could be candidates for part of the topologically degenerate GS.
%Further, the low energy second excited triplet is in the $+1$ inversion representation and certainly does not provide candidates.
% (Similar situation will be discussed below in the case of larger samples.)

Turning to the Hubbard model, the GS energy on this sample is shown in Fig.~\ref{fig:Hubbard_energy_per_site}, in comparison to results for the 32-site sample obtained using the DMRG method. Using ED we find a doubly degenerate ground state in the regime $0<U/t<61.31$, matching the irrep shown in Table~\ref{tab:irreps}b. It is well known that the small-$J$ regime in the t-J model and the large-$U$ regime in the Hubbard model are related by perturbative analysis. Indeed we find that at $U/t>61.31(1)$ the ground state forms a one-dimensional trivial irrep of the symmetry group, which is consistent with the related level crossing in the t-J model at $J/t=0.089(1)$.
%Using ED we find a doubly degenerate ground state, with low lying excitations at $M$-points.The inversion symmetry quantum number of the GS is confirmed using DMRG and it matches Table~\ref{tab:irreps}b for $U/t$ in a very wide range.

Finally, we use ED on this small sample as a benchmark for DMRG calculations which perfectly matched the ED energies.

\subsection{Variational Monte Carlo calculations of the d+id superconductor phase in the t-J model}\label{sec:VMC}
\begin{figure}
 \includegraphics[width=0.48\textwidth]{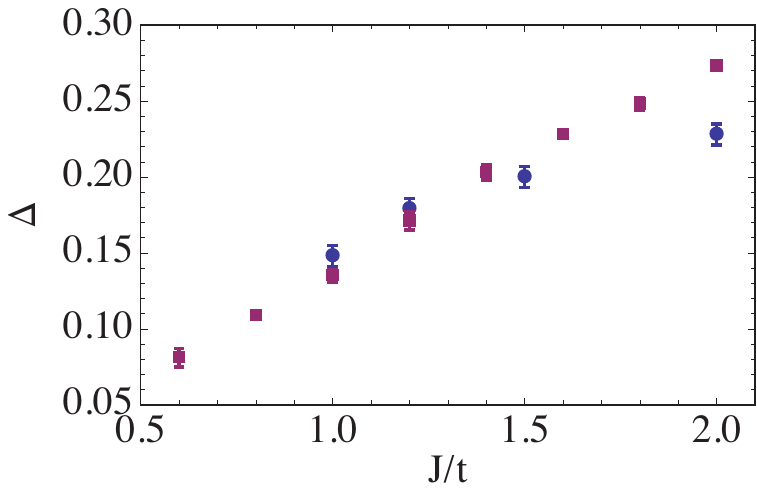}
 \caption{The optimal value of pairing amplitude $\Delta$, which is the only variational parameter in the projected d+id wavefunction, for 24-site (blue disks) and 32-site (red squares) samples.}
 \label{fig:optpairing32}
\end{figure}

\begin{table*}
\begin{minipage}{1.\textwidth}
  \centering
  \begin{minipage}{0.25\linewidth}
    \begin{figure}[H]
      \includegraphics[width=\linewidth]{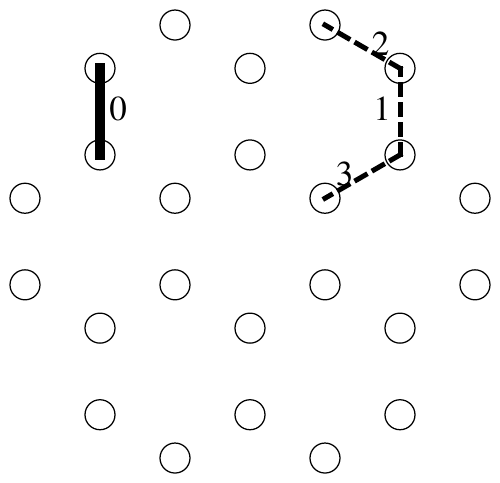}
    \end{figure}
  \end{minipage}
\begin{minipage}{0.6\linewidth}
  \begin{table}[H]
  \begin{tabular}{c|c|c|c|c|c|c}
    $J/t$ &$\frac{Arg(\Delta_{01})}{2\pi}$ & $\frac{Arg(\Delta_{02})}{2\pi}$ & $\frac{Arg(\Delta_{03})}{2\pi}$& $|\Delta_{01}|$ & $\frac{|\Delta_{02}|}{|\Delta_{01}|}$ & $\frac{|\Delta_{03}|}{|\Delta_{01}|}$ \\
    \hline\hline
\multirow{2}{*}{$2.0$}&-0.0025(9)& 0.327(1)& -0.324(1)& 0.00447(2)& 1.04(1)& \
1.05(1)\\
&-0.022(5)& 0.313(6)& -0.311(7)& 0.00073(2)& 0.92(7)& \
0.88(7)\\
    \hline
\multirow{2}{*}{$1.5$}&0.001(1)& 0.322(2)& -0.324(2)& 0.00329(2)& 0.97(2)& \
0.97(2)\\
&0.000(7)& 0.311(7)& -0.310(7)& 0.00062(2)& 0.96(7)& \
0.97(7)\\
\hline
 \multirow{2}{*}{$1.0$}&0.001(2)& 0.296(2)& -0.304(2)& 0.00212(2)& 0.97(2)& \
0.94(2)\\
&0.000(7)& 0.305(8)& -0.306(8)& 0.00050(2)& 1.00(9)& \
0.98(9)\\
\hline
\multirow{2}{*}{$0.78$}&-0.001(2)& 0.289(2)& -0.283(2)& 0.00163(2)& 0.93(3)& \
0.94(3)\\
&0.003(7)& 0.304(9)& -0.296(8)& 0.00048(2)& 1.0(1)&1.0(1)\\
\hline
\multirow{2}{*}{$0.5$}&-0.005(3)& 0.245(3)& -0.235(4)& 0.00116(2)& 0.91(3)& \
0.78(3)\\
&0.02(2)& 0.28(1)& -0.28(1)& 0.00020(2)& 1.5(3)&1.5(3)\\
\hline
$0.2$&-0.008(4)& 0.119(6)& -0.127(6)& 0.00084(2)& 0.75(5)& 0.76(5)\\
\hline
$0.1$&-0.011(4)& 0.065(5)& -0.067(5)& 0.00081(2)& 0.89(5)&0.84(5)\\
\hline\hline
\end{tabular}
\end{table}
\end{minipage}
\end{minipage}
\caption{Pair-pair correlation function in t-J model on 24-site sample, comparing DMRG ground state projected into the $\exp(-i2\pi/3)$ eigenspace of $C_6$ rotation (top values) to VMC result on d+id variational wavefunction (bottom values) in each row. Last two rows are DMRG only. The correlation function $\Delta_{bb'}=\langle\hat{B}_{ij}^\dagger\hat{B}_{kl}\rangle$ is considered for nearest neighbor bond $b=ij$ (labeled 0 in figure) and nearest neighbor bond $b'=kl$ being one of $b'=1,2,3$. To reduce statistical error, the presented value for any of these bond pairs $bb'$ is obtained by averaging over all bond pairs related by translation symmetry.}
  \label{tab:pp}
\end{table*}

The Hilbert space on which $|\Psi_{d+id}(\chi,\Delta)\rangle$ (Eq.~\eqref{eq:psi_did}) is defined is too large for direct computation. Therefore we use the Variational Monte Carlo (VMC) technique, within which the expectation values of observables in this state are calculated using:\cite{Gros:1989p7551,Horsch:2014p7574}
\begin{equation}
  \label{eq:1}
  \langle\phi|\hat{O}|\phi\rangle=\sum_R\frac{|\langle R|\phi\rangle|^2}{\sum\limits_{R'}|\langle R'|\phi\rangle|^2}\frac{\langle R|\hat{O}|\phi\rangle}{\langle R|\phi\rangle},
\end{equation}
where $|\phi\rangle$ is the considered many-body state, while $|R\rangle$ are states in the appropriate Hilbert space which are probabilistically sampled using the first fraction in Eq.~\eqref{eq:1} as the distribution in a Metropolis algorithm. Concretely, the states $|R\rangle$ in the t-J model Hilbert space are given by the spin-occupation basis:
\begin{equation}
  \label{eq:4}
|R(s_1,s_2,...s_N)\rangle_{tJ}\equiv\prod^{N_F}_{s_{i}=\alpha}c_{i\alpha}^{\dagger}|0\rangle,
\end{equation}
where  $N$ is the number of sites, $s_i=\uparrow,\downarrow,0$ depending on whether the site-$i$ is spin-up, spin-down or empty, $c_{i\alpha}$ annihilates electron of spin $\alpha=\uparrow,\downarrow$ at site $i$, and $|0\rangle$ is the vacuum. There are exactly $N_F$ non-empty sites, enforcing the fixed fermion number, and obviously there is no double occupancy. We choose to order the $c_{i\alpha}^\dagger$ operators according to site label $i$, thereby fixing the fermion signs in the $|R(s_1,s_2,...s_N)\rangle_{tJ}$ basis. Similarly, in the Hubbard model we have $s_i=\uparrow\downarrow,\uparrow,\downarrow,0$, and 
\begin{equation}
  \label{eq:5} |R(s_1,s_2,...s_N)\rangle_{Hubbard}\equiv\prod_{s_{j}=\uparrow\downarrow}c_{j\uparrow}^{\dagger}c_{j\downarrow}^{\dagger}\prod_{s_{i}=\alpha}c_{i\alpha}^{\dagger}|0\rangle,
\end{equation}
where again there are in total exactly $N_F$ operators $c_{i\alpha}^\dagger$, and in the obtained $|R(s_1,s_2,...s_N)\rangle_{Hubbard}$ we order them according to site label $i$, keeping the $c_{i\uparrow}^\dagger$ before the $c_{i\downarrow}^\dagger$ for each doubly occupied site $i$.

We focus on the total $S_z$ equal to zero sector (in both models), by additionally choosing an equal number of spin-up and spin-down electrons. Note that the DMRG calculation conserves this spin quantum number of a state, so we can work in an $S_z$ sector. As discussed in detail in the following, we also measured quantities after projecting the wavefunction to a certain symmetry sector using a projector $P$, and note here that both the action of the operator $\hat{O}$ and $P$ are dealt with by acting directly on the $\langle R|$ in Eq.~\eqref{eq:1}.

The optimal value of the single variational parameter, the pairing $\Delta/\chi\in\mathbb{R}$, which minimizes the variational energy, is shown in Fig.~\ref{fig:optpairing32}. For smaller $J/t$ the pairing is too small and harder to determine precisely.

The energy of wavefunction with optimal pairing is compared to DMRG ground state on 24-site and 32-site samples in Fig.~\ref{fig:did_relative_energy}, showing that the d+id variational state captures between $97\%$ and $99\%$ of DMRG GS energy throughout the d+id phase.

The main signature of the d+id phase is the complex phase of pairing, Fig.~\ref{fig:cSDW_dpid_real_space}b. We therefore calculate the pair-pair correlation function:
\begin{equation}
  \label{eq:2}
  \langle\hat{B}^\dagger_{ij}\hat{B}_{kl}\rangle,\text{ with } \hat{B}_{ij}\equiv c_{i\uparrow}c_{j\downarrow}-c_{i\downarrow}c_{j\uparrow}
\end{equation}
the singlet pairing. The pattern from Fig.~\ref{fig:cSDW_dpid_real_space}b should be revealed in the long-range physics, so the most interest lies in pairs of nearest-neighbor bonds $ij$ and $kl$ which are as far from each other as possible. Table~\ref{tab:pp} reveals that the pattern indeed occurs and becomes weaker with decreasing $J/t$.

The spin-spin correlation function is very short-ranged as expected, so we do not present it in detail.\footnote{For instance, on the 32-site sample the ratio of correlation between farthest sites and nearest neighbor sites is typically around 5 times smaller than in the DMRG state in c-SDW/SCCL state (Fig.~\ref{fig:szszfar}).}
%\begin{figure}
% \includegraphics[width=0.5\textwidth]{SzSzdidJ1VMC}
% \caption{Spin-spin correlation in variational d+id state calculated using VMC for $J/t=1$.}
% \label{fig:szszdidJ1}
%\end{figure}

\subsection{DMRG simulations on the 32-site sample}
\label{sec:32site_DMRG}

\begin{table*}
  \begin{minipage}{0.9\linewidth}
  \begin{table}[H]
  \begin{tabular}{c|c|c|c|c|c|c|c|c|c}
    $J/t$ &0.1&0.2& 0.25& 0.5& 0.78& 0.82& 1.0& 1.5& 2.0\\
    \hline\hline
$\langle\textrm{Inv}\rangle$&-1.0000(3)&-1.0000(6)& -1.0000(4)& -1.0000(5)&-0.9998(5) &1.0000(6)& 0.9999(4)& 1.0000(4)&0.9998(6)\\
%\hline\hline
\end{tabular}
%\caption{\label{tab:inv32} Expectation value of inversion operator in DMRG ground state of t-J model on the 32-site sample.}
\end{table}
\end{minipage}
\\
%\begin{minipage}{0.75\textwidth}
  \begin{minipage}{0.9\linewidth}
  \begin{table}[H]
  \begin{tabular}{c|c|c|c|c|c|c|c}
    $J/t$ &0.1& 0.2& 0.5& 0.78& 1.0& 1.5& 2.0\\
    \hline\hline
$\langle\textrm{Inv}\rangle$&0.9996(3)& 0.9986(6)& 0.999(5)&0.9995(3)& 0.9978(7)& 0.9990(4)& 0.9996(3)\\
%\hline\hline
\end{tabular}
%\caption{\label{tab:inv24} Expectation value of inversion operator in DMRG ground state of t-J model on the 24-site sample.}
\end{table}
\end{minipage}
\\
\begin{minipage}{0.9\linewidth}
  \begin{table}[H]
    \begin{tabular}{c|c|c|c|c|c|c|c|c}
%      \hline\hline
    $U/t$ &1.0&2.0&3.0& 4.0& 6.0& 8.0& 16.0& 40.0\\
    \hline\hline
$\langle\textrm{Inv}\rangle$&-0.997(2)&-0.9994(6)&-0.994(4)& -0.9999(7)& -1.0000(3)&-1.0000(4)& -1.0000(2) & -0.9996(5)\\
%\hline\hline
\end{tabular}
%\caption{\label{tab:inv32U} Expectation value of inversion operator in DMRG ground state of Hubbard model on the 32-site sample.}
\end{table}
\end{minipage}
\caption{\label{tab:inv} Expectation value of inversion operator in DMRG ground state. (Top) t-J model on the 32-site sample; (Middle) t-J model on the 24-site sample; (Bottom) Hubbard model on 32-site sample.}
%\end{minipage}
\end{table*}

\begin{figure*}
\includegraphics[width=0.9\linewidth]{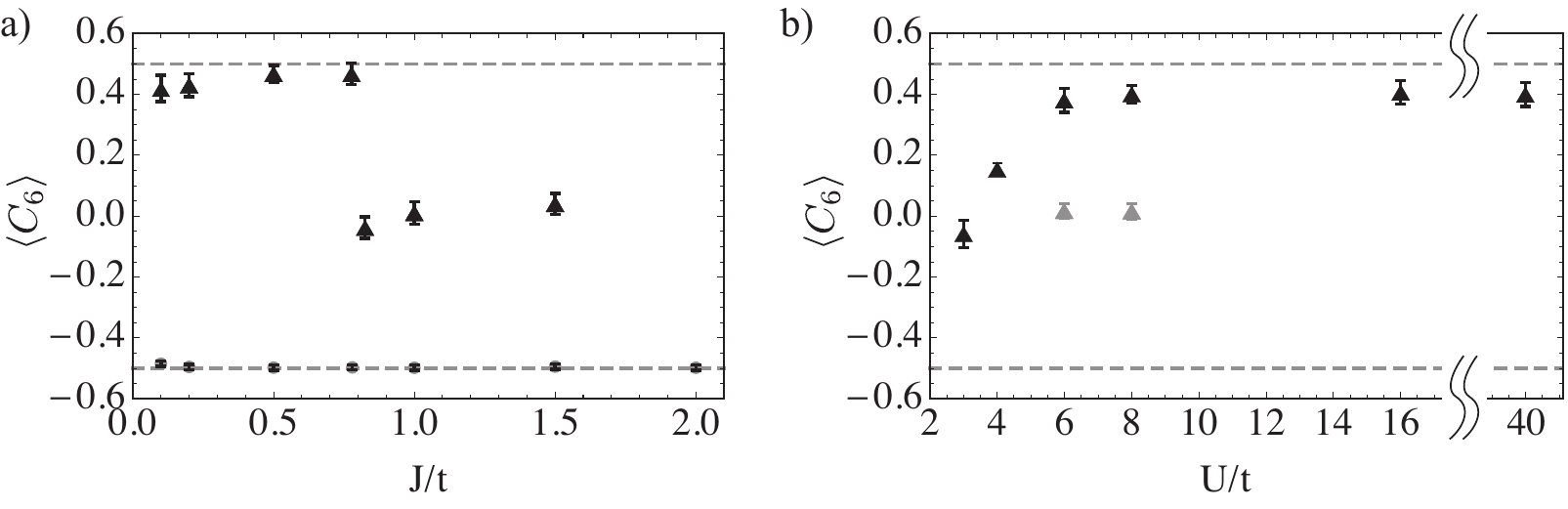}
%\begin{minipage}{1.\textwidth}
%  \centering
%  \begin{minipage}{0.45\linewidth}
%\begin{figure}[H]
%\includegraphics[width=\linewidth]{C6J}
%\end{figure}    
%\end{minipage}
%\hspace{0.08\textwidth}
 % \begin{minipage}{0.45\linewidth}
%\begin{figure}[H]
% \includegraphics[width=\linewidth]{C6U}
%\end{figure}
%\end{minipage}
%\end{minipage}
 \caption{ \label{fig:c6} Expectation value of 60-degree rotation operator ($C_6$) in DMRG ground state: a) t-J model on 32-site sample (triangles) with projection of wavefunction to center of mass momentum $\Gamma$, and on 24-site sample (disks) without projection. b) Hubbard model on the 32-site sample, with (black triangles) and without (smaller gray triangles) projection to $\Gamma$ momentum.}
\end{figure*}

We have used DMRG to obtain the ground state (GS) of the periodic 32-site sample (Fig.~\ref{fig:samples}). Our calculation is based on the open-source DMRG software package ITensor,\cite{White:1992p7946,itensor} where the periodic two-dimensional shape of our samples was implemented simply by introducing long-range hopping (of same size $t$) in the native DMRG one-dimensional representation of the system. The limit on dimension of MPS matrices was between 10.000 and 11.000. We find truncation errors around $(2\sim 7)\cdot 10^{-4}$, depending on model and parameter regimes. Although such error values seem too large in view of general DMRG performance, in this work we found it appropriate to apply a different physical criterion for convergence, namely, that the expectation values of symmetry transformations allow a clear assignment of quantum numbers to the ground state; additionally, when appropriate, in measurements we projected the GS to a sector having some quantum numbers fixed, to effectively get closer to the true GS. This approach will be described in detail below. Appendix~\ref{app:dmrg} presents further details on our DMRG setup and convergence.

Focusing first on the t-J model, we find a very precise quantization of the inversion operator expectation value in the GS, as shown in Table~\ref{tab:inv}. For this sample there is a sharp transition at $J/t=0.80(2)$ at which the low-$J$ ground states (blue phase in Fig.~\ref{fig:phase_diagrams}), having inversion $-1$, switch to high-$J$ ground states (red in Fig.~\ref{fig:phase_diagrams}), which are in the $+1$ representation of inversion. Due to change of symmetry quantum numbers, we expect this to be a first order phase transition in thermodynamic limit. Given that GS is in a representation having inversion $+1\,(-1)$, and since there is no reason for additional degeneracy except due to time-reversal, the 60-degree rotation operator ($C_6$, with $C_6^3=$Inversion) should be represented by one of numbers $\{1,\exp(i 2\pi/3),\exp(-i 2\pi/3)\}\, (\{-1,\exp(i\pi/3),\exp(-i\pi/3)\})$.

A crucial subtlety here is that the DMRG calculation automatically provides a \textit{real-valued} wavefunction for our real Hamiltonians. This DMRG wavefunction will be denoted as $|\psi\rangle$ in the following discussion. If $|\psi\rangle$ gives the converged true ground state, it must be an equal superposition of two conjugate partners in a two-dimensional irrep when $C_6$ is represented by a complex number. Simple calculation shows that generally the $C_6$ expectation value for a converged real ground state wavefunction must be one of $\{-1,-1/2,1/2,1\}$, corresponding to the four possible irreps of the symmetry group respectively: the $C_6$-odd one-dimensional irrep, the two-dimensional irrep as shown in Table~\ref{tab:irreps}b, the two-dimensional irrep as the c-SDW/SCCL shown in Table~\ref{tab:irreps}a, and the trivial one-dimensional irrep. Note that the DMRG we applied here can be viewed as a variational wavefunction technique in real space, in which lattice symmetry is not implemented at all.

Therefore we use the $C_6$ expectation value as a physical criterion for successful convergence of the DMRG wavefunction. Namely, if $\langle\Psi|C_6|\Psi\rangle$, with $|\Psi\rangle$ defined shortly, is found to be one of the four values: $\{-1,-1/2,1/2,1\}$, the DMRG has successfully converged. On the 24-site sample (see Sec.\ref{sec:24site_DMRG}), we find that using $|\Psi\rangle=|\psi\rangle$ the $C_6$ expectation value is well converged in the parameter regimes of interest. However on the 32-site sample, in order to improve convergence, we project $|\psi\rangle$ to the sector with center of mass momentum equal to $\Gamma$; namely, we use $|\Psi\rangle=P_{\Gamma}|\psi\rangle$ as the wavefunction in MC measurement of the $C_6$ expectation value, Eq.~\eqref{eq:1}, where $P_{\Gamma}$ is the projection operator into the $\Gamma$-sector. (We also check that $|\psi\rangle$ has a big portion in the $\Gamma$-sector for all parameter values so that this projection is not creating unphysical artifacts.)

Fig.~\ref{fig:c6}a demonstrates the result that in the low-$J$ regime ($0.1<J/t<0.8$), the rotation expectation value is indeed consistent with $1/2$ on the 32-site sample. Therefore the GS irrep in this regime is the same as the c-SDW/SCCL phase as shown in Table~\ref{tab:irreps}a. However, for the lowest values, $0<J/t< 0.1$, the $\langle C_6\rangle$ does not converge to either of $\{-1,-1/2,1/2,1\}$, and this also happens for the 24-site sample for $0<J/t< 0.07$; on the other hand, the 8-site exact diagonalization shows a singlet ground state for $0<J/t<0.089(1)$. All this evidence suggests the existence of a different quantum phase in this lowest $J$ regime. Given that such lowest $J$ regime is not the most interesting for correlated materials, we leave it for future work, and focus exclusively on values $J/t>0.1$.

In the high-$J$ regime ($J/t>0.8$) unfortunately the $\langle C_6\rangle$ is close to zero and far from any of the $\{-1,-1/2,1/2,1\}$, which indicates that the 32-site DMRG GS for $J/t>0.8$ has not converged well enough; it cannot give reliable information about correlations. Nevertheless the inversion quantum number for $J/t>0.8$ is found to be accurately $+1$, consistent with the d+id SC and sharply distinguished from the $0.1<J/t<0.8$ value $-1$ (see Table\ref{tab:inv}). In the following discussion and in the next Section, the $J/t>0.8$ phase is actually confirmed to be the d+id SC using complementary variational Monte Carlo results as well as DMRG on the smaller 24-sample which has no such issues with convergence.

%\begin{figure}
% \includegraphics[width=0.3\textwidth]{energypersite}
% \caption{DMRG ground states energy per site of t-J model on all three samples, compared to VMC calculation on the projected d+id wavefunction. Black, blue and red lines represent the DMRG results for 8-site, 24-site and 32-site sample, respectively. The yellow line is the VMC result.}
%\label{fig:energy_per_site}
% \end{figure}

\begin{figure}
 \includegraphics[width=0.5\textwidth]{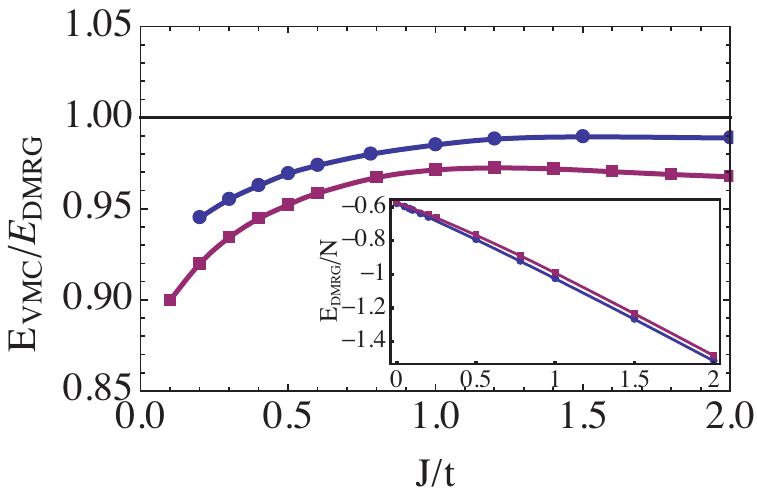}
 \caption{The lowest ground state energy of projected d+id variational wavefunction, obtained by VMC, is shown as a fraction of the DMRG ground state energy on the same sample. Blue line is for the 24-site, red line is for 32-site sample. Inset: The DMRG energy per site of 24-site (blue) and 32-site (red) samples.}
 \label{fig:did_relative_energy}
\end{figure}

Energetics of the DMRG GS of t-J model are shown in Fig.~\ref{fig:did_relative_energy}. The energy of the single-parameter variational wavefunction discussed in Section~\ref{sec:did} is quantitatively compared to the DMRG energy, showing that the d+id candidate wavefunction captures more than $97\sim 99\%$ of DMRG GS energy throughout the high-$J$ phase. In addition, the energy of the d+id variational state deviates significantly in the low-$J$ phase.

To further identify the nature of the DMRG GS, we consider spin-spin and pair-pair correlation functions. The expectation values are obtained using the Monte Carlo (MC) technique, Eqs.~\eqref{eq:1},~(\ref{eq:4}),~(\ref{eq:5}), using between 300 and 1000 MC measurements with 40 MC steps between each measurement and with a 500 MC step thermalization. Further, the measurements are averaged across 64 independent MC runs, and the measurement errors in this paper represent the error of the mean. To correctly calculate observables we need to choose a particular rotation sector from the $|\Psi\rangle$, since the DMRG mixes rotation sectors by selecting a real wavefunction as discussed above. According to Table~\ref{tab:irreps} this projection to a rotation eigenstate means breaking the time-reversal symmetry, which should naturally happen in the thermodynamic limit. For all measurements on this 32-site sample, in the phase with inversion $-1$ we choose the $\exp(-i\pi/3)$ sector of $C_6$. (Note that for the $+1$ inversion phase ($J/t>0.8$) in the t-J model the $\langle C_6\rangle$ is not converged, so we do not use it.) More precisely, the correlation functions we study next are obtained as $\langle\psi|P_\Gamma P_{C6}\hat{O}P_\Gamma P_{C6}|\psi\rangle$ in Eq.~\eqref{eq:1}, with $P_{C6}$ projecting into the desired rotation eigenspace.
%We note when additional averaging using lattice symmetries was applied.

%Since the GS is expected to have $\Gamma$ momentum, we performed calculations of observables with including a projection to $\Gamma$ momentum, i.e., using $\langle\psi|P_\Gamma \hat{O}P_\Gamma|\psi\rangle$ in Eq.~\eqref{eq:1}.

\begin{figure*}
\includegraphics[width=1.0\textwidth]{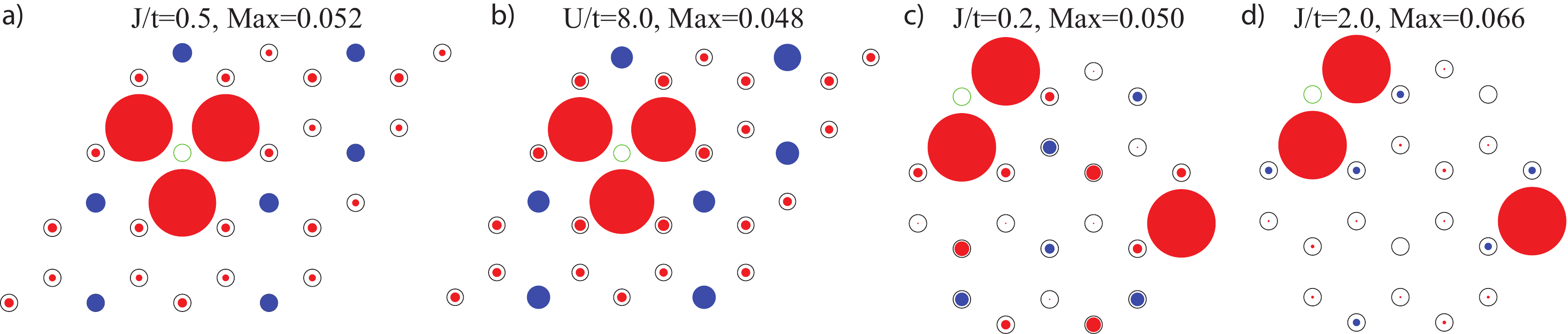}
\caption{Spin-spin correlation function $\langle S_z(i)S_z(j)\rangle$ measurement. Blue is positive, red negative, and disk radius is proportional to amplitude. Site $i$ is fixed at green circle, while every bond $i,j$ is averaged over translations and rotations to increase the number of sampled observable values in MC and thereby reduce statistical error. ``Max'' labels absolute amplitude of largest shown disk. The measurement on 32-site sample uses the DMRG ground state projected into sector with center of mass momentum $\Gamma$ and $C_6$ rotation eigenvalue $\exp(-i\pi/3)$: (a) t-J model, $J/t=0.5$, (b) Hubbard model, $U/t=8$. The spin-spin correlation is longer-ranged and consistent with tetrahedral pattern throughout c-SDW/SCCL phase. On 24-site sample the t-J model DMRG ground state is projected into sector with $C_6$ eigenvalue $\exp(-i2\pi/3)$. (c) Same correlation behavior is found deep in small-$J$ regime of 24-site sample, while (d) Spin pattern is lost in large-$J$ regime, even as short-range correlations grow.}
\label{fig:szszJcolor}
\end{figure*}

\begin{figure*}
 \includegraphics[width=1.0\textwidth]{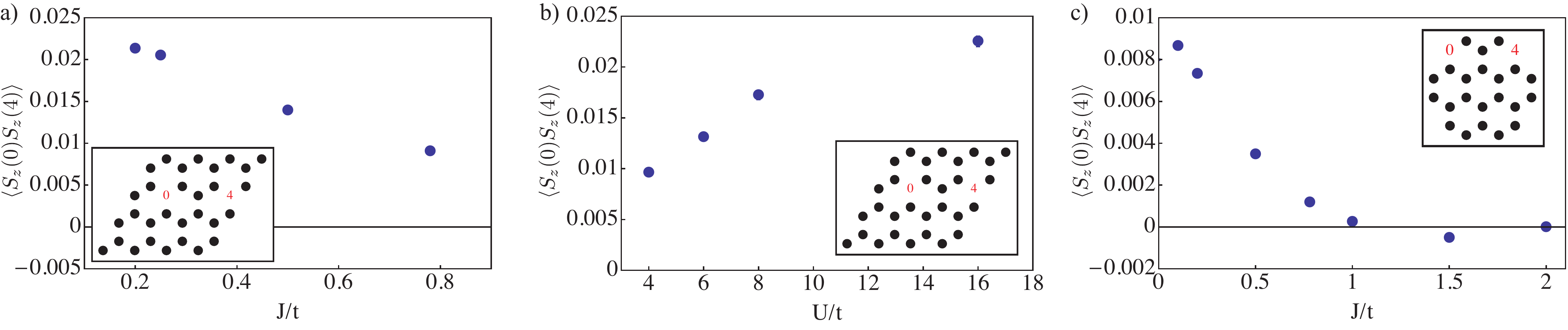}
 \caption{Value of spin-spin correlation function $\langle S_z(i)S_z(j)\rangle$ for farthest pair of sites $i,j$ (see insets), with DMRG ground state projected into sector having center of mass momentum $\Gamma$, and $C_6$ eigenvalue $\exp(-i\pi/3)$ on 32-site sample: (a) t-J model, (b) Hubbard model. (c) On 24-site sample the t-J model DMRG ground state is projected to $\exp(-i2\pi/3)$ eigenvalue sector of $C_6$: the spin correlation vanishes with crossover to d+id-like regime. Averaging over translationally and rotationally related pairs is included in all measurements to reduce the statistical error. The correlation consistently grows throughout c-SDW/SCCL phase with larger $U$ (smaller $J$).}
 \label{fig:szszfar}
\end{figure*}

\begin{figure*}
 \includegraphics[width=0.9\textwidth]{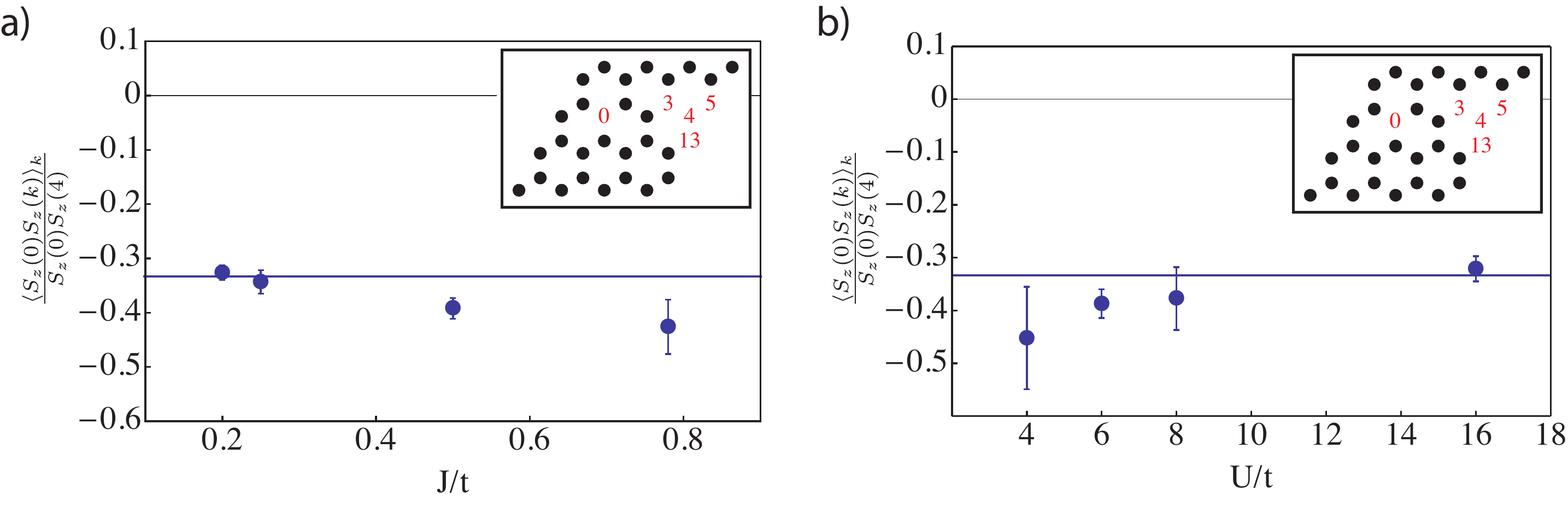}
 \caption{Nature of spin-spin correlation pattern $\langle S_z(i)S_z(j)\rangle$ in the c-SDW/SCCL phase, with DMRG ground state of 32-site sample projected into sector having center of mass momentum $\Gamma$, and $C_6$ eigenvalue $\exp(-i\pi/3)$: a) t-J model, b) Hubbard model. Three values of correlation with sites chosen as $i=0$, $j=3,5,13$ (see inset) are averaged, and that average is divided by correlation between $i=0,j=4$. Tetrahedral pattern predicts this ratio to be $-1/3$ (blue line). Correlation for each site pair $i,j$ is averaged over all translationally related pairs to reduce statistical error.}
 \label{fig:szszratios}
\end{figure*}

We calculate the spin-spin correlation function by setting the observable $\hat{O}=S_z(i) S_z(j)$ with some sites $i,j$. Although the short-range physics dictates that nearest neighbor $i,j$ correlations grow with $J/t$ (this is indeed observed), we are interested in long-range physics and therefore choose the farthest pair $i,j$, Fig.~\ref{fig:szszfar}a, finding that this correlation grows with going deeper into the low-$J$ phase.
%The correlation function is additionally averaged over all translation and rotation images of the $i,j$ bond.

Fig.~\ref{fig:szszJcolor} demonstrates the spin-spin correlation pattern for $i,j$ bonds of all lengths, revealing a pattern consistent with tetrahedral spin correlations (Fig.~\ref{fig:cSDW_dpid_real_space}a) in the low-$J$ phase. The overlap of spin vectors in the tetrahedron predicts a ratio of $-1/3$ in the correlation when the spins at sites $i,j$ are parallel compared to when they are not. Our measurement of this ratio for the farthest possible site pairs $i,j$ is consistent with the prediction, Fig.~\ref{fig:szszratios}a.

\begin{figure}
 \includegraphics[width=0.45\textwidth]{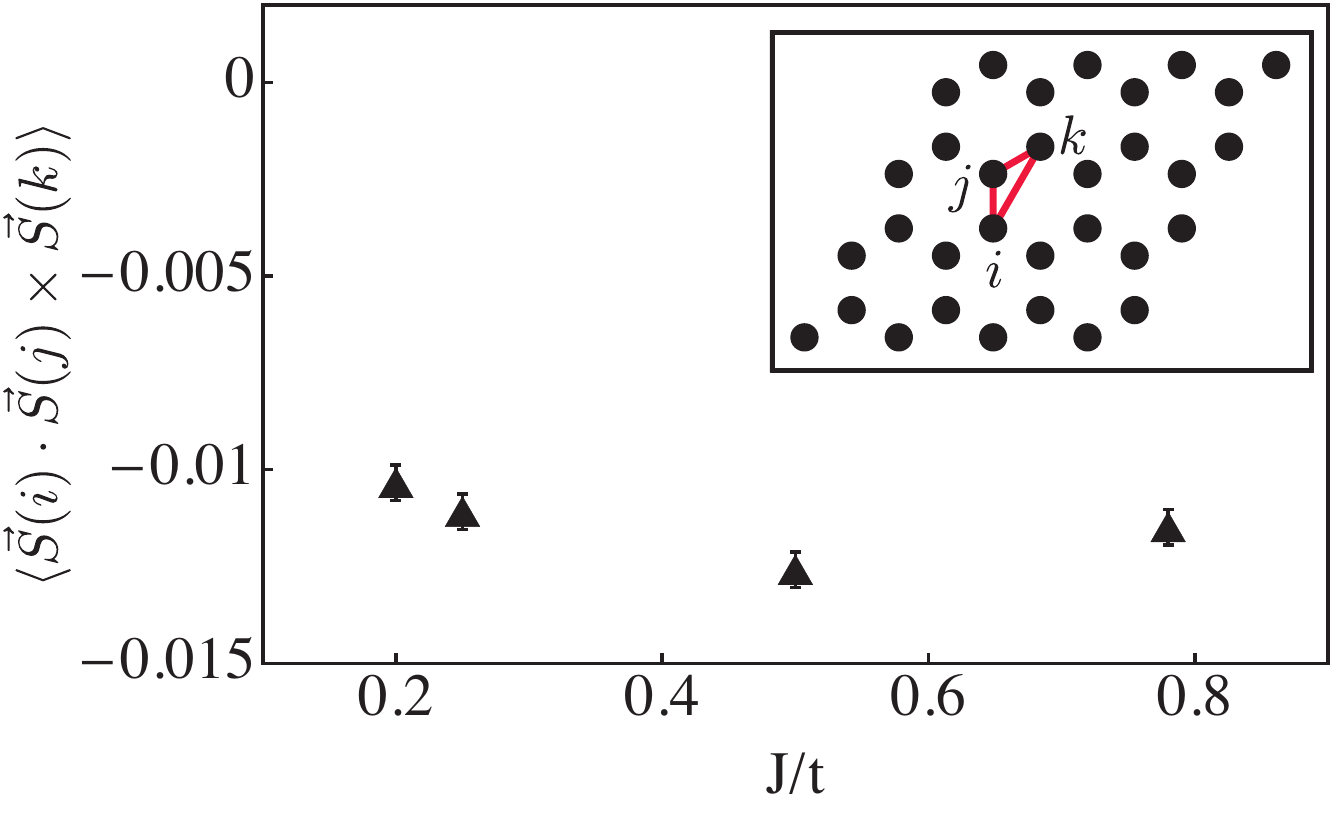}
 \caption{Spin chirality of a characteristic triangle (inset) in 32-site sample, from t-J model DMRG ground state projected into sector having center of mass momentum $\Gamma$, and $C_6$ eigenvalue $\exp(-i\pi/3)$. The chirality is averaged over all translations and rotations of the triangle to reduce statistical error.}
 \label{fig:chir32}
\end{figure}

% We also calculate the spin chirality for two characteristic triangles in the honeycomb lattice, see inset of Fig.~\ref{fig:chir32}. The magnitude of chirality $\vec{S}_i\cdot\vec{S}_j\times\vec{S}_k$ is consistent with magnitude of spin-spin correlation, leading to the very small chirality value for the larger triangle. The triangles are averaged over translations and rotations on the lattice. We also check that projecting to opposite rotation sector (e.g. $\exp(+i\pi/3)$ in low-$J$ phase) reverses the sign of chirality.
We also calculate the spin chirality $\vec{S}_i\cdot(\vec{S}_j\times\vec{S}_k)$ for the smallest triangle in the honeycomb lattice, see Fig.~\ref{fig:chir32}. The magnitude of chirality of around $0.01$ is consistent with magnitude of nearest neighbor spin-spin correlation of $\sim0.05$.\footnote{We also check that projecting to opposite rotation sector (e.g., $\exp(+i\pi/3)$ in low-$J$ phase) reverses the sign of chirality.}
% The triangles are averaged over translations and rotations on the lattice.

In the low-$J$ phase, where spin indicates the c-SDW/SCCL state, the pair-pair correlation function is extremely short-ranged and beyond nearest bond pairs hard to distinguish from zero within our numerical precision (see Appendix~\ref{app:dmrg}).

Let us now turn to the Hubbard model on the 32-site sample, having ground state energy presented in Figure~\ref{fig:Hubbard_energy_per_site}. Table~\ref{tab:inv} demonstrates our result that for a very wide range of parameters $1<U/t<40$ the expectation value of inversion operator is very accurately quantized to $-1$. Figure~\ref{fig:c6}b shows that for all $U/t\gtrsim 6$ we find a satisfying agreement of 60-degree rotation expectation value $\langle C_6\rangle$ with $+1/2$.
% , indicating an equal superposition of $\exp(\pm i\pi/3)$ rotation sectors.
The same figure shows the influence of projection to $\Gamma$ momentum, i.e., using $|\Psi\rangle=P_{\Gamma}|\psi\rangle$, which significantly improves this agreement.
% With the help of $\Gamma$ momentum projection, we can consider even these low $U$ DMRG states as well converged to the true GS.
It is not surprising that convergence worsens for low $U/t$, due to the existence of many low energy states, but we believe it is limited by our maximal available $m$. For instance, at $U/t=4$ the $C_6$ expectation with projection to $\Gamma$ momentum improves from $0.10(2)$ at $m=8.000$ to $0.16(2)$ at $m=10.500$ (see Appendix~\ref{app:dmrg}).
In fact, using degenerate perturbation theory on the 32-site sample around free electron state $t=1,U=0$ (Appendix~\ref{app:perturbation}), we find the same quantum numbers as for the c-SDW/SCCL state. We therefore expect that the c-SDW/SCCL ground state quantum number persists through the whole range $0<U/t<40$ on this sample.

%Therefore, to ensure our criterion for convergence is obeyed even for the low $U/t<5$, the following correlation measurements were obtained using projection to $\Gamma$ momentum together with projection to $\exp(-i\pi/3)$ sector, i.e., by using $\langle\psi|P_\Gamma P_{C6=-\pi/3}\hat{O}P_\Gamma P_{C6=-\pi/3}|\psi\rangle$.

The spin-spin correlation function (again, for this sample we take $P_\Gamma P_{C6}|\psi\rangle$ in the $\exp(-i\pi/3)$ sector of $C_6$) throughout the entire well-converged and physically interesting regime $U/t\gtrsim 4$ is qualitatively the same as in the c-SDW/SCCL phase of t-J model, Fig.~\ref{fig:szszJcolor}a (see also Appendix~\ref{app:dmrg}). Quantitatively, Figs.~\ref{fig:szszfar}c,~\ref{fig:szszratios}b show how the long-range tetrahedral spin pattern describes this phase very well, and strengthens with growing $U/t$. This is consistent with the mapping between low-$J$ and large-$U$ models, confirming the c-SDW/SCCL nature of the phase in both models.

\subsection{DMRG simulations on the 24-site sample}
\label{sec:24site_DMRG}

The fully symmetric 24-site sample, Fig.~\ref{fig:samples}b, is large enough to provide some longer-range physics information, but small enough to allow excellent DMRG convergence and precise measurements (see general discussion of our DMRG convergence criteria in previous Section). It may even be suitable for exact diagonalization numerical simulations using currently available computing power. We therefore investigated the quantum numbers of the three competing states, c-SDW, SCCL and d+id superconductor, on this sample, and found that unfortunately all these phases share the same quantum numbers as in Table \ref{tab:irreps}b. %Therefore exact diagonalizations on this sample cannot sharply distinguish them.
Therefore, a smooth crossover takes place in the t-J model. To support the claim that the high-$J$ phase observed on the 32-site sample is the d+id SC, in this section we will consider the t-J model on the 24-site sample and show that it clearly exhibits a change in its correlation properties from the characteristic c-SDW/SCCL to the d+id SC behavior as $J/t$ is increased within the $0.1<J/t<2$ parameter region. (We will not discuss the Hubbard model on this sample.)
% even though our 32-site DMRG result is less reliable in this regime

Table~\ref{tab:inv} shows the very precise quantization of inversion to $+1$ in the DMRG GS in the entire region $0.1<J/t<2$ (as explained in previous section, we do not further discuss the $0<J/t< 0.1$). The GS is almost entirely in the $\Gamma$ momentum sector, so we use $|\Psi\rangle=|\psi\rangle$ and find that $\langle C_6\rangle$ is very close to $-1/2$ (Fig.~\ref{fig:c6}a) in the entire considered parameter region. This corresponds to quantum numbers in Table~\ref{tab:irreps}b. The energetics in Fig.~\ref{fig:did_relative_energy} shows $99\%$ agreement with variational d+id wavefunction at larger $J/t$, which significantly worsens as we go to lower $J/t$, indicating the crossover to c-~SDW/SCCL state.

Due to smaller sample size and the fact that $\Gamma$ momentum projection is unnecessary, we could use 10.000 MC measurements in correlation functions, significantly reducing the statistical error. The correlation measurements are all done in the $\exp(-i 2\pi/3)$ rotation sector, corresponding to the $+1$ value of inversion.

Figs.~\ref{fig:szszJcolor}c,d contrast the spin-spin correlation at $J/t=0.2$ and $2$, respectively. The former is clearly consistent with tetrahedral spin correlations. On the other hand, the $J/t=2$ case exemplifies a completely different, and much shorter ranged, spin correlation pattern. Fig.~\ref{fig:szszfar}c quantifies the weakening of the tetrahedral pattern, which rapidly drops to zero with $J/t$ growing towards 1, indicating the existence of the crossover.

Complementary information is found in the pair-pair correlation function (see Eq.~\eqref{eq:2}), presented in Table~\ref{tab:pp}. At largest value, $J/t=2$, the correlation pattern matches the ideal pattern of Fig.~\ref{fig:cSDW_dpid_real_space}b with percent precision. By the time we reach the lowest value $J/t=0.1$, the overall correlation amplitude drops fivefold, the different pairs' correlation varies in amplitude significantly, and their relative phase of $2\pi/3$ drops to $0.07\cdot 2\pi$. The table shows that these results match the evolution of pair-pair correlation in the variational d+id wavefunction, up to an overall amplitude difference in the correlation function. Altogether, the existence of crossover between c-SDW/SCCL and d+id SC in the t-J model on this sample is clearly confirmed.

\section{The Spin-charge-Chern liquid}\label{sec:SCCL}
\subsection{Low energy effective theory: Parton construction and the K-matrix formulation}\label{sec:SCCL_parton_K_matrix}

In two spatial dimensions, a description of Abelian topological order can be given by Abelian $U(1)^N$ Chern-Simons theory.\cite{Wen:1992p6753,Girvin:1987p8027,Zhang:1989p4231} The low energy effective Lagrangian relevant for us has the following generic form
\begin{align}
  \mathcal{L}_{CS}=\frac{\varepsilon_{\mu\nu\lambda}}{4\pi}\sum_{I,J=1}^{N} a_{\mu}^IK_{I,J}\partial_{\nu}a_{\lambda}^{J}
  \label{K_matrix_Lagrangian}
\end{align}
where $\mu,\nu,\lambda = 0,1,2$ in 2+1D and summation over repeated indices is implied. $\mathbf{K}$ is a symmetric $N\times N$ matrix with integer entries. A quasiparticle in this theory is described by an $N$ component integer vector $\mathbf{l}$, whose components determine the $N$ $U(1)$ gauge charges of the excitation. The particle couples to internal gauge field $a_{\mu}$ as $-a_{\mu}^Il_Ij^{\mu}$. Here, $j_{\mu}$ is the 3-current for a single quasiparticle.

The quasiparticle statistics can be easily read out by integrating out $a_{\mu}^I$. The self(exchange) statistics of a quasiparticle $\mathbf{l}$ is given by its statistics angle
\begin{align}
  \theta_\mathbf{l}=\pi \mathbf{l}^{t}\mathbf{K}^{-1}\mathbf{l},
  \label{self_theta}
\end{align}
while the mutual(braiding) statistics of a quasiparticle $\mathbf{l}$ and $\mathbf{l'}$ is characterized by
\begin{align}
  \theta_{\mathbf{l},\mathbf{l'}}=2\pi \mathbf{l}^t\mathbf{K}^{-1}\mathbf{l'}.
  \label{mutual_theta}
\end{align}

Quasiparicles generally have anyonic statistics and are thus nonlocal. However, there is a special type of quasiparticle $\mathbf{\tilde{l}}=\mathbf{Kl}$, where $\mathbf{l}\in\mathbb{Z}^N$. $\mathbf{\tilde{l}}$ is mutual boson to all other quasiparticles, so it can be viewed as a local excitation, in the topologically trivial sector. Examples include electron excitations of fractional quantum Hall systems and spin-1 magnons in $Z_2$ spin liquids. Two quasiparticles whose difference is in the trivial topological sector should be considered as being in the same topological sector. %Self statistics of $\mathbf{\tilde{l}}$ depends on the diagonal entries of $\mathbf{K}$. If diagonal elements are all even numbers, $\mathbf{\tilde{l}}$ must be self boson and (\ref{K_matrix_Lagrangian}) describes a bosonic system, such as a $Z_2$ spin liquid. If some elements are odd numbers, $\mathbf{\tilde{l}}$ can be a fermion, which means (\ref{K_matrix_Lagrangian}) is a fermion theory, such as a fractional quantum Hall state.
Further, the ground state degeneracy (GSD) on a torus is\cite{Wen:1990p7458,Wen:1990p5870}
\begin{align}
  GSD=|det \mathbf{K}|,
%  \label{GSD}
\end{align}
which is equal to the number of topological sectors (quasiparticle types).

In the following we will construct the effective field theory for SCCL state. In the slave-fermion approach (\ref{slave_fermion}), the electron is separated into a bosonic spinon and a fermionic holon. The fermionic holons fill a $C=1$ Chern band, which can be described by a Chern Simons term
\begin{align}
  \mathcal{L}_f=\frac{\varepsilon_{\mu\nu\lambda}}{4\pi}a_{\mu}^f\partial_{\nu}a_{\lambda}^f,
  \label{CS_holon}
\end{align}
where a $2\pi$ flux (vortex) of gauge field $a_{\mu}^f$ is a holon particle. On the other hand, a pair of bosonic spinons can be described as a $2\pi$ flux of an internal gauge field $a_{\mu}^p$. (In the liquid phase, there is a superfluid of spinon pairs, not of spinons.) Finally, the holon and spinon are glued together to form the electron by a $U(1)$ gauge field $a_{\mu}^c$. This $U(1)$ gauge field acts as a constraint in the Lagrangian
\begin{align} \mathcal{L}_c=\frac{\varepsilon_{\mu\nu\lambda}}{2\pi}a_{\mu}^c\partial_{\nu}(-a_{\lambda}^f+2a_{\lambda}^p),
  \label{CS_constraint}
\end{align}
where the factor $2$ accounts for pair of spinons having twice the internal gauge charge of a single spinon. Now, we define $a_{\mu}^I = (a_{\mu}^f,a_{\mu}^p,a_{\mu}^c)$, leading to
\begin{align}
\mathcal{L}_{eff}=\mathcal{L}_f+\mathcal{L}_c=\frac{\varepsilon_{\mu\nu\lambda}}{4\pi}\sum_{I,J=1}^3 a_{\mu}^IK_{0 I,J}\partial_{\nu}a_{\lambda}^J,
\end{align}
and we find
\begin{equation}
  \mathbf{K}_0=
  \begin{pmatrix}
    1 & 0 & -1 \\
    0 & 0 & 2 \\
    -1 & 2 & 0 \\
  \end{pmatrix},\;
    \mathbf{K}_0^{-1}=
  \begin{pmatrix}
    1 & 1/2 & 0 \\
    1/2 & 1/4 & 1/2 \\
    0 & 1/2 & 0 \\
% 1 & \frac{1}{2} & 0 \\
%    \frac{1}{2} & \frac{1}{4} & \frac{1}{2} \\
%    0 & \frac{1}{2} & 0 \\
  \end{pmatrix}.
  \label{K0_3x3}
\end{equation}
We get $GSD=4$ from this $K$-matrix description. Let us identify the four different quasiparticle types. Inspecting $\mathbf{K}_0^{-1}$ and values of statistics angles, Eqs.~\eqref{self_theta}, \eqref{mutual_theta}, we can identify the electron $e=(1,0,1)$, vison $v=(0,1,0)$, spinon $b=(0,0,1)$ and the bound state of spinon and vison $bv=(0,1,1)$. Notice that the holon $f=(1,0,0)$ and spinon $b$ differ by an electron, so they belong to the same topological sector.
%It is easy to check that the exchange and braiding statistics are as expected.
%Notice that holon $f=(1,0,0)$ and spinon $b$ differ by an electron, so they belong to the same topological sector and cannot be distinguished without information about their spin and charge.

SCCL is however not fully described by its topological properties. Symmetry interplays with topological order, leading to symmetry fractionalization (see, e.g., Refs.\onlinecite{Wen:2002p6309,Essin:2013p7585,Chen:2014p8001,Hung:2012p7537,Mesaros:2012p7535}). Within the $\mathbf{K}$-matrix formulation, it is possible to assign quantum numbers of onsite symmetries, e.g., charge and spin, to quasiparticles.\cite{Lu:2013p7858}
%In the following, we will show how this assignment can be done in the above $K$-matrix description, and that it is consistent with the physical properties of quasiparticles.

Namely, we define the charge vector $\mathbf{t}_c=(1,0,0)$ and $S_z$ vector $\mathbf{t}_{S_z}=(1/2,-1,0)$, so that $a_{\mu}^I$ couples to external test gauge fields as
\begin{align} \mathcal{L}_{ext}=\frac{\varepsilon_{\mu\nu\lambda}}{2\pi}t_{c,I}A_{\mu}^c\partial_{\mu}a_{\lambda}^{I}+\frac{\varepsilon_{\mu\nu\lambda}}{2\pi}t_{S_z,I}A_{\mu}^{S_z}\partial_{\mu}a_{\lambda}^{I},
\end{align}
where $A_{\mu}^c$ is the gauge field that couples to electric charge, while $A_{\mu}^{S_z}$ couples to $S_z$. Quasiparticle $\mathbf{l}$ carries electric charge $\mathbf{t}_c^t\mathbf{K}_0^{-1}\mathbf{l}$ and carries $S_z=\mathbf{t}_{S_z}^t\mathbf{K}_0^{-1}\mathbf{l}$. We can now identify spinons $b_{\uparrow/\downarrow}$ as $(0,0,\pm 1)$, while holon $f$ remains just $(1,0,0)$. It is straightforward to see that holon indeed carries electric charge $1$ and $S_z=0$, while $b_{\uparrow}$($b_{\downarrow}$) carries no charge and $S_z=\frac{1}{2}(-\frac{1}{2})$. Electron $e_{\uparrow (\downarrow)}$ is simply the bound state of $f$ and $b_{\uparrow (\downarrow)}$, and it is in the topologically trivial sector. The vison, expressed by $(0,\pm1,0)$, carries charge $\pm\frac{1}{2}$, and since it has statistical angle $\frac{\pi}{4}$, the vison can be viewed as 'half holon'. Bound state of spinon and vison carries both charge $\pm\frac{1}{2}$ and spin $\pm\frac{1}{2}$, with statistical angle $\frac{5\pi}{4}$.

There exists another state, described by $\bar{\mathbf{K}}_0\equiv-\mathbf{K}_0$, which is related to the above state by time reversal. In this state, vison excitation has statistical angle $-\frac{\pi}{4}$ while bound state of spinon and vison has statistical angle $-\frac{5\pi}{4}$.

\subsection{Modular Transformations and Rotation Quantum Numbers}
\label{sec:modular}
$\mathbf{S}$ and $\mathbf{T}$ matrices obtained from modular transformations of ground states on torus are believed to encode quasiparticle braiding and exchange statistics.\cite{Wen:1990p7458} Additionally, as pointed out by Refs.\onlinecite{Zhang:2012p7534,Wen:2012p8018}, it seems that if system has $C_6$ rotation symmetry the ground state quantum numbers of $C_6$ equal the eigenvalues of $\mathbf{ST}$.

The relation between modular $\mathbf{S}$,$\mathbf{T}$ matrices and the rotational symmetry of a topologically ordered phase may be understood as follows. First note that $\mathbf{S}$,$\mathbf{T}$ matrices are in principle measurable quantities in practical model Hamiltonians. In particular, given a topologically ordered phase in 2+1D with its topologically degenerate ground sector on torus $T^2$, one can firstly find a minimally entangled state (MES) basis\cite{Zhang:2012p7534}. For instance, for the $S$-matrix element between two MES $|\Xi_i\rangle$ and $|\Xi_j\rangle$: $\mathbf{S}_{ij}$, one can perform the following thought numerical measurement. Because the topological properties do not depend on local geometry, we can assume that these ground states live on a square with periodic boundary conditions. Then one can consider the state rotated by 90$^\circ$ around the square center: $R_{90^\circ}|\Xi_i\rangle$. Because $R_{90^\circ}|\Xi_i\rangle$ and $|\Xi_j\rangle$ belong to the same topological phase, in the absence of symmetry there 
should exist a Hamiltonian path $H(\tau)$ ($\tau\in[0,1]$) such that $|\Xi_j\rangle$($|\Xi_j\rangle$) are the ground state of $H(0)$($H(1)$), and the ground state sectors of $H(\tau)$ are adiabatically connected. One can then define a projection operator $\hat P_{\tau}$ into the ground state sector of $H(\tau)$ for any given $\tau$.

The many-body quantum amplitude related to the adiabatic time-evolution process of the $S$-transformation can be computed as $s_{ij}\equiv\langle\Xi_j|\hat P_{(N-1)/N}\cdot...\cdot\hat P_{2/N}\cdot\hat P_{1/N}R_{90^\circ}|\Xi_i\rangle$ as $N\rightarrow \infty$. This computation is a realization of the topological quantum field theory time-evolution. In particular, if the system has a $90^\circ$ rotational symmetry, the Hamiltonian path $H(\tau)$ can be conveniently chosen to be a constant: $H(\tau)=H(0)$. In this case, $s_{ij}$ can be simply computed as the $R_{90^\circ}$ transformation matrix in the MES basis: $s_{ij}\equiv\langle\Xi_j|R_{90^\circ}|\Xi_i\rangle$. 

We expect that this quantum amplitude $s_{ij}$ is related to the  $\mathbf{S}$-matrix elements $\mathbf{S}_{ij}$ at most by an overall ambiguity $U(1)$ phase $e^{i\theta}$, which is due to the nonuniversal local physics in the time-evolution, and a phase $e^{i\phi_i-i\phi_j}$ which is due to the gauge choice of $|\Xi_i\rangle$,$|\Xi_j\rangle$. Even with these ambiguities, based on the above argument, it is clear that in a $90^\circ$ rotational symmetric system, the $R_{90^\circ}$ eigenvalues in the topologically degenerate ground state sector can be determined by the eigenvalues of the $\mathbf{S}$-matrix, up to an overall U(1) phase factor. Similar consideration for a $60^\circ$ rotational symmetric system leads to the conclusion that the $C_6$ eigenvalues in the topologically degenerate ground state sector can be determined by the eigenvalues of the matrix product $\mathbf{S}\mathbf{T}$, up to an overall U(1) phase factor. In addition, it has been proposed that this U(1) phase factor is simply unity \cite{Zhang:2012p7534} which is consistent with numerical simulations on several model Hamiltonians\cite{Cincio:2013p7854,He:2013p8022,Gong:2013p7942}.

However, we find in the SCCL phase on the honeycomb lattice, the $C_6$ eigenvalues in the ground state sector and the eigenvalues of $\mathbf{S}\mathbf{T}$ differ by an overall U(1) phase factor that is system-size dependent. In particular, one can obtain $\mathbf{S}$ and $\mathbf{T}$ matrices from our $\mathbf{K}$-matrix. According to Ref\cite{Wen:2012p8018}, using Eq.~\eqref{K0_3x3} and choosing four quasiparticle vectors as $(0,0,0), (0,0,1), (0,1,0), (0,1,1)$, one obtains
\begin{align}
  \mathbf{S}=\frac{\xi}{2}
  \begin{pmatrix}
    1 & 1 & -1 & -1 \\
    1 & 1 & 1 & 1 \\
    1 & -1 & i & -i \\
    1 & -1 & -i & i \\
  \end{pmatrix},\;
  \mathbf{T}=\eta
  \begin{pmatrix}
    1 & 0 & 0 & 0 \\
    0 & 1 & 0 & 0 \\
    0 & 0 & -e^{-\frac{i\pi}{4}} & 0 \\
    0 & 0 & 0 & e^{-\frac{i\pi}{4}} \\
  \end{pmatrix},
  \label{ST_matrix}
\end{align}
where $\xi$,$\eta$ are $U(1)$ phase factors. Although in Ref.\onlinecite{Wen:2012p8018} these phase factors are fully determined using modular transformations on fractional quantum Hall liquid analytic wavefunctions on torus, their values are not important for the following discussion.

\begin{figure}
 \includegraphics[width=0.35\textwidth]{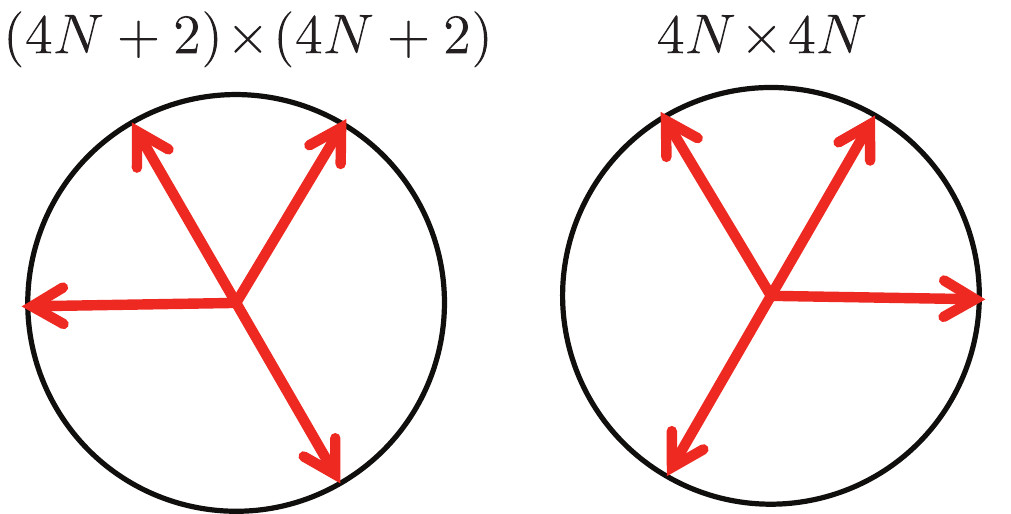}
 \caption{Symmetry quantum numbers for $C_6$ rotation, calculated analytically (Appendix~\ref{app:wf_quantum_number}) for the fourfold topologically degenerate ground state sector of the SCCL phase on different lattice sizes. $C_6$ eigenvalues are plotted on the unit-circle in complex plane. The sets of eigenvalues differ by overall phase between different system sizes ($N$ is integer, and $X\times Y$ labels number of unit-cells along $a_1,a_2$), although all systems are topologically a torus.}
 \label{fig:STc6}
\end{figure}

The eigenvalues of $\mathbf{ST}$ are found to be $\xi\cdot\eta\cdot(1, -1, e^{i\pi/3}, e^{-i\pi/3})$. On the other hand, the analytical construction of SCCL projective wavefunctions allows us to compute the symmetry properties in the ground state sector (see Appendix~\ref{app:wf_quantum_number}), and we find that the $C_6$ quantum numbers of topological ground states differ for $4N\times4N\times2$ and $(4N+2)\times (4N+2)\times2$ lattice sizes by an overall $U(1)$ phase, see Fig.~\ref{fig:STc6}. On $4N\times4N\times2$ systems the $C_6$ eigenvalues are found to be $e^{i\pi/3}\cdot(1, -1, e^{i\pi/3}, e^{-i\pi/3})$, while on $(4N+2)\times (4N+2)\times2$ systems these are $e^{i2\pi/3}\cdot(1, -1, e^{i\pi/3}, e^{-i\pi/3})$. In contrast to previous understanding, our example of SCCL explicitly shows that $C_6$ quantum numbers and eigenvalues of $\mathbf{ST}$ are related by a lattice size dependent phase factor.

\subsection{Gapless edge states and experimental signatures}\label{sec:SCCL_edge}
We will first derive the edge theory of SCCL using the effective field theory from previous subsection. We consider two cases of symmetry on the edge: 1) Charge conservation and spin-rotations around $S_z$ (group $U(1)_c\times U(1)_z$); and to capture more of the spin-rotation symmetry 2) Charge conservation, $S_z$ rotations, and $\pi$-rotation around a perpendicular axis. The second case is detailed in in Appendix~\ref{app:edgesym}, but in both cases we find a gapless chiral holon edge mode, which differs from the gapless chiral electron mode of the c-SDW state. We therefore propose several experimental signatures for distinguishing c-SDW and SCCL states in the last two subsections.

The effective action describing edge excitations of Abelian Chern-Simons theory can be derived from gauge invariance of Lagrangian Eq.~\eqref{K_matrix_Lagrangian} expanded by higher order (Maxwell) terms, on a manifold with boundary.\cite{Wen:1995p8031} The edge physics is captured by $N$ chiral boson fields $\{\phi_I\simeq \phi_I+2\pi|1\le I\le N\}$:
\begin{align}
  S_{edge}^{0}=\frac{1}{4\pi}\int \dd t\mathrm{d}x\sum_{I,J}(K_{I,J}\partial_t \phi_I\partial_x \phi_J-V_{I,J}\partial_x\phi_I\partial_x\phi_J).
\end{align}
Here, $V_{I,J}$ is positive definite constant matrix, which depends on system details. The number of right movers $n_+$ and left movers $n_-$ are given by the signature of $\mathbf{K}$. The commutation relations between these chiral boson fields are fixed by the first term, and describe the following Kac-Moody algebra\cite{Wen:1995p8031}:
\begin{align}
  [\partial_x\phi_I(x),\partial_y\phi_J(y)]=2\pi iK_{I,J}^{-1}\partial_x \delta(x-y).
  \label{KM_algebra}
\end{align}

There is a one-to-one correspondence between quasiparticles in the bulk and chiral boson fields living on the edge. Operator $V_{\mathbf{l}}=\exp(i\sum_Il_I\phi_I)$ creates quasiparticle $\mathbf{l}$ on the edge.
% The edge theory generically contains scattering, and the edge modes may be gapped out.
Generic action for scattering takes the form of Higgs terms:
\begin{align}
  S_{edge}^1=\sum_{\mathbf{\bar{l}}}C_{\mathbf{\bar{l}}}\int\dd t\mathrm{d}x \cos(\sum_I\bar{l}_I\phi_I+\alpha_I),
\end{align}
where $\mathbf{\bar{l}}$ are local bosonic excitations, which can be expressed as $\mathbf{Kl}$ for some integer vector $\mathbf{l}$. However, in the presence of symmetry, chiral boson fields may transform nontrivially under symmetry operations, and some Higgs terms may be forbidden in the symmetry-preserving edge.\cite{Lu:2012p7468}

\subsubsection{Edge modes with $U(1)_c\times U(1)_z$ symmetry}

Now let us turn to edge theory for SCCL. As the set of independent local excitations $\mathbf{\bar{l}}$ we choose the columns of $\mathbf{K}_0$. The 1st column of $\mathbf{K}_0$ matrix is an electron, carrying charge 1 and spin 1/2, while the 2nd column is boson pair with spin 1. They both transform nontrivivally under $S_z$ rotations. The 3rd column is bound state of two vison and a charge -1 holon, which is a trivial boson carrying trivial quantum number of $S_z$ and charge. Thus, the only Higgs term allowed by this symmetric boundary is
\begin{align}
  \mathcal{L}_{Higgs}=C\cos(\phi_1-2\phi_2).
\end{align}
This term gaps out two counter-propagating edge modes, and leaves the gapless chiral boson mode $\phi_1$ on the edge. Thus, the edge theory of SCCL can be modeled as 1d chiral fermion liquid of spinless holons.

Even with added $\pi$-rotation around an axis perpendicular to $S_z$, as shown in Appendix~\ref{app:edgesym}, $\phi_1$ remains the only gapless edge mode. We therefore found that the edge with charge conservation and any of above spin-rotation symmetries has a chiral fermion liquid of spinless holons. The c-SDW edge on the other hand has a chiral fermion liquid of electrons. We therefore next propose tunneling experiments to distinguish these two phases.

\subsubsection{Point junction}
\begin{figure}
\begin{minipage}{0.48\textwidth}
  \centering
  \begin{minipage}{0.6\linewidth}
    \begin{figure}[H]
      \includegraphics[width=0.8\textwidth]{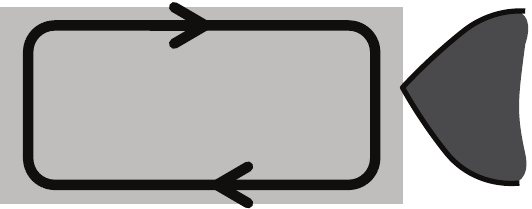}
    \end{figure}
  \end{minipage}
  \\
\begin{minipage}{0.6\linewidth}
  \begin{table}[H]
  \center
  \begin{tabular}{|c|c|c|}
    \hline
     & Chiral SDW & SCCL \\\hline
    Metal & Const & $T^4$ \\\hline
    SC & $T^2$ & $T^2$ \\\hline
  \end{tabular}
\end{table}
\end{minipage}
\end{minipage}
\caption{(Top) Point contact measurement with weak tunneling ($G\ll e^2/h$) into edge states can distinguish between the c-SDW and SCCL phase. (Bottom) Temperature dependence of tunneling conductance through point junction between Metal/SC lead and c~-SDW/SCCL, exhibiting different values of power law exponent $\alpha$.}
\label{fig:pc}
 \label{tab:conduct}
\end{figure}

In this part, we will discuss the experimental signature of transport through a weak tunnel junction connecting a metallic/singlet SC lead to c-SDW/SCCL (Fig.~\ref{fig:pc}). %We find that tunneling conductance c-SDW and SCCL has different scaling dimension.
Our results of tunneling conductance are listed in a table in Fig.~\ref{tab:conduct}. Below we will find the same exponents for the voltage dependence of the conductance.

These scaling forms, and therefore the experimental signatures, should hold in the regime of weak tunneling, $G\ll e^2/h$. More formally, the weak-tunneling condition corresponds to the assumption $T,V\ll T_K$, where $T_K$ is a characteristic energy scale of the junction depending on details of the point contact.
%; for the special case of c-SDW/metallic lead junction the formal condition is $c_K\gg 1$, where $c_K$ is the dimensionless constant characterizing the junction ($c_K^{-1}$ is proportional to $G$ in this regime).

The total Hamiltonian can be modeled as a sum of three pieces
\begin{align}
  H_{tot}=H_0+H_{lead}+H_{tunn},
\end{align}
where $H_0$ is Hamiltonian for c-SDW/SCCL, $H_{lead}$ is Hamiltonian for SC/Metal lead, and $H_{tunn}$ describes tunneling through point contact. For the most general case, we can write
\begin{align}
  \label{eq:tunnelingHamiltonian}
  H_{tunn}=t[O_0^{\dagger}O_{lead}+h.c.],
\end{align}
where $O_0$ is electron or singlet pair annihilation operator on c-SDW/SCCL side, while $O_{lead}$ is the corresponding operator in the lead.

Before calculating tunneling conductance, it is instructive to consider a simple renormalization group (RG) transformation, which tells us how the tunneling amplitude $t$ varies with the energy (or temperature) scale.\cite{Kane:1992p8028} Assume $O_0\sim\tau^{-\delta_0}$ and $O_{lead}\sim\tau^{-\delta_{lead}}$, where $\tau$ is imaginary time. Consider an RG step which integrates out Matsubara frequencies between $\Lambda/b$ and $\Lambda$, where $\Lambda$ is a high frequency cut-off. Then the RG equation for $t$ is given to leading order by
\begin{align}
  \frac{\partial t}{\partial l}=(1-\delta)t
\end{align}
where $\delta=\delta_0+\delta_{lead}$. At nonzero temperature, the RG flows are cut off by $T$ ($T\gg V$), leading to $t_{eff}\sim t T^{\delta-1}$. One expects tunneling conductance to vary as $t_{eff}^2$, which gives the result
\begin{align}
  G(T)\sim t^2T^{2\delta-2}
  \label{conduct_scale}
\end{align}

We now present the case of metal/SCCL junction in detail, referring the reader to Appendix~\ref{app:tunneling} for the other cases listed in table of Fig.~\ref{tab:conduct}. (Note that the scaling for c-SDW/SC junction follows directly from Ref.\onlinecite{Fisher:1994p8066}.)
Due to the spin gap on the boundary of SCCL, single electron tunneling will be exponentially suppressed at low temperatures. So, the leading contribution is from singlet pair tunneling, and in Eq.~\eqref{eq:tunnelingHamiltonian} we have:
\begin{align}
  &O_0\equiv f^{\dagger}(x=\xi)f^{\dagger}(x=0)\\
  &O_{lead}\equiv\psi_{M,\uparrow}(x=0)\psi_{M,\downarrow}(x=0),
\end{align}
where the product of holon operators $f^\dagger$ in $O_0$ represents annihilation of a local singlet pair of electrons on SCCL edge due to the presence of bosonic spinon pairing (see Eqs.~\eqref{slave_fermion},\eqref{eq:sf_MF}), while coherence length $\xi$ appears due to the Pauli principle. So, $\delta_{lead}=2\delta_{FL}=1$, where we used that the scaling dimension $\delta_{FL}=1/2$ for Fermi liquid system in any dimension.\cite{Shankar:1994p7862} The operator $f(x=\xi)f(x=0)$ has the same scaling dimension as operator $f(x=0)\partial_xf(x=0)$, giving $\delta_0=1+2\delta_{FL}=2$, where the holon operator on the edge scales with $\delta_{FL}$ since it forms a chiral fermion liquid analogous to the one on the edge of integer quantum Hall systems.\cite{Stone:1994p8029} This leads to the announced $G(T)\sim T^4$ for this junction.

One generally expects that the voltage-dependent conductance $G(V)$ scales in the same way as $G(T)$. We checked that this is true using a perturbative calculation (i.e., the Fermi golden rule) of the nonlinear current-voltage ($I-V$) characteristic in the regime $T\ll V$. The calculation details for all junctions are presented in Appendix~\ref{app:tunneling}.

\subsubsection{Line junction}
\begin{figure}
 \includegraphics[width=0.35\textwidth]{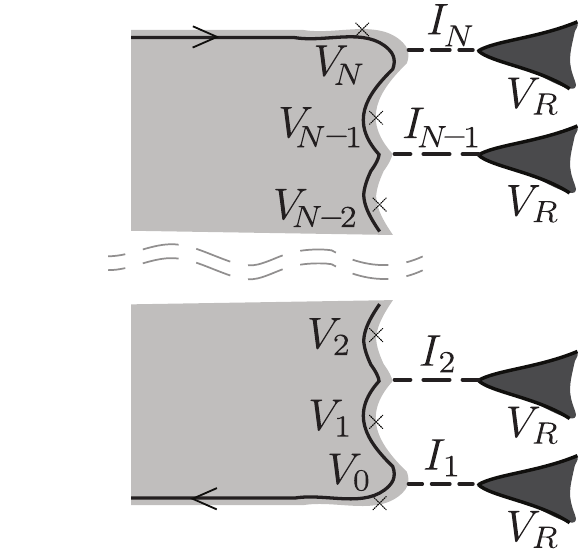}
 \caption{Line junction modeled as the large$-N$ limit of array of point contacts. In weak tunneling regime having $T\ll V$ or $V\ll T$, with $T$---temperature, $V$---voltage, the conductance can distinguish between the c-SDW and SCCL phases.
%In regime $T\ll V\ll T_K$, with $T$---temperature, $V$---voltage, $T_K$---effective total tunneling energy scale, the conductance can distinguish between the c-SDW and SCCL phases.
}
 \label{fig:lc}
\end{figure}

A more common setup in experiments is the line junction,\cite{Chamon:1997p7909} which can be viewed as a large number of weakly coupled point junctions, as sketched in Fig.~\ref{fig:lc}. Here, and throughout our discussion on the line junction, weakly coupled means
\begin{align}
  \label{eq:3}
  T,V&\ll T_K^{(1)},\dots,T_K^{(N)},
\end{align}
where $N$ is the total number of point junctions, while $T_K^{(n)}$ is the characteristic energy scale determined by details of the $n$-th junction\cite{Chamon:1997p7909}. Physically, this weak-coupling condition in the line junction means that the regime of weak tunneling, $G\ll e^2/h$, is available, at least for low enough voltage (see further below). The special case of c-SDW/metallic lead junction is left for the end of this subsection, since it is much simpler to analyze and does not require such assumptions.

The number $\alpha\neq 0$ which appears below is simply the value of exponent in Fig.~\ref{tab:conduct} for the considered combination of quantum state and lead. The special case of c-SDW/metallic lead junction has exponent $\alpha=0$, and is discussed at the end.

%It turns out that as $T/V$ increases the conductance $G$ reaches the universal value $e^2/h$, so we consider the $T\to 0$ case, and only vary $V$.
First, let us consider the $T\ll V$ regime. We will find that for small voltages, the scaling of conductance can distinguish the c-SDW and SCCL in the same way as table in Fig.~\ref{tab:conduct}.

The expression for current-voltage characteristic we obtain (see Appendix~\ref{app:tunneling}) is
\begin{align}
  I/V=\frac{e^2}{h}\left[1-\frac{T_K}{(\alpha V^{\alpha}+T_K^{\alpha})^{1/\alpha}}\right],\quad\text{$T\ll V$},
%I/V=\left\{\begin{aligned}
%&\frac{e^2}{h}\left[1-\frac{T_K}{(\alpha V^{\alpha}+T_K^{\alpha})^{1/\alpha}}\right],& \text{if %$\alpha\neq 0$},\\
%&\frac{e^2}{h}\left[1-\exp(-1/c_K)\right],& \text{if $\alpha=0$},
%\end{aligned}\right.
  \label{line_junction_conductance}
\end{align}
where the voltage difference $V\equiv V_R-V_0$ (Fig.~\ref{fig:lc}), $\alpha\neq0$ is the exponent in the point junction scaling $G\sim V^\alpha$ (table in Fig.~\ref{tab:conduct}), and the effective $T_K$ is the single parameter describing the line junction and incorporating all the $T_K^{(n)}$ as well as their fluctuations:
\begin{align}
    T_{K}^{-\alpha}\equiv\sum_{n=1}^N(T_K^{(n)})^{-\alpha},
%  T_{K}^{-\alpha}&\equiv\sum_{n=1}^N(T_K^{(n)})^{-\alpha},\\\notag
%  c_{K}^{-1}&\equiv\sum_{n=1}^N(c_K^{(n)})^{-1}.
\end{align}
(Note that the definition of $T_K$ also depends on scaling exponent $\alpha$.)

The above $\alpha\neq 0$ conductance result holds for all values of $V,T_K$ at $T\ll V$, as long as the assumptions used to derive the expression hold, namely, each individual point contact is weakly coupled. This just means $V\ll T_K^{(n)}$ for all $n$.
% The $\alpha=0$ result actually holds for regime $T\ll V$ as long as each individual point contact is weakly coupled, i.e., $1\ll c_K^{(n)}$ for all $n$.
However, the effective $T_K$ can be much smaller than all $T_K^{(n)}$ in a long line junction (large $N$). Therefore, let us examine the tunneling conductance $G$ in two regimes: $T\ll V\ll T_K$ and $T,T_K\ll V$.
%$T\ll V\ll T_K$ ($T\ll V;1\ll c_K$) and $T,T_K\ll V$ ($T\ll V;c_K\ll 1$) for an $\alpha\neq 0$ ($\alpha=0$) junction.

For the first regime we get:
\begin{align}
   G(V)\approx \frac{e^2}{h}\frac{V^{\alpha}}{T_K^{\alpha}},\quad\text{$T\ll V\ll T_K$,}
%  G(V)\approx\left\{
  %\begin{aligned}
%  \frac{e^2}{h}\frac{V^{\alpha}}{T_K^{\alpha}},\quad&\text{if $\alpha\neq 0$,\;  $T\ll V\ll T_K$,}\\
%\frac{e^2}{h}\frac{1}{c_K},\quad&\text{if $\alpha=0$,\;  $T\ll V ;1\ll c_K$,}
%  \end{aligned}\right.
\end{align}
manifesting the same scaling form as that in point contact junction. (Note that still $\alpha\neq0$.)

On the other hand, in the regime $T,T_K\ll V$ we get
\begin{align}
    \label{eq:GV_eq}
  G(V)\approx \frac{e^2}{h},\quad\text{$T,T_K\ll V$}.
\end{align}

The derivation for $V\ll T$ regime is similar, and we reach the same final conclusions as for previous case. The current-voltage characteristic in this regime is:
\begin{align}
  I/V=\frac{e^2}{h}\left[1-e^{-\frac{T^{\alpha}}{(T'_K)^{\alpha}}}\right],\quad\text{$V\ll T$},
%  I/V=\left\{\begin{aligned}
%&\frac{e^2}{h}\left[1-e^{-\frac{T^{\alpha}}{(T'_K)^{\alpha}}}\right],& \text{if $\alpha\neq 0$},\\
%&\frac{e^2}{h}\left[1-\exp(-1/c'_K)\right],& \text{if $\alpha=0$},
%\end{aligned}\right.
\end{align}
see Appendix~\ref{app:tunneling}. In this regime a characteristic energy scale $T'_K$, analogous but different from $T_K$, describes a point junction having $\alpha\neq0$. For a given combination of quantum phase and lead forming the junction, we expect the ratio $T'_K/T_K$ to be a universal number of order 1. With this in mind, we again consider two regimes: $V\ll T\ll T'_K$ and $V,T'_K \ll T$.
% ($V\ll T ;1\ll c_K'$) ($V\ll T;c_K'\ll 1$)

We get
\begin{align}
  G(T)\approx\frac{e^2}{h}\frac{T^{\alpha}}{{T'}_K^{\alpha}},\quad\text{$V\ll T\ll T'_K$},
%G(T)\approx\left\{
%\begin{aligned}
%\frac{e^2}{h}\frac{T^{\alpha}}{{T'}_K^{\alpha}},\quad&\text{if $\alpha\neq 0$,\;$V\ll T\ll T'_K$},\\\notag
%\frac{e^2}{h}\frac{1}{c'_K},\quad&\text{if $\alpha=0$,\;$V\ll T;1\ll c_K'$},
%\end{aligned}\right.
\end{align}
while in the other regime:
\begin{align}
  \label{eq:GT_eq}
  G(T)\approx \frac{e^2}{h},\quad\text{$V,T'_K \ll T$},
\end{align}
recalling that $\alpha\neq0$ is the same exponent found for the point junction, Fig.~\ref{tab:conduct}. Details are in Appendix~\ref{app:tunneling}.

Concerning the c-SDW/metallic lead junction, which has scaling exponent $\alpha=0$ in the point junction, a simple calculation reveals that the line junction conductance is constant, $G=\frac{e^2}{h}\cdot c$, with $0<c\leq 1$ a non-universal constant describing the line junction (see Appendix~\ref{app:tunneling}). $c=1$ corresponds to the regime in which the chiral electron edge modes are equilibriated with the lead.

We conclude that the line junction tunneling conductance $G(V)(G(T))$ can distinguish between c-SDW and SCCL in the regime where $G\ll e^2/h$ and $T\ll V ~(V\ll T)$, which corresponds to weakly coupled line junctions. In this case the edge modes are not thermally equilibriated with the lead. For example, in this regime, the zero bias tunneling conductance for the c-SDW/metallic lead line junction is temperature-independent while for the SCCL/metallic lead line junction it is expected to scale as $T^4$. On the other hand, although $G=e^2/h$ is one experimental signature of the quantum anomalous Hall effect in the c-SDW phase, we find that even in the SCCL phase the universal value $G=e^2/h$ can be realized, e.g., in the regime of Eqs.~\eqref{eq:GV_eq},~\eqref{eq:GT_eq}, where it represents the thermal equilibriation of chiral holon edge modes with the lead. Therefore the $G=e^2/h$ is not a unique property of the c-SDW phase.

\section{Discussion and conclusions}\label{sec:conclusion}
%SET
%Finite temp.
%Other lattices.
This paper studies the phase diagrams of correlated electronic models on the honeycomb lattice at 1/4 doping, using a combination of analytical construction of quantum wavefunctions and various numerical simulations. Interestingly, all phases appearing in our main results, the phase diagrams in Fig.\ref{fig:phase_diagrams}, are interaction-driven topological phases. In particular, we find that either the c-SDW state or the SCCL phase occupies the majority of the realistic parameter regimes for correlated materials. In the present study, due to the limitation of sample sizes, we cannot sharply distinguish these two phases in the phase diagrams. Distinguishing them in numerical simulations requires careful finite size scaling, which we leave as a subject of future investigation. However, we study the sharp signatures of c-SDW and SCCL phases in transport experiments, which can be used to identify and distinguish these phases in candidate materials.

The method applied here, namely using lattice quantum numbers to sharply distinguish competing quantum phases, is not limited to the models studied in this paper. In particular, in time-reversal symmetry breaking phases, the ground states often form non-trivial multi-dimensional irreps of the lattice symmetry groups. When this happens, the analytical understanding of the non-trivial irreps can be used to identify/distinguish candidate quantum phases in numerical simulations.

From a general point of view, what are the possible candidate phases in correlated electronic systems at generic fillings? First, charge inhomogeneity is always a possibility. For instance, stripe-like charge modulations have been observed in numerical simulations of the t-J model on the square lattice.\cite{White:2000p8048} Assuming charge being homogeneous, incommensurately filled systems and commensurately filled systems are quite different at the conceptual level. Generally speaking, in order to accommodate an incommensurate filling, the system could either develop superconductivity, or the doped charges could form a Fermi surface.\footnote{The Fermi surface is only a mean-field level description. For example, one could imagine the situation that the Fermi surface is strongly coupled with dynamical gauge fields.\cite{Kaul:2008p8050}} In any case the system is expected to be a charge conductor in the bulk. However, at commensurate fillings, the system has a third option: The charges could condense into 
many-body states without causing a charge inhomogeneity or superconductivity, and a bulk energy gap of charge-excitations can be generated. We term this third scenario as the charge-insulator scenario.

In conventional quantum phases in which Luttinger's theorem\cite{Luttinger:1960p8033} is valid, the charge-insulator scenario must be accompanied with translational symmetry breaking such as long-range magnetic ordering. The c-SDW phase belongs to this situation. However in exotic quantum phases in which fractionalization occurs, translational symmetry does not need to be broken. For instance, the SCCL phase is a translationally invariant charge insulator. Other examples include the recently studied Fractional Chern insulators,\cite{Sheng:2011p8046,Regnault:2011p8044,Wang:2012p5933,Lu:2012p6109,Liu:2012p8034,Kourtis:2012p8039,Jian:2013p8035,Grushin:2012p8036} which are symmetric many-body states that exist in models with commensurately filled nearly-flat bands in the presence of strong interactions.

One goal of this paper is to investigate the competition between the superconductivity and the charge-insulator phases in commensurately doped correlated systems. Exactly at the 1/4 doping, we find that the charge-insulator phase(c-SDW/SCCL) occupies the majority of realistic regimes of the models that we investigated. Meanwhile, although the d+id superconductor phase is found only at $J/t>0.8$ in the t-J model, as a variational state, it captures $>90\%$ of the ground state energy even in the regime $0.1<J/t<0.8$(see Fig.~\ref{fig:did_relative_energy}). Therefore the d+id state serves as a nearby competing phase.

As doping deviates away from 1/4 slightly, the extra electric charges need to be absorbed by excitations in a charge-insulator. In the SCCL phase, these charge excitations form a finite density of anyons: $v$, $b_{\alpha}v$ or fermionic chargeon-$f$; while in the c-SDW phase, these excitations would be a finite density of electronic quasiparticles. However, the d+id superconductor state, as a charge superfluid, can absorb extra electric charges without causing excitations. In the regimes in which c-SDW/SCCL phase is realized at 1/4 doping, we expect that the ground state is likely to be the d+id state as the doping is tuned away from 1/4 by a finite amount.

One may wonder that due to the Mermin-Wagner theorem, the long-range magnetic order cannot be observed at finite temperatures in the c-SDW phase. In addition, our discussion of the low temperature tunneling conductance in the c-SDW phase did not consider this thermal fluctuation effect. However, in an ideal $SU(2)$ symmetric system the correlation length of the magnetic order diverges exponentially at low temperatures. Thus even a tiny spin-orbit coupling strength would pin the magnetic order at low temperatures in realistic materials, which justifies our treatment.

Recently there has been a lot of interest in the understanding of interplays between global symmetry and topological order, which have been named as ``symmetry enriched'' phenomena (see, e.g., Refs.\onlinecite{Wen:2002p6309,Essin:2013p7585,Chen:2014p8001,Hung:2012p7537,Mesaros:2012p7535,Maciejko:2010p7431,Levin:2012p7429,Cho:2012p7426,Swingle:2011p7432}). In the present study, the SCCL phase serves as a new example of a symmetry enriched topological phase which could be realized in materials. In the SCCL phase, the symmetry enriched phenomena include the charge-$1/2$ spin-neutral anyons with statistical angle $\pi/4$ and the gapless chargeon chiral edge states. And the latter one has direct experimental signature as $G\sim T^4$ in tunneling conductance experiments.

Our results are relevant for many correlated materials on the honeycomb lattice. Doped graphene, in which the long-range Coulomb interaction is screened, is an intermediately correlated material that may be modeled by the Hubbard model on a honeycomb lattice with $U/t=2\sim 3$.\cite{CastroNeto:2009p3917} InCu$_{2/3}$V$_{1/3}$O$_3$ is a strongly correlated spin-1/2 antiferromagnet on the honeycomb lattice.\cite{Kataev:2005p8065} However, doping these systems up to 1/4 remains experimentally challenging but may be achievable in a foreseeable future due to the progress of experimental techniques on thin films.\cite{Das:2008p8054,McChesney:2010p7899} In addition, recently a new route for realization of honeycomb lattice thin films was proposed, based on growth of (111) bilayers of perovskites.\cite{1Xiao:2011p7761,Middey:2012p7893,Okamoto:2013p8067} For instance, after trigonal lattice distortion is included, $a_{1g}$-active compounds may be realizations of single-band correlated systems on the honeycomb lattice.\cite{Doennig:2013p8051}  In addition, cold-atom optical lattices can be used to realize the Hubbard model on the honeycomb lattice.\cite{Polini:2013p8068,Uehlinger:2013p8070}

We thank Fa Wang, Yuan-Ming Lu and Satoshi Okamoto for helpful discussions. The DMRG calculations were performed using the ITensor library, http://itensor.org/. This study is supported by the Alfred P. Sloan foundation and National Science Foundation under Grant No. DMR-1151440. We thank Boston College Research Service for providing the computing facilities where the numerical simulations were performed.

\begin{appendix}
\section{Symmetry group of the honeycomb lattice model}\label{app:symmetry_group}

The symmetry group (SG) of our honeycomb lattice model is generated by the following symmetry operations (shown in Fig.\ref{fig:fermi_surface}):(1) Translations $T_{1,2}$ by Bravais lattice vectors $\vec{a}_{1,2}$; (2) The $\frac{\pi}{3}$-rotation $C_6$ around the $\hat{z}$ axis through the honeycomb plaquette center; (3) Mirror reflection with respect to the $\hat{x}-\hat{z}$ plane combined with the time-reversal operation, labeled as $\bar{\sigma}$. Note that $\bar{\sigma}$ is an antiunitary symmetry since it includes time-reversal operation. It acts on the Hamiltonian through a combination of a unitary symmetry operation and complex conjugation $\mathcal{C}$.

We label a lattice site by coordinates $(x,y,s)$, where $\vec{r}=x\vec{a}_1+y\vec{a}_2+\vec{r}_s$ is its position vector. $\vec{a}_1=a(\sqrt{3},0)$ and $\vec{a}_2=a(\sqrt{3},3)/2$ are two Bravais lattice vectors, $s=u,v$ is the sublattice index, and in our coordinate system $\vec{r}_u=-a(\sqrt{3},1)/2$ and $\vec{r}_v=a(-\sqrt{3},1)/2$. Under symmetry operations, the $(x,y,s)$ coordinates transform as
  \begin{align}
    \begin{split}
      T_1 & : (x,y,s) \to (x+1,y,s),\\
      T_2 & : (x,y,s) \to (x,y+1,s),\\
      \bar{\sigma} & : (x,y,u) \to (x+y,-y,v),\\
       & : (x,y,v) \to (x+y,-y,u),\\
      C_6 & : (x,y,u) \to (1-y,x+y-1,v),\\
       & : (x,y,v) \to (-y,x+y,u).
    \end{split}
    \label{sym_transf}
  \end{align}

  The multiplication rules of the above SG are completely determined by the following algebraic relations:
  \begin{align}
    \begin{split}
      T_1^{-1}T_2T_1T_2^{-1} & =\mathbf{e},\\
      T_2^{-1}C_6T_1C_6^{-1} & =\mathbf{e},\\
      T_1^{-1}C_6T_1T_2^{-1}C_6^{-1} & =\mathbf{e},\\
      T_1^{-1}\bar{\sigma}T_1\bar{\sigma}^{-1} & =\mathbf{e},\\
      T_2^{-1}\bar{\sigma}T_1T_2^{-1}\bar{\sigma}^{-1} & =\mathbf{e},\\
      \bar{\sigma}C_6\bar{\sigma}C_6 & =\mathbf{e},\\
      C_6^6=\bar{\sigma}^2 &=\mathbf{e},
    \end{split}
    \label{sym_constraint}
  \end{align}
where $\mathbf{e}$ represents the identity element of SG.

\section{Parton construction of symmetric quantum wavefunctions}\label{app:wavefunction_construction}

In this section, we use the slave-fermion method to construct the projective wavefunction of c-SDW/SCCL, and a slave-boson method to describe d+id SC.

\subsection{c-SDW/SCCL states}

In this Section we will consider all mean-field Ans\"atze allowed by the projective symmetry group construction, and pick out one that correctly describes the c-SDW/SCCL states.

\subsubsection{Projective symmetry group analysis}
The projective symmetry group (PSG)\cite{Wen:2002p8024,Wen:2002p6309,Wang:2006p6704} classifies different mean field Ans\"atze, and we will briefly review it and apply it here.
Although projective wavefunctions are invariant under the symmetry group action (listed in Appendix \ref{app:symmetry_group}), the mean field wavefunction before projection can still explicitly break symmetry. In fact, due to the $U(1)$ gauge field that glues spinon and holon together, a mean-field wavefunction only needs to be invariant under a combined symmetry and gauge transformation. Also, there is a many-to-one correspondence between mean-field states and physical electron states: Any two parton mean-field states related to each other by a $U(1)$ gauge transformation ${e^{i\phi(\mathbf{r})}}$ correspond to the same electron state. 

More precisely, we associate a $U(1)$ gauge group element $e^{i\phi_X(j)}$, dependent on site $j$, to each element $X$ of the lattice symmetry group, and the mean-field Ansatz is invariant under the PSG operation
\begin{align}
  \begin{split}
    b_{j\alpha}\to e^{i\phi_X(j)}b_{X(j)\alpha},\\
    f_j\to e^{i\phi_X(j)}f_{X(j)}. 
  \end{split}
\end{align}
Therefore, the mean field Ansatz satisfies
\begin{align}
  \begin{split}
  A_{X(i)X(j)}=& e^{i(\phi_X(i)+\phi_X(j))}A_{ij},\\
  B_{X(i)X(j)}=& e^{-i(\phi_X(i)-\phi_X(j))}B_{ij}.\\
  \chi_{X(i)X(j)}=& e^{-i(\phi_X(i)-\phi_X(j))}\chi_{ij}.
  \end{split}
\end{align}
The low energy gauge fluctuations of mean-field Ansatz are controlled by the invariant gauge group (IGG)\cite{Wen:2002p6309},
\begin{align}
  \begin{split}
  A_{ij}=& e^{i(\phi_{\mathbf{e}}(i)+\phi_{\mathbf{e}}(j))}A_{ij},\\
  B_{ij}=& e^{-i(\phi_{\mathbf{e}}(i)-\phi_{\mathbf{e}}(j))}B_{ij},\\
  \chi_{ij}=& e^{-i(\phi_{\mathbf{e}}(i)-\phi_{\mathbf{e}}(j))}\chi_{ij}.
  \end{split}
  \label{IGG}
\end{align}
For reasons discussed in Sec.\ref{sec:sym_wavefunctions},  IGG is $Z_2$ in our case ($\phi_{\mathbf{e}}=0,\pi \mod 2\pi$). The algebraic relations (\ref{sym_constraint}) put constraints on the gauge transformation $\phi_X(x,y,s)$. Following a procedure similar to the one in Ref.\onlinecite{Wang:2010p6724}, we find the solution:
\begin{align}
  \phi_{T1}(x,y,s)&=0,\\
  \phi_{T2}(x,y,s)&=p_1\pi x,\\
  \phi_{C_6}(x,y,s)&=\frac{1}{6}(p_1+p_6)\pi+p_1\pi(\frac{x(x-1)}{2}+xy),\\
  \phi_{\bar{\sigma}}(x,y,u)&=p_1\pi(x+y+\frac{y(y-1)}{2}),\\
  \phi_{\bar{\sigma}}(x,y,v)&=p_1\pi(x+y+\frac{y(y-1)}{2})+p_7\pi
  \label{PSG_solution}
\end{align}
where $p_1,p_7=0,1$, and $p_6=0,1,\dots,5$. In total, there are $24$ solutions for PSG with $IGG=Z_2$ in honeycomb lattice for symmetry group defined in Appendix~\ref{app:symmetry_group}.

\subsubsection{Wavefunction for c-SDW/SCCL}
There are further constraints on a mean field Ansatz of the c-SDW/SCCL. First, we want quarter doped holons to fill a Chern band, which will lead to the anomalous quantum Hall response. This requires at least doubling of the unit cell. So we only consider the $\pi$-flux states having $p_1=1$. In this case, we double the unit-cell in $x$ direction. When two Ans\"atze are a time reversal pair, we only need to consider one of them.

It turns out that $p_1=1, p_6=0, p_7=0$ gives the mean-field Ansatz for c-SDW/SCCL. We first construct the mean-field Hamiltonian with NN and NNN hopping/pairing. This particular PSG solution partially fixes the phases of mean-field parameters. The pattern is shown in Fig.~(\ref{fig:SCCL_pattern}). After solving Bogoliubov equations for bosons (spinons), we find that the boson band minima lie at $\pm(\pi/2,\pi)$ of the reduced Brillouin zone.

Now, we are able to construct the wavefunction from the mean-field Hamiltonian. Let us consider the most general form of the Hamiltonian in momentum space. Spinon Hamiltonian has BCS form
\begin{align}
  H_b^{MF}&=\sum_k \beta_k^{\dagger}D(k)\beta_k + const,\\
  D(k)&=
  \begin{pmatrix}
    B(k)-\mu & A(k) \\
    A^{\dagger}(k) & B^{t}(-k)-\mu \\
  \end{pmatrix}
\end{align}
where $\beta_k=(b_{1k\uparrow},\dots,b_{nk\uparrow},b_{1-k\downarrow}^{\dagger},\dots,b_{n-k\downarrow}^{\dagger})^t$ is Nambu spinor in momentum space, and $n$ is the number of sublattices. $A(k)$ and $B(k)$ are $n\times n$ matrices, the Fourier transforms of pairing and hopping, respectively.

We can use $M(k)\in SU(n,n)$ for diagonalizing $D(k)$ to get the spectrum of spinons. Expressing
\begin{align}
  M(k)=
  \begin{pmatrix}
    u(k) & w(k)\\
    v(k) & x(k)\\
  \end{pmatrix},
  \label{boson_spinon_eigvec}
\end{align}
it is not hard to derive the BCS-type wavefunction for bosonic spinons as
\begin{align}
  |\Psi_b^{MF}\rangle=\prod_k \exp(\phi_{ij}(k)b_{ik\uparrow}^{\dagger}b_{j-k\downarrow}^{\dagger})|0\rangle,
  \label{spinon_wf}
\end{align}
where $\phi(k)=[u^{\dagger}(k)]^{-1}v^{\dagger}(k)=w(k)[x(k)]^{-1}$.

For the c-SDW/SCCL ansatz in Fig.\ref{fig:SCCL_pattern}, there are four sites in one unit cell so $n=4$ in this case.  The boson condensation occurs (i.e., long-range magnetic order is established) when the boson band minima at $\pm(\pi/2,\pi)$ touch zero. When this happens, the zero energy modes satisfying $D(\pm(\pi/2,\pi))\Psi(\pm(\pi/2,\pi))=0$ determine the magnetic ordering pattern. They are found to be (in one of the two degenerate ground states):
\begin{widetext}
\begin{align} \Psi[(\pi/2,\pi)]&=\frac{1}{2\sqrt{3+\sqrt{3}}}(-e^{i\pi/4},e^{i\pi/4},i\sqrt{2+\sqrt{3}},i\sqrt{2+\sqrt{3}},\frac{-1+i}{2}(1+\sqrt{3}),\frac{-1+i}{2}(1+\sqrt{3}),-1,1)\notag\\
 \Psi[(-\pi/2,\pi)]&=\frac{1}{2\sqrt{3-\sqrt{3}}}(e^{i\pi/4},e^{i\pi/4},i\sqrt{2-\sqrt{3}},-i\sqrt{2-\sqrt{3}},\frac{-1-i}{2}(-1+\sqrt{3}),\frac{-1+i}{2}(-1+\sqrt{3}),1,1).
\end{align}
\end{widetext}

The general boson condensate takes the form: $\langle\beta_{(\pi/2,\pi)}\rangle=c_1\Psi[(\pi/2,\pi)]$ and $\langle\beta_{-(\pi/2,\pi)}\rangle=c_2\Psi[(\pi/2,\pi)]$, where $c_1,c_2$ are two complex numbers. Here among the four real parameters in $c_1,c_2$, one of them, $|c_1|^2+|c_2|^2$, controls the magnitude of the magnetic order parameter. A different choice of the other three real parameters can be shown to generate a global $SU(2)$ spin rotation in the spin space. The real space magnetic order pattern is nothing but the tetradedral pattern with the chirality shown in Fig.\ref{fig:cSDW_dpid_real_space}. The other degenerate state can be obtained by time-reversal transformation.

Now let us look at the fermionic holon part. Hamiltonian of holons is free fermion hopping model,
\begin{align}
  H_f^{MF}=\sum_k \psi^{\dagger}(k)h(k)\psi(k),
\end{align}
where $\psi(k)=(f_{1k},\dots,f_{nk})^t$, and $n$ is band index. Using $W(k)\in SU(n)$ to diagonalize $h(k)$, we get
\begin{align}
  |\Psi_f^{MF}\rangle=\prod_{i,k} d_{ik}^{\dagger}|0\rangle,
\end{align}
where $d_{ik}=W_{ij}(k)^\dagger f_{jk}$. Fermions fill bands from the lowest to the $i$-th, depending on doping. In the case of c-SDW/SCCL phases, the doped holon fills the lowest band. On the mean-field level, it is straightforward to show that the holon real hoppings on the nearest neighbor and second neighbor give a band structure with Dirac points located at $\pm(\pi/2,\pi)$ in the lowest two bands. The imaginary hoppings (see Fig.\ref{fig:SCCL_pattern}) on the second neighbor open energy gaps at the two Dirac points and the resulting lowest band carries Chern number one. The wavefunction of c-SDW/SCCL is obtained from projection to physical Hilbert space as shown in Eq.(\ref{c-SDW_projection}). The wavefunction of c-SDW/SCCL is obtained from projection to phyical Hilbert space as shown in Eq.(\ref{c-SDW_projection}).

Finally, there is an important subtlety in the PSG construction related to finite samples. Although we explicitly construct a mean-field Ansatz which is invariant under a combination of symmetry and local gauge transformations, it is possible that we can not achieve this consistently on some finite lattice samples with PBC, i.e., having no open boundary. In c-SDW/SCCL case only $4N\times4N$ lattice sample supports the PSG pattern. However, when $\pi$-flux is included in both directions, the resulting Ansatz is symmetric (up to a gauge transformation) in $(4N+2)\times(4N+2)$ lattice samples. Wavefunctions obtained by $\pi$-flux insertion are related to topologically degenerate ground states in thermodynamic limit. We discuss this further in Appendix \ref{app:wf_quantum_number}.

\subsection{d+id SC state}

Construction of d+id SC state is much simpler. Mean field Ansatz is given in Sec.\ref{sec:sym_wavefunctions}. Consider the bosonic holon part first. For the uniform hopping model, bosons will condense at $\Gamma$ point, and only contribute a constant number after projection. For the fermionic spinon part, the mean field wavefunction is of BCS type:
\begin{align} |\Psi_{d+id,f}^{MF}\rangle=|k=0\rangle\otimes\prod_{k\neq0}\exp(\phi_{d+id,ab}(k)f_{ak\uparrow}^{\dagger}f_{b-k\downarrow}^{\dagger})|0\rangle.
\end{align}
Here, $\phi_{d+id}(k)=-[u^{\dagger}(k)]^{-1}v^{\dagger}(k)$, where $u(k)$ and $v(k)$ are $2\times2$ matrices and 
$(\begin{smallmatrix}
  u(k)\\
  v(k)
\end{smallmatrix})$
are eigenvectors corresponding to positive eigenvalues of $H_{d+id,f}^{MF}(k)$ in Eq.~\eqref{eq:Hmf_did}. Note that due to vanishing of pairing at the $\Gamma$ point, $|k=0\rangle=c_{k=0,\uparrow}^{\dagger}c_{k=0,\downarrow}^{\dagger}|0\rangle_{k=0}$ is not a BCS type wavefunction, and only contributes a constant number (similarly to the bosonic part), so we can omit it in the following analysis.

\section{Understanding quantum numbers}
\label{app:wf_quantum_number}
In this section, we use projective wavefunctions to analytically understand quantum numbers of c-SDW/SCCL and d+id SC on different lattice samples. The results are not limited to projective wavefunctions but hold throughout the quantum phase.

\subsection{c-SDW/SCCL state}
We will consider four wavefunctions formed from the considered Ansatz by flux insertion, as they represent the topologically degenerate ground state manifold (the flux is inserted through the handles of the torus formed by the periodic system). To understand quantum numbers for various lattice sizes, it is convenient to use momentum space. The mean-field Ansatz of c-SDW/SCCL already has a doubled unit-cell in $x$ direction, and to make the Brillouin zone more symmetric we double the unit-cell in the other direction too. This enlarged unit-cell contains 8 sites and in this entire Section we will call it the ``quadrupled UC'' to avoid any confusion (see Fig.~\ref{fig:SCCL_pattern}b). Thus Brillouin zone becomes a hexagon, and it is simpler to consider $C_6$ rotation in momentum space.

It turns out that all further calculations are greatly simplified if we immediately insert a $\pi$-flux through both directions of every quadrupled UC in the c-SDW/SCCL Ansatz. Then we consider two types of samples analogous to Fig.~\ref{fig:samples}a: The $4N\times4N\times2=2N\times2N\times8$, to which the 32-site sample belongs; and the $(4N+2)\times(4N+2)\times2=(2N+1)\times(2N+1)\times8$, to which the 8-site belongs. (Note that the latter family experiences the above $\pi$-flux insertion as an insertion through the entire system, and the Ansatz is changed to a topologically degenerate one; for the former family the flux insertion is a simple redefinition of gauge.) All PSG transformations can be performed consistently on all above samples in this redefined Ansatz. We label the state described by the redefined Ansatz as \rm{[0,0]}. The other three topologically degenerate states are obtained by adding $\pi$-flux through entire system in different directions, and the states are labeled as [0,$\pi$], [$\pi$,0] and [$\pi$,$\pi$]. In the following, we will analyze the quantum numbers of these four states.

\subsubsection{\rm{[0,0]} state}
Because the quadrupled unit-cell is doubled comparing to unit-cell of mean-field Ansatz, we get double degeneracy for every band. Boson band minimum is moved to $\Gamma$ point due to the insertion of $\pi$-flux through every quadrupled unit-cell. The special property of this \rm{[0,0]} state is that the mean-field Ansatz is indeed invariant up to a gauge transformation defined by PSG on all $2N\times2N\times 2$ lattice sizes. Further, gauge transformation $G_U$ associated with symmetry operation $U$ turns out to be independent of unit-cell, but only depends on sublattice index. For the other three states, we find that it is impossible to write a consistent mean field Ansatz invariant under all PSG operations (especially the $C_6$ rotation). In other words, the other three states break (rotation) symmetry explicitly.

In momentum space, PSG transformation is defined as
\begin{align}
  \begin{split}
    \mathbf{b}_{k\alpha}&\to G_U\cdot S_U(k)\cdot \mathbf{b}_{U\circ k\alpha}, \quad \alpha=\uparrow,\downarrow\\
    \mathbf{f}_{k}&\to G_U\cdot S_U(k)\cdot \mathbf{f}_{U\circ k},
  \end{split}
\end{align}
where $\mathbf{b}_{k\alpha}=(b_{1k\alpha},\dots,b_{nk\alpha})^t$, $\mathbf{f}_{k}=(f_{1k},\dots,f_{nk})^t$, while $n=8$ is number of bands(sublattices). $U\circ k$ is symmetry transformation for k points while $S_U(k)$ is an $n\times n$ unitary matrix which represents action of symmetry on sublattice. $G_U$ is the associated gauge transformation, with $(G_U)_{ij}=\delta_{ij}\exp(i\phi_U(i))$. Note that in general the gauge transformation of fermions has more freedom, and we can choose a different $G_U$ than for bosons. Here, for simplicity, we assume fermions have the same PSG as bosons. The mean-field Hamiltonian is invariant under PSG.

First we analyze the contribution to quantum numbers from fermionic (holon) part. For symmetry $U$ and associated gauge transformation $G_U$, the invariance of holon Hamiltonian can be expressed as
\begin{align}
  H_f(k)=G_US_U(k)H_f(U\circ k)S_U^{\dagger}(k)G_U^{\dagger}.
\end{align}
Setting $\alpha(k)$ as an eigenvector of $H_f(k)$ with eigenvalue $\lambda$, we can define $\alpha(U\circ k)\equiv S_U^{\dagger}(k)G_U^{\dagger}\alpha(k)$. It is easy to see that $\alpha(U\circ k)$ is indeed an eigenvector of $H_f(U\circ k)$ with eigenvalue $\lambda$. In this way, we can generate
\begin{align}
  \alpha(U^{i}\circ k)=S_U^{\dagger}(U^{i-1}\circ k)G_U^{\dagger}\alpha(U^{i-1}\circ k),\\\notag i=1,\dots,m_U-1,
\end{align}
where we assume $U^{m_U}\circ k=k$, and $m_U$ can vary for different $k$. Note that 
\begin{align}
  \alpha(k)&=\alpha(U^{m_U}\circ k)\\\notag
  &\neq S_U^{\dagger}(U^{m_U-1}\circ k)G_U^{\dagger} \alpha(U^{m_U-1}\circ k).
\end{align}
However, since there is a two-fold degeneracy, it is always possible to choose appropriate $\alpha(k)$ such that 
\begin{align}
 \alpha(k)=e^{i\theta_{\alpha(k)}} S_U^{\dagger}(U^{m_U-1}\circ k)G_U^{\dagger} \alpha(U^{m_U-1}\circ k).  
\end{align}

Now we apply symmetry on this set of states $\alpha(U^{i-1}\circ k), i=0,\dots,m_U-1$. By definition,
\begin{align}
  \mathbf{U}[\alpha(U^i\circ k)]=S_U(U^i\circ k)\alpha(U^{i+1}\circ k).
\end{align}
Using the definition of $\alpha(U^{i}\circ k)$, it is straightforward to derive
\begin{align}
  \mathbf{U}[\alpha(U^i\circ k)]=\left\{
    \begin{array}{l l}
      G_U^{\dagger}\alpha(U^i\circ k) & \quad i=0,\dots,m_U-2\\
      e^{i\theta_{\alpha(k)}}G_U^{\dagger}\alpha(U^{i}\circ k) & \quad i=m_U-1
    \end{array}.
    \right.\
\end{align}
So under symmetry operation, this set of eigenstates will pick up a $\theta_{\alpha(k)}$ phase plus a gauge transformation. It is clear that $\theta_{\alpha(k)}$ is directly related to Berry phase of symmetry operation, which is independent of our choice of basis. (To be more precise, this phase is invariant under $U(1)$ phase choice of $\alpha(U^i\circ k)$).
From the above transformation law, it is not hard to get the contribution to quantum numbers from holons. Examples will be presented below.

Let us now do a similar analysis on bosonic (spinon) part. For BCS-type Hamiltonian, the invariance of Hamiltonian under PSG transformation can be expressed as
\begin{align}
  H_b(k)=&
  \begin{pmatrix}
    G_US_U(k) & 0\\
    0 & G_U^*S_U^*(-k)\\
  \end{pmatrix}
  \cdot \\\notag 
  & H_b(U\circ k) \cdot
  \begin{pmatrix}
    S_U^{\dagger}(k)G_U^{\dagger} & 0\\
    0 & S_U^t(-k)G_U^t\\
  \end{pmatrix}.
\end{align}
Assuming $H_b(k)\beta(k)=\lambda \beta(k)$, and using a similar method to above, we can generate $\beta(U\circ k),\dots,\beta(U^{m_U-1}\circ k)$ as eigenvectors of $H_b(U\circ k),\dots,H_b(U^{m_U-1}\circ k)$ with eigenvalue $\lambda$. By appropriately choosing these vectors, it is possible to make
\begin{align}
  U[\beta(U^{i}\circ k)]=\exp[i\theta_{\beta(U^i\circ k)}]
  \begin{pmatrix}
    G_U^{\dagger} & 0\\
    0 & G_U^{t}\\
  \end{pmatrix}
  \cdot \beta(U^{i}\circ k).
\end{align}
In the following, we will show that the additional $U(1)$ phase $\exp[i\theta_{\beta(U^i\circ k)}]$ is unimportant for the BCS-type wavefunction. We only need to focus on $G_U$ in the BCS-type wavefunction.

Applying symmetry $U$ on $M(k)$ defined in Eq.(\ref{boson_spinon_eigvec}), we get
\begin{align}
  \mathbf{U}[M(k)]=&
  \begin{pmatrix}
    G_U^{\dagger} & 0\\
    0 & G_U^t\\
  \end{pmatrix}
  \cdot
  \begin{pmatrix}
    u(k) & w(k) \\
    v(k) & x(k) \\
  \end{pmatrix}
  \cdot
  \begin{pmatrix}
    \Theta_1(k) & 0 \\
    0 & \Theta_2(k) \\
  \end{pmatrix},
\end{align}
where $\Theta_1(k)$ and $\Theta_2(k)$ are $n\times n$ diagonal matrices, and their elements are additional $U(1)$ phases for different eigenvectors discussed above. Particularly,
\begin{align}
  \mathbf{U}[w(k)]=G_U^{\dagger}w(k)\Theta_2(k),\quad \mathbf{U}[x(k)]=G_U^{t}x(k)\Theta_2(k).
\end{align}
According to Eq.(\ref{spinon_wf}), Cooper pair creation operator is $\phi_{ij}(k)b_{ik\uparrow}^{\dagger}b_{j-k\downarrow}^{\dagger}$, where $\phi(k)=w(k)\cdot [x(k)]^{-1}$. So under symmetry transformation
\begin{align}
  \phi_{ij}(k)b_{ik\uparrow}^{\dagger}b_{j-k\downarrow}^{\dagger}\to [G_U^{\dagger}\phi G_U^*]_{ij}(k)b_{ik\uparrow}^{\dagger}b_{j-k\downarrow}^{\dagger},
\end{align}
only picking up a gauge transformation defined by PSG. We can view this as $b_{ik\alpha}^{\dagger}\to G_U(i)^{*}b_{ik\alpha}^{\dagger}$ under symmetry transformation $U$, where $G_U(i)$ is the $i$-th diagonal element of $G_U$. Since BCS-type wavefunction is formed by condensation of Cooper pairs, when acted on by symmetry, the only contribution comes from gauge transformation $G_U$. It is worth mentioning that this result also applies to fermionic singlet superconductor, which appears in the case of fermionic spinon in d+id SC.

In the following, we will apply the above results to symmetry group defined in Appendix~\ref{app:symmetry_group}.
First, let us consider the quantum number of $T_1$. Written in momentum space, its gauge transformation can be expressed as a diagonal matrix
\begin{align}
  G_{T_1}=\mathrm{Diag}[-1,1,-1,1,1,-1,1,-1],
\end{align}
depending only on sublattice index, while
\begin{align}
  S_{T_1}(k)=
  \begin{pmatrix}
    0 & \mathbf{I}_{4\times4} \\
    e^{ik_1}\mathbf{I}_{4\times4} & 0\\
  \end{pmatrix}.
\end{align}
Assuming $\alpha(k)$ is eigenstate of $H_f(k)$, then
\begin{align}
  \mathbf{T_1}[\alpha(k)]&=S_{T_1}(k)\cdot\alpha(k) \\\notag
  &=e^{i\theta_{\alpha}(k)}G_{T_1}^{\dagger}\alpha(k)
\end{align}
(after choosing a convenient $\alpha(k)$). It is easy to show that $\theta_{\alpha(-k)}=-\theta_{\alpha(k)}$. Thus the phase apart from $G_{T_1}^{\dagger}$ will always cancel. The holon wavefunction will transform as
\begin{align}
  T_1|\psi_f\rangle&=\prod_{i,k}f_j^{\dagger}(T_1^{\dagger}(k)W(k))_{ji}|0\rangle\\\notag
  &=\prod_{i,k}f_j^{\dagger}(G_{T_1}^{\dagger}W(k))_{ji}|0\rangle,
\end{align}
where $i=1,2$ for case of one quarter doping. We can view this as $f_j^{\dagger}\to G_{T_1}^{*}(j)f_j^{\dagger}$. 

Now we turn to the spinon wavefunction. According to previous analysis, spinon $b_{ik\alpha}^{\dagger}$ picks up phase $G_{T_1}^{*}(i)$ under $T_1$. For the total projective wavefunction, we have a constraint on Hilbert space: There is only one spinon or holon per site. Due to this constraint, the total phase obtained from $T_1$ is simply the product of $G_{T_1}^*(i)$ for all lattice sites. So the $T_1$ quantum number of c-SDW/SCCL is $1$.

For translation $T_2$, we do a similar procedure as for $T_1$, and find that the quantum number of $T_2$ also equals $1$. So, we can conclude that the center of mass of \rm{[0,0]} state is at $\Gamma$ point for $2N\times2N\times2$ lattice size, i.e., for both sample families introduced in this Section.

Let us turn to $C_6$ symmetry. It is straightforward to get the sublattice transformation matrix:
\begin{align}
  S_{C_6}(k)=
  \begin{pmatrix}
    0 & 0 & 0 & 0 & 0 & 0 & 0 & e^{-ik_2} \\
    1 & 0 & 0 & 0 & 0 & 0 & 0 & 0 \\
    0 & 1 & 0 & 0 & 0 & 0 & 0 & 0 \\
    0 & 0 & 0 & 0 & 0 & 0 & e^{-ik_1} & 0 \\
    0 & 0 & 0 & 0 & 0 & 1 & 0 & 0 \\
    0 & 0 & 1 & 0 & 0 & 0 & 0 & 0 \\
    0 & 0 & 0 & 1 & 0 & 0 & 0 & 0 \\
    0 & 0 & 0 & 0 & e^{i(-k_1+k_2)} & 0 & 0 & 0 \\
  \end{pmatrix}.
\end{align}
For fermionic holon, the gauge transformation can be chosen as 
\begin{align}
  G_{f,C_6}=\mathrm{Diag}[1,1,1,-1,1,-1,1,1],
\end{align}
while for bosonic spinon,
\begin{align}
  G_{b,C_6}=\beta G_{f,C_6},
\end{align}
where $\beta=\pi/6$. Note that although $G_{b,C_6}$ is also a consistent gauge transformation for fermion, we choose $G_{f,C_6}$ different from $G_{b,C_6}$ for simplicity.

We have three classes of $k$ points in Brillouin zone according to their transformation rule under $C_6$: 1) $\Gamma$ point, which transforms back to itself under $C_6$, so $m_{C_6}=1$; 2) Three $M$ points, which transform back to themselves under $C_6^3$(inversion), so $m_{C_6}=3$; 3) Other $k$ points, which are invariant only under $C_6^6$, so $m_{C_6}=6$. Using the method developed above, we calculate the additional $U(1)$ phase under $C_6$ for the 1st and 2nd holon band (in quarter doped case, holons always fill these 2 bands). The result is listed below:
\begin{center}
\begin{tabular}{|c|c|c|}
  \hline
  & $e^{i\theta_1}$ & $e^{i\theta_2}$ \\ \hline
  $\Gamma$ point & $-i$ & $e^{i5\pi/6}$ \\\hline
  $M$ points & $i$ & $-i$ \\\hline
  \text{Others} & $-1$ & $-1$ \\\hline
\end{tabular}
\end{center}
We checked this numerically for various mean-field parameter values.

It is easy to see that only $\Gamma$ point contributes additional phase, which equals $e^{i\pi/3}$. Under $C_6$ symmetry, holon wavefuncion transforms as
\begin{align}
  C_6|\psi_f\rangle&=\prod_{i,k}f_j^{\dagger}(C_6^{\dagger}(k)W(k))_{ji}|0\rangle\\\notag
  &=e^{i\pi/3}\prod_{i,k}f_j^{\dagger}(G_{f,C_6}^{\dagger}W(k))_{ji}|0\rangle.
\end{align}
For spinon part, the transformation law is
\begin{align} C_6|\psi_b\rangle&=\prod_{k}\exp[(G_{b,C_6}^{\dagger}\phi(k)G_{b,C_6}^*)_{ij}b_{ik\uparrow}^{\dagger}b_{j-k\downarrow}^{\dagger}]|0\rangle\\\notag
  &=\prod_{k}\exp[\beta^2(G_{f,C_6}^{\dagger}\phi(k)G_{f,C_6}^*)_{ij}b_{ik\uparrow}^{\dagger}b_{j-k\downarrow}^{\dagger}]|0\rangle.
\end{align}
We can view this as if every spinon picks up factor $\beta=e^{i\pi/6}$ (plus fermion gauge transformation) under $C_6$.
% Note that contribution from $G_{f,C_6}$ is trivial due to constraint of Hilbert space.
It is straightforward to calculate that for $4N\times4N\times2$ lattice size $C_6$ quantum number equals $e^{i\pi/3}$, while for $(4N+2)\times(4N+2)\times2$ lattice size $C_6$ quantum number is $e^{-2i\pi/3}$. For the state related by time reversal, quantum numbers are obtained by conjugation.

\subsubsection{Other three states}
Using the method developed above, we calculated translation quantum numbers of the three other states. It turns out that the center of mass of these three states are three $M$ points ((0,$\pi$), ($\pi$,0) and ($\pi$,$\pi$)). While calculation details will not be presented in this paper, there is a simple physics picture. Consider adding $\pi$-flux in $x$ direction to the {\rm{[0,0]}} state, and then translating in the same direction. This corresponds to every fermion hopping one lattice spacing in $x$ direction, and they will see this additional $\pi$-flux. Thus, compared to original state, the translation quantum number in $x$ direction is multiplied by $-1$, so the center of mass will change from $\Gamma$ point to $M$ point $[\pi,0]$.

For rotation, we note that $C_6$ is not a symmetry for these three states. But the three states are symmetric under inversion symmetry $C_6^3$. Applying the above method, we find that these three states have opposite inversion quantum number to {\rm{[0,0]}} state, which is consistent with our field theory analysis in Section~\ref{sec:modular}.

\subsection{d+id SC state}
Understanding the quantum numbers of d+id SC is much simpler. Firstly, holons always condense at $\Gamma$ point, and contribute an overall constant, thus can be neglected. Secondly, two spinons that occupy $\Gamma$ point will also have no contribution, as discussed in Appendix~\ref{app:wavefunction_construction}. For other spinons, which have a BCS-type wavefunction, the analysis of quantum numbers is similar to bosonic spinon part above: Under lattice symmetry, only gauge transformations contribute to quantum numbers.

For translation $T_1$ and $T_2$, associated gauge transformations $G_{T_1}$ and $G_{T_2}$ are trivial. So, the center of mass is $\Gamma$ for any lattice size.

Under $C_6$ rotation, mean-field wavefunction changes as
\begin{align}
  C_6|\psi_f\rangle =\prod_{k\neq0}\exp[e^{-i2\pi/3}\phi_{d+id,ij}(k)f_{ik\uparrow}^{\dagger}f_{j-k\downarrow}^{\dagger}]|0\rangle.
\end{align}
We can view this as if every fermion picked up $e^{-i\pi/3}$ after $C_6$ (except for fermions at $\Gamma$ point). So $C_6$ quantum number for $2N\times2N\times2$ lattice size is
\begin{align} \left(\frac{3}{4}\times2N\times2N\times2-2\right)\times\left(-\frac{\pi}{3}\right)=\frac{2\pi}{3}\ mod \ 2\pi,
\end{align}
independent of lattice size.

Next consider the inversion $C_6^3$ quantum number.  For d+id SC, it is always 1. For c-SDW/SCCL, on $4N\times4N\times2$ lattice size, inversion quantum number equals -1, while on $(4N+2)\times(4N+2)\times2$, inversion quantum number is 1. This provides a sharp signature to distinguish c-SDW/SCCL state and d+id SC in finite samples.

% \section{Quantum numbers of competing phases on the 24-site sample}\label{app:24_site_sample}

%Although the 32-site sample is likely to be too large, a fully symmetric 24-site sample, as shown in Fig.\ref{fig:24_site_sample}, may have a reasonable size suitable for exact diagonalization numerical simulations in the currently available computing power. We therefore investigated the quantum numbers of the three competing phases, c-SDW/SCCL and d+id superconductor, on this sample. We find that all these phases share the same quantum numbers as in Table \ref{tab:irreps}(b). Therefore exact diagonalizations on this sample cannot sharply distinguish them.

\section{Edge theory of SCCL with added $\pi$-spin-rotation about perpendicular axis}
\label{app:edgesym}

To consider spin rotation symmetry in x and y directions, one must enlarge the $\mathbf{K}$ matrix by adding degrees of freedom that are in a topologically trivial phase
$\bigl(\begin{smallmatrix}
  0 & 1 \\
  1 & 0 \\
\end{smallmatrix}\bigr)$. Then we get
\begin{align}
  \mathbf{K'}=
  \begin{pmatrix}
    1 & 0 & -1 \\
    0 & 0 & 2 \\
    -1 & 2 & 0 \\
  \end{pmatrix}
  \oplus
  \begin{pmatrix}
    0 & 1 \\
    1 & 0 \\
  \end{pmatrix}
  =
  \begin{pmatrix}
    1 & 0 & -1 & 0 & 0 \\
    0 & 0 & 2 & 0 & 0 \\
    -1 & 2 & 0 & 0 & 0 \\
    0 & 0 & 0 & 0 & 1 \\
    0 & 0 & 0 & 1 & 0 \\
  \end{pmatrix}.
\end{align}
Performing a transformation on $\mathbf{K'}$ gives us
\begin{align}
  \mathbf{K}=\mathbf{X}^t\mathbf{K'X}=
  \begin{pmatrix}
    1 & 0 & 0 & 1 & 0 \\
    0 & 0 & 0 & -1 & 1 \\
    0 & 0 & 0 & 1 & 1 \\
    1 & -1 & 1 & 0 & 0 \\
    0 & 1 & 1 & 0 & 0 \\
  \end{pmatrix},
\end{align}
where we used
\begin{align}
  \mathbf{X}=
  \begin{pmatrix}
    1 & 0 & 0 & 0 & 0 \\
    0 & 1 & 0 & 0 & 0 \\
    0 & 0 & 0 & -1 & 0 \\
    0 & 1 & 1 & 0 & 0 \\
    0 & 0 & 0 & 1 & 1 \\
  \end{pmatrix}.
\end{align}
Any such transformation by a matrix $X$ in $GL(5,\mathbb{Z})$, the group of unimodular $N\times N$ matrices, can be seen as a relabeling of topological degrees of freedom since $\mathbf{l}\to\mathbf{l'}=\mathbf{X^{t}}l$, and the physics remains unchanged. However, the particular choice of $X$, inspired by Ref. \onlinecite{Lu:2013p7858}, allows an easier identification of physical properties. The charge vector $\mathbf{t}_c=(1,0,0,0,0)$ and $S_z$ vector $\mathbf{t}_{S_z}=(1/2,-1,0,0,0)$ are direct extensions of the original ones. We identify holon $f$ as $(1,0,0,0,0)$, spinon $b_{\uparrow}$ as $(0,0,0,1,0)$ and spinon $b_{\downarrow}$ as $(0,0,0,0,1)$. Electron $e_{\uparrow (\downarrow)}$ is simply the bound state of $f$ and $b_{\uparrow (\downarrow)}$, and it is in the topologically trivial sector. Vison can be viewed as `half holon', and is expressed as $(0,1,0,0,0)$ with charge $\frac{1}{2}$ and $(0,0,1,0,0)$ with charge $-\frac{1}{2}$. It is easy to check that the statistical angles and quantum numbers of these quasiparticles are correct.

The general consideration of symmetry in this $\mathbf{K}$-matrix formulation has been considered in Ref.\onlinecite{Lu:2013p7858}. Under symmetry $g\in G_s$, chiral boson field $\phi_I$ will transform as
\begin{align}
  &\phi_I\to\sum_J W_{I,J}^g\phi_J+\delta\phi_I^g \\\notag
  &\mathbf{K}=(\mathbf{W}^g)^t\mathbf{K}\mathbf{W}^g, \quad \mathbf{W}^g\in GL(N,\mathbb{Z}).
\end{align}
Notice that the above symmetry transformations $\{\mathbf{W}^g,\delta\phi^g|g\in G_s\}$ must be compatible with group structure of symmetry group $G_s$. More precisely, the local bosonic degree of freedom $\{\hat{M}_I\equiv e^{il_I\sum_JK_{I,J}\phi_J}\}$ must form a linear representation of symmetry group $G_s$ while nonlocal quasiparticles can transform projectively.

It is not yet known how to incorporate the full $SU(2)$ spin rotation symmetry in the $\mathbf{K}$-matrix formulation. However, we can choose a subgroup of $SU(2)$, which is generated by $g_1$, the $\pi$ rotation around $S_x$ direction, and rotations around $S_z$ direction, $U_{\theta}, \theta\in[0,4\pi)$. They satisfy the following algebra:
\begin{align}
  g_1^4=e\\\notag
  U_{\theta_1}U_{\theta_2}=U_{\theta_1+\theta_2\,mod\,4\pi}\\\notag
  U_{\theta}g_1=g_1U_{-\theta}
\end{align}
In fact, we can view this group as a projective representation of a $SO(2)_z\rtimes Z_2$ subgroup of $SO(3)$.

Following Ref. \onlinecite{Lu:2013p7858}, we find a consistent solution for $\{\mathbf{W}^g,\delta\phi^g|g\in G_s\}$ that describes SCCL, namely
\begin{align}
  \mathbf{W}^{g_1}=
  \begin{pmatrix}
    1 & 0 & 0 & 0 & 0 \\
    1 & -1 & 0 & 0 & 0 \\
    0 & 0 & 1 & 0 & 0 \\
    0 & 0 & 0 & 0 & 1 \\
    0 & 0 & 0 & 1 & 0 \\
  \end{pmatrix},
  \quad \delta\vec{\phi}^{g_1}=
  \begin{pmatrix}
    0\\
    0\\
    0\\
    \pi/2\\
    \pi/2\\
  \end{pmatrix},
  \\\notag
  \mathbf{W}^{U_{\theta}}=\mathbf{1}_{5\times 5}, \quad \delta\vec{\phi}^{U_{\theta}}=(0,0,0,\theta/2,-\theta/2)^t.
\end{align}
Explicitly, the quasiparticles transform as
\begin{align}
  \vec{\phi}=
  \begin{pmatrix}
    \phi_1\\
    \phi_2\\
    \phi_3\\
    \phi_4\\
    \phi_5\\
  \end{pmatrix}
  \xrightarrow{g_1}
  \begin{pmatrix}
    \phi_1\\
    \phi_1-\phi_2\\
    \phi_3\\
    \phi_5+\pi/2\\
    \phi_4+\pi/2\\
  \end{pmatrix},\\\notag
  \vec{\phi}=
  \begin{pmatrix}
    \phi_1\\
    \phi_2\\
    \phi_3\\
    \phi_4\\
    \phi_5\\
  \end{pmatrix}
  \xrightarrow{U_{\theta}}
  \begin{pmatrix}
    \phi_1\\
    \phi_2\\
    \phi_3\\
    \phi_4+\theta/2\\
    \phi_5-\theta/2\\
  \end{pmatrix}.
\end{align}
Notice that $\phi_2$ and $\phi_1-\phi_2$ only differ by a trivial boson, so they are actually the same quasiparticle with the same quantum numbers. Further, each spinon ($\phi_{4,5}$) acquires $-1$ Berry phase after $2\pi$-spin-rotation, while holon $\phi_1$ and vison $\phi_{2,3}$ transform trivially, as expected.

There are several Higgs terms allowed on a symmetric boundary by this transformation law. However, one should consider the largest subset such that all terms can condense simultaneously, meaning that the arguments of these terms commute. Furthermore, the condensed fields must not break the symmetry. Thus we arrive at the following Higgs terms:
\begin{align}
  \mathcal{L}_{Higgs}&=C_1\cos(\phi_1+2\phi_3)+C_2\cos(\phi_1-2\phi_2).
%  \mathcal{L}_{Higgs,2}&=C_1'\cos(2\phi_4+2\phi_5)+C_2'\cos(\phi_1-2\phi_2),
\end{align}
Define $\theta_1=\phi_1+2\phi_3, \theta_2=\phi_1-2\phi_2$ and their conjugate variables are $\varphi_1=\phi_4+\phi_5, \varphi_2=\phi_4-\phi_5$. It is easy to show that $\{\theta_1,\varphi_1\},\{\theta_2,\varphi_2\}$ form two decoupled Luttinger liquids, so they can be gapped by the Higgs term $\mathcal{L}_{Higgs} =C_1\cos\theta_1+C_2\cos\theta_2$. The only remaining gapless degree of freedom is $\phi_1$. So the edge of SCCL is chiral fermion liquid of holons.

\section{Tunneling conductance calculation for different junctions}
\label{app:tunneling}

\subsection{Point junctions}

For completeness we repeat the metal lead/SCCL case from the main text here.

\begin{itemize}
  \item {\bf{Metal and c-SDW}}
\end{itemize}
Tunneling Hamiltonian is 
\begin{align}
  H_{tunn}=[t\psi_{c-SDW}^{\dagger}(x=0)\psi_M(x=0)+h.c.].
\end{align}
For Fermi liquid systems, the scaling dimension $\delta_{FL}=1/2$ in any dimension.\cite{Shankar:1994p7862} So, $\delta=2\delta_{FL}=1$. Using Eq.(\ref{conduct_scale}), we get that $G(T)$ is constant in this case.
\begin{itemize}
  \item {\bf{Metal and SCCL}}
\end{itemize}
Due to the spin gap on the boundary of SCCL, single electron will decay exponentially when tunneling to edge of SCCL. So, the major contribution is from singlet pair tunneling. 
\begin{align} H_{tunn}=[tf^{\dagger}(x=\xi)f^{\dagger}(x=0)\psi_{M,\uparrow}(x=0)\psi_{M,\downarrow}(x=0)+h.c.],
\end{align}
where superconducting coherence length $\xi$ appears due to the Pauli principle. So, $\delta_M=2\delta_{FL}=1$, where we used $\delta_{FL}=1/2$. The operator $f(x=\xi)f(x=0)$ has the same scaling dimension as operator $f(x=0)\partial_xf(x=0)$, giving $\delta_0=1+2\delta_{FL}=2$. We get $G(T)\sim T^4$.
\begin{itemize}
  \item {\bf{SC and c-SDW}}
\end{itemize}
The tunneling Hamiltonian is
\begin{align}
  H_{tunn}=[t\psi_{c-SDW}^{\dagger}(x=\xi)\psi_0^{\dagger}(x=0)\hat{c}(x=0)+h.c.]
\end{align}
where Cooper pair operator $\hat{c}$ is a complex number inside SC. Therefore $\delta=2$ in this case, and we get $G(T)\sim T^2$.
\begin{itemize}
  \item {\bf{SC and SCCL}}
\end{itemize}
Since singlet Cooper pairs are not influenced by spin gap, the result should be the same as for {\bf{SC and c-SDW}}, namely $G(T)\sim T^2$.

We next present the perturbative Fermi golden rule calculations for various point junctions.

The tunneling current is 
\begin{align}
  I(V)=2\pi t^2\int_0^V\rho_0(V)\rho_{M/SC}(V)\dd V,
\end{align}
where
\begin{align}\notag
  \rho_0(V)=&\sum_{N}|{}_0\langle N|O_0^{\dagger}(x=0)|\tilde{0}\rangle_0|^2\delta(E_{N}^0-V-E_{\tilde{0}}^0)\sim\\
  \sim& \int_{-\infty}^{\infty}e^{iVt'}\langle O_0(x=0,t')O_0^{\dagger}(x=0,0)\rangle\dd t',
\end{align}
while
\begin{align}\notag
&\rho_{M/SC}(V)=\sum_{N}\mid{}_{M/SC}\langle N|O_{M/SC}(x=0)|\tilde{0}\rangle_{M/SC}|^2\times\\
&\qquad\qquad\times\delta(E_N^{M/SC}+V-E_{\tilde{0}}^{M/SC})\sim\\\notag
&\sim \int_{-\infty}^{\infty}e^{iVt'}\langle O_{M/SC}^{\dagger}(x=0,t')O_{M/SC}(x=0,0)\rangle\dd t'.
\end{align}
Here, $O_0(O_{M/SC})$ is electron or electron pair annihilation operator of c-SDW/SCCL(M/SC). The scaling dimension of $I(V)$ is encoded in the long-time correlator of $O_0$ and $O_{M/SC}$.

For tunneling junction between metal and c-SDW, $O_0=\psi_{c-SDW}$ and $O_{M/SC}$ is $\psi_M$. Then, 
\begin{align}
  \langle\psi_{c-SDW}(x,t)\psi_{c-SDW}^{\dagger}(x,0)\rangle\sim t^{-1}\\\notag
  \langle\psi_M^{\dagger}(x,t)\psi_M(x,0)\rangle\sim t^{-1}
\end{align}
So, $\rho_0(V)$ and $\rho_M(V)$ are constant numbers. We get $I(V)\sim \int_0^V\rho_0(V)\rho_M(V) \dd V \sim V$, and tunneling conductance $G(V)=dI/dV$ is constant.

For tunneling junction between metal and SCCL, only singlet pairs can tunnel. The above formulas give $\rho_0(V)\sim V^3$, while $\rho_M(V)\sim V$. So, $I\sim \int_0^V V^3\dd V\sim V^5$, and conductance $G(V)\sim V^4$.

For SC lead, the main contribution is from tunneling of singlet Cooper pairs. Therefore, $G(V)$ scales in the same way for c-SDW and SCCL, and we get $G(V)\sim V^2$. Comparing with the above results, the perturbative calculation is indeed consistent with simple RG analysis.

\subsection{Line junction}

Voltage on the metal/SC side is a constant number, labeled by $V_R$ (Fig.~\ref{fig:lc}). Electron scattered from lead will lose its phase and always keep at the same voltage. On the c-SDW/SCCL side, voltage is maintained between scattering events and is accumulated, as shown in Fig.~\ref{fig:lc}.

We first completely derive the case of junctions for which the point contact scaling exponent $\alpha\neq0$, and deal with the $\alpha=0$ case (c-SDW/metallic lead) at the end.

The voltage at $n$-th point junction is labeled by $V_n$, while the tunneling current is $I_n$. Due to anomalous quantum Hall response of electron/holon, we get
\begin{align}
  \label{eq:edgecurrent}
  V_n-V_{n-1}=\frac{I_n}{e^2/h}.
\end{align}
According to the result for point contact having $\alpha\neq0$, in the regime $T\ll V$,
\begin{align}
  I_n=\frac{e^2}{h}\frac{(V_R-V_{n-1})^{\alpha+1}}{(T_K^{(n)})^{\alpha}},
  \label{line_junction_In}
\end{align}
where $\alpha$ is the scaling exponent for tunneling conductance obtained in point junction case. Eq.(\ref{line_junction_In}) can also be viewed as definition of $T_K^{(n)}$. Define $x_n=V_R-V_n$, to get
\begin{align}
  (x_{n-1}-x_{n})\cdot\frac{e^2}{h}=\frac{e^2}{h}\frac{x_{n-1}^{\alpha+1}}{(T_K^{(n)})^{\alpha}},
\end{align}
which we can transform into a differential equation:
\begin{align}
  -\frac{\mathrm{d}x}{\mathrm{d}n}=\frac{x^{\alpha+1}}{(T_K^{(n)})^{\alpha}}.
\end{align}
Integrating the above equation from the initial $x_0$ to the final $x_N$ yields
\begin{align} -\int_{x_0}^{x_N}\frac{\mathrm{d}x}{x^{\alpha+1}}=\sum_{n=1}^N\frac{1}{(T_K^{(n)})^{\alpha}}=\frac{1}{T_{K}^{\alpha}},
\end{align}
in which we defined the effective $T_K$ from the individual $T_K^{(n)}$. It is much smaller than $T_K^{(n)}$ for large $N$ (given the positive values of $\alpha$). Here $T_K$ becomes the only important parameter which incorporates $T_K^{(n)}$ as well as their fluctuations.

After integration, one obtains
\begin{align}
  x_N=\frac{VT_K}{(\alpha x_0^{\alpha}+T_K^{\alpha})^{1/\alpha}}
\end{align}
where we define $V\equiv x_0=V_R-V_0$. The total current flowing from metal/SC to c-SDW/SCCL is obtained from the voltage difference $V_N-V_0$, and is given by 
\begin{align}
  I=\frac{e^2}{h}(V_N-V_0)=\frac{e^2}{h}(V-x_N),
\end{align} 
so the tunneling conductance is
\begin{align}
  G(V)=\frac{e^2}{h}\left[1-\frac{T_K}{(\alpha V^{\alpha}+T_K^{\alpha})^{1/\alpha}}\right].
\end{align}
The result expressed holds for all values of $V,T_K$ at $T\to 0$, as long as the assumptions used to derive the expression holds, namely, each individual point contact junction is weakly coupled, $V\ll T_K^{(n)}$ for all $n$. Note that the effective $T_K$ for a long line junction (large $N$) can be very small compared to all $T_K^{(n)}$.

Now, let us consider the small voltage regime, namely, $V$ much smaller than temperature $T$. However, we still require the weak coupling condition for single point junctions, namely, $T\ll {T'}_K^{(n)}$. Notice that in general ${T'}_K^{(n)}\neq T_K^{(n)}$, but we expect they have similar magnitudes. In this case, according to point junction result
\begin{align}
  I_n=\frac{e^2}{h}\frac{T^{\alpha}}{({T'}_K^{(n)})^{\alpha}}\cdot(V_R-V_{n-1}).
\end{align}
Following similar steps as above, we get
\begin{align}
  -\frac{\mathrm{d}x}{\mathrm{d}n}=\frac{T^{\alpha}}{({T'}_K^{(n)})^{\alpha}}\cdot x
\end{align}
By solving this equation, it is straightforward to get the tunneling conductance as a function of $T$:
\begin{align}
  G(T)=\frac{e^2}{h}\left[1-e^{-\frac{T^{\alpha}}{(T'_K)^{\alpha}}}\right],
\end{align}
where we define
\begin{align}
  \frac{1}{({T'}_K)^{\alpha}}\equiv\sum_{n=1}^{N}\frac{1}{({T'}_K^{(n)})^{\alpha}}.
\end{align}

Finally we consider the c-SDW/metallic lead line junction, i.e., the case of $\alpha=0$. The derivation procedure is the same as for the above case, and starts from the point junction result:
\begin{align}
  I_n=\left\{\begin{aligned}
&\frac{e^2}{h}\frac{1}{c_K^{(n)}}(V_R-V_{n-1}),& \text{if $T\ll V$},\\
&\frac{e^2}{h}\frac{1}{{c'}^{(n)}_K}(V_R-V_{n-1}),& \text{if $V\ll T$},
\end{aligned}\right.
\label{eq:a0edgecurrent}
\end{align}
with $c_K^{(n)},{c'}^{(n)}_K$ dimensionless constants characterizing the $n$-th point junction. In fact, $c_K^{(n)},{c'}^{(n)}_K$ are defined by these equations, and the expression are valid in the weak-coupling regime of the point junction, i.e., $1\ll c_K^{(n)},{c'}^{(n)}_K$, which physically corresponds to low enough temperatures and voltages. Using Eqs.~\eqref{eq:edgecurrent},~\eqref{eq:a0edgecurrent}, and the same procedure as above, we get:
\begin{align}
  I/V=\left\{\begin{aligned}
&\frac{e^2}{h}\left[1-\exp(-1/c_K)\right],& \text{if $T\ll V$},\\\notag
&\frac{e^2}{h}\left[1-\exp(-1/c'_K)\right],& \text{if $V\ll T$},
\end{aligned}\right.
\end{align}
where we defined
\begin{align} \frac{1}{{c}_K}\equiv\sum_{n=1}^{N}\frac{1}{{c}^{(n)}_K},\quad\frac{1}{c'_K}\equiv\sum_{n=1}^{N}\frac{1}{{c'}^{(n)}_K}.
\end{align}

%\section{Variational Monte Carlo calculations of the c-SDW/SCCL Ansatz in the t-J model}
%[{\color{red} SJ: please fill details, what we need: energy curve, magnetic moment as a function of J/t(showing strong quantum fluctuation)} ]

\section{DMRG data and convergence}
\label{app:dmrg}
\begin{figure}
 \includegraphics[width=0.48\textwidth]{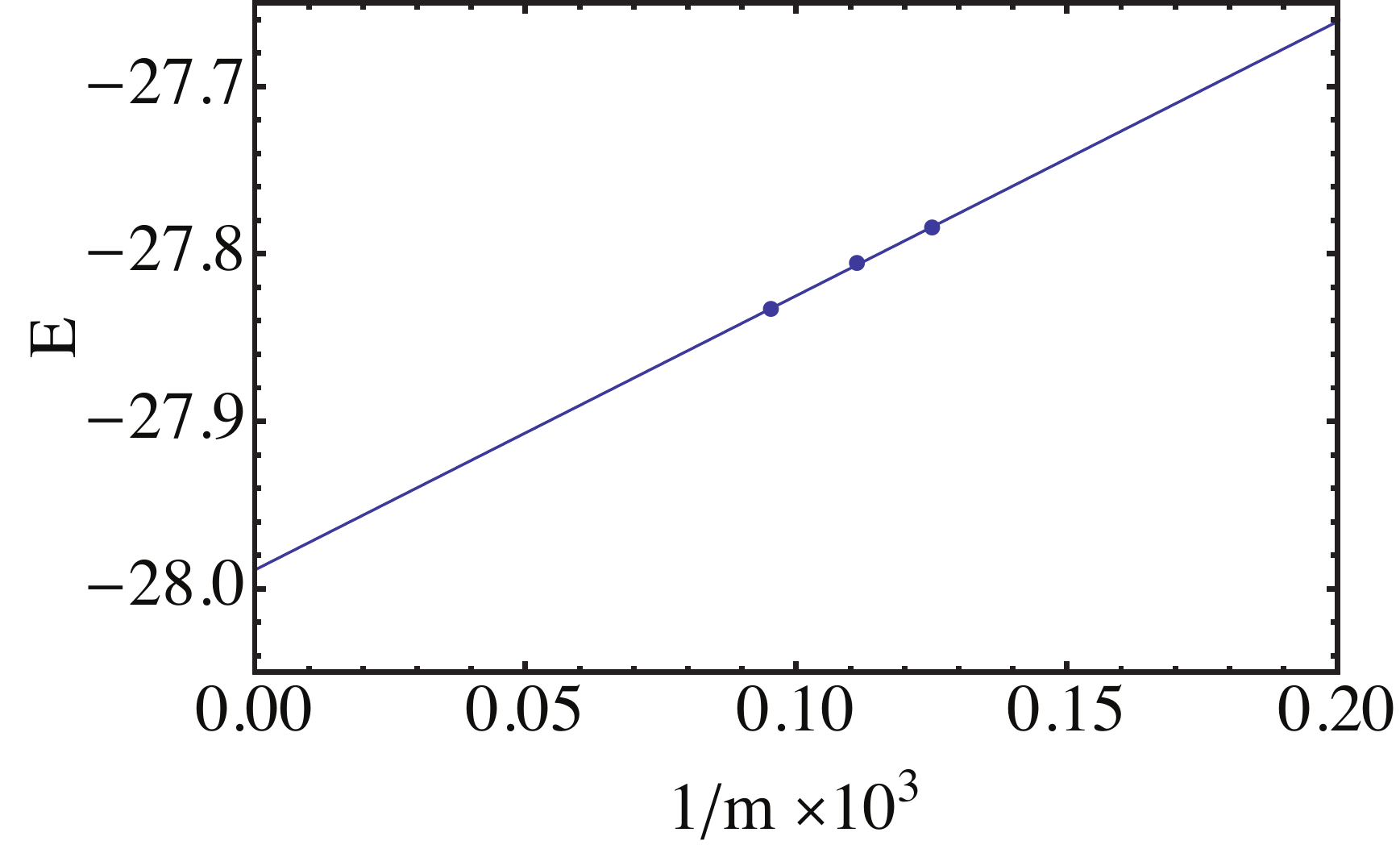}
 \caption{DMRG energy of Hubbard model with $U/t=6$ on 32-site sample, as function of limiting MPS matrix size $m$.}
 \label{fig:mconv}
\end{figure}

\begin{figure*}
 \includegraphics[width=0.96\textwidth]{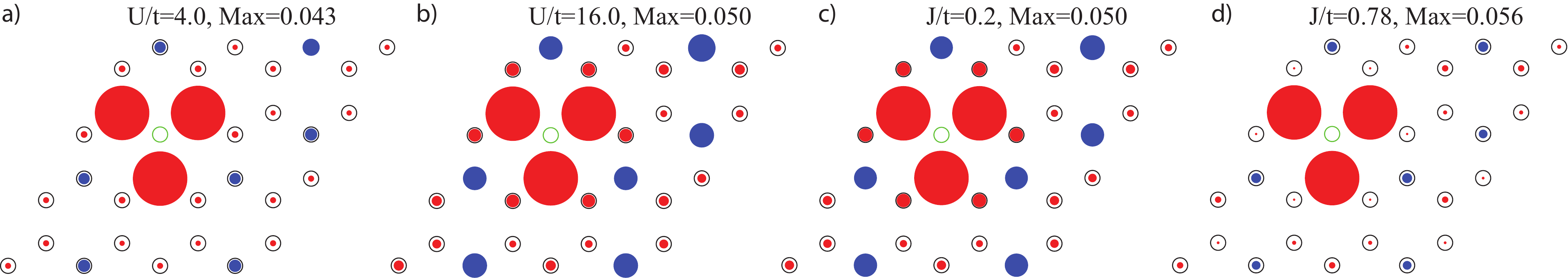}
 \caption{Spin-spin correlation function $\langle S_z(i)S_z(j)\rangle$ in the DMRG ground state projected into sector with center of mass momentum $\Gamma$ and $C_6$ rotation eigenvalue $\exp(-i\pi/3)$. Blue is positive, red negative, and disk radius is proportional to amplitude. Site $i$ is fixed at green circle, while every bond $i,j$ is averaged over translations and rotations to reduce statistical error by increasing the number of sampled observable values in MC. ``Max'' labels absolute amplitude of largest shown disk. All four parameters are chosen within the c-SDW/SCCL phase: (a),(b) Hubbard model; (c),(d) t-J model.}
 \label{fig:SzSzcolorAPP}
\end{figure*}

Here we discuss the precise DMRG setup, convergence to true ground state with limiting MPS matrix size, and also present some measurements for parameter values not shown explicitly in the main text.

To represent the two-dimensional periodic samples in the DMRG in a way that eases convergence, we labeled the sites $1\ldots N$ such that the longest necessary hopping range is minimized. For present samples, which have aspect ratio of 1, it was sufficient to sequentially order site labels $1\ldots N$ from, say, left to right within each row and then from one row to the next. With larger two-dimensional samples in lattices with higher coordination, it is advantageous to avoid labeling rows sequentially, but instead, starting from one row, sequence the one below it, then the one above it, and so on in an alternating fashion. We have checked for some parameter values that the labeling which minimizes the longest range hopping indeed allows faster sweeps and better convergence in the same amount of time.

The convergence of DMRG energy is however limited in practice by the maximal size of MPS matrices, $m$, which does not exceed 11.000 in our calculations. In Fig.~\ref{fig:mconv} we present a typical convergence of DMRG energy as a function of $1/m$, with a linear fit extrapolation towards infinite $m$. This is not the common way of considering DMRG convergence, but it is informative given our $m$ value limitations.

As discussed in Section~\ref{sec:32site_DMRG}, in this paper we quantify the DMRG convergence to the true ground state by using the expectation value of $C_6$ symmetry operation ($60^\circ$ rotation), which should be one of $\{-1,1/2\}$ ($\{1,-1/2\}$) when the inversion is $-1$ ($+1$). (The inversion is always numerically very precisely quantized.) As shown in Fig.~\ref{fig:c6}, the $C_6$ measurement on the 32-site sample indicates the convergence failure in the $J/t>0.8$ phase; in the Hubbard model, the convergence progressively worsens with lowering $U/t$ below the value 5. It is not surprising that convergence worsens for low $U/t$, but we believe it is mainly due to our $m$ limitation. For instance, at $U/t=4$ the $C_6$ expectation with projection to $\Gamma$ momentum improves from $0.10(2)$ at $m=8.000$ to $0.16(2)$ at $m=10.500$.

Next, we present additional details about correlation functions on the 32-site sample.

In Fig.~\ref{fig:SzSzcolorAPP} we show the spin-spin correlation function in the c-SDW/SCCL phase for several values of parameters, as addition to Fig.~\ref{fig:szszJcolor}. The values are chosen to demonstrate how the longer-range spin correlations match the tetrahedral pattern even better as $U/t$ grows and as $J/t$ decreases. On the other hand, the magnitude of short-range spin correlations grows with both $U/t$ and $J/t$ as expected. Let us here emphasize again that we use the total $S_z$ equal to zero sector in both models throughout this paper. The DMRG calculation conserves this quantum spin number of a state, as well as the total number of fermions.

In Section \ref{sec:32site_DMRG} we claimed that the pair-pair correlation function on 32-site sample in the c-SDW/SCCL phase is very short ranged. Here we provide a numerical example to compare to 24-site sample results in Table~\ref{tab:pp}. On 32-site sample we consider the $J/t=0.78$ DMRG GS projected to $\exp(-i\pi/3)$ eigenspace of $C_6$, and pairs of nearest-neighbor bonds separated exactly as in figure of Table~\ref{tab:pp}. Every correlation value is obtained using 64 MC runs of $10.000$ measurements, and averaged over translations of the bond pair to additionally reduce statistical error. (The usual $500$ measurements give a statistical error that overwhelms the value of correlations.) Out of the three bond pairs, the maximal correlation magnitude is $0.00059(2)$, to be compared with $0.00163(2)$ and $0.00447(2)$, the values for $J/t=0.78$ and $J/t=2.0$, respectively, for 24-site sample from Table~\ref{tab:pp}. The complex phases of the three bond-pair correlations in d+id state are $0,1/3,-1/3$ in units of $2\pi$, but in the considered 32-site measurement we find $0.5(1),0.40(5),-0.45(4)$.

Finally, we explained in Section \ref{sec:32site_DMRG} that on the 32-site sample the DMRG GS did not converge well in the large-$J$ phase $J/t>0.8$, so the correlation measurements are not trustworthy, but we note for completeness that in that regime the obtained DMRG GS with projection to $\Gamma$ center of mass momentum and either $\exp(-i\pi/3)$ or $\exp(-i2\pi/3)$ eigenvalue of $C_6$, completely loses resemblance to tetrahedral spin pattern without developing a d+id pair-pair correlation pattern.

\section{Exact perturbative results in the Hubbard model on the 32-site sample}
\label{app:perturbation}
First let us set $t=1$ and tune $U=0$. There are 16 momentum points in the Brillouin Zone. Sorted by the distance to the $\Gamma$ point, we have: one $\Gamma$ point, 6 points related to $(\pi/2,0)$ by $C_6$ rotations, 6 points related to $(\pi/2,\pi)$ by $C_6$ rotations (these are the midpoints between adjacent $M$-points), and 3 $M$-points. Each momentum point has two energy levels (the valence and conduction band) with each level having two-fold spin degeneracy. At $3/4$-filling, the 24 electrons should fully fill the conduction band at the $\Gamma$ point and the 6 points related to $(\pi/2,0)$ by $C_6$ rotations. The remaining 10 electrons will partially fill conduction band at the 6 mid-points between the $M$-points and the 3 $M$-points. Note that due to the hexagonal shape of the Fermi surface, these 9 momentum points have the same energy.

We will consider the $S_z=0$ sector only. This means that one fills 5 spin-up(spin-down) electrons in the 9 states, which gives a total of $\begin{pmatrix}9\\5\end{pmatrix}^2=15876$ degenerate many-body states with $E_0=-42.8328$. We also only focus on the sector with center of mass momentum being $\Gamma$. This further reduces the number of degenerate ground states down to $1002$.

Next we turn on a small $U$ and perform the standard degenerate perturbation calculation by diagonalizing the $1002$x$1002$ matrix of the $U$-term. To the first order in $U$ we find that the ground state becomes two-fold degenerate, with energy given by $E=E_0+4.11095\cdot U$. This two-fold ground state forms the same irrep of the symmetry group as the c-SDW/SCCL phase on this 32-site sample (see Table~\ref{tab:irreps}).
\end{appendix}

\bibliography{onequarterhoneycomb}

%merlin.mbs 2010-03-15 4.21a (PWD, AO, DPC)
\begin{thebibliography}{105}%
\makeatletter
\providecommand \@ifxundefined [1]{%
 \@ifx{#1\undefined}
}%
\providecommand \@ifnum [1]{%
 \ifnum #1\expandafter \@firstoftwo
 \else \expandafter \@secondoftwo
 \fi
}%
\providecommand \@ifx [1]{%
 \ifx #1\expandafter \@firstoftwo
 \else \expandafter \@secondoftwo
 \fi
}%
\providecommand \natexlab [1]{#1}%
\providecommand \enquote  [1]{``#1''}%
\providecommand \bibnamefont  [1]{#1}%
\providecommand \bibfnamefont [1]{#1}%
\providecommand \citenamefont [1]{#1}%
\providecommand \href@noop [0]{\@secondoftwo}%
\providecommand \href [0]{\begingroup \@sanitize@url \@href}%
\providecommand \@href[1]{\@@startlink{#1}\@@href}%
\providecommand \@@href[1]{\endgroup#1\@@endlink}%
\providecommand \@sanitize@url [0]{\catcode `\\12\catcode `\$12\catcode
  `\&12\catcode `\#12\catcode `\^12\catcode `\_12\catcode `\%12\relax}%
\providecommand \@@startlink[1]{}%
\providecommand \@@endlink[0]{}%
\providecommand \url  [0]{\begingroup\@sanitize@url \@url }%
\providecommand \@url [1]{\endgroup\@href {#1}{\urlprefix }}%
\providecommand \urlprefix  [0]{URL }%
\providecommand \Eprint [0]{\href }%
\@ifxundefined \urlstyle {%
  \providecommand \doi  [0]{\begingroup \@sanitize@url \@doi}%
  \providecommand \@doi [1]{\endgroup \@@startlink {\doibase
  #1}doi:\discretionary {}{}{}#1\@@endlink }%
}{%
  \providecommand \doi  [0]{doi:\discretionary{}{}{}\begingroup
  \urlstyle{rm}\Url }%
}%
\providecommand \doibase [0]{http://dx.doi.org/}%
\providecommand \Doi [0]{\begingroup \@sanitize@url \@Doi }%
\providecommand \@Doi  [1]{\endgroup\@@startlink{\doibase#1}\@@Doi}%
\providecommand \@@Doi [1]{#1\@@endlink}%
\providecommand \selectlanguage [0]{\@gobble}%
\providecommand \bibinfo  [0]{\@secondoftwo}%
\providecommand \bibfield  [0]{\@secondoftwo}%
\providecommand \translation [1]{[#1]}%
\providecommand \BibitemOpen [0]{}%
\providecommand \bibitemStop [0]{}%
\providecommand \bibitemNoStop [0]{.\EOS\space}%
\providecommand \EOS [0]{\spacefactor3000\relax}%
\providecommand \BibitemShut  [1]{\csname bibitem#1\endcsname}%
%</preamble>
\bibitem [{\citenamefont {Shankar}(1994)}]{Shankar:1994p7862}%
  \BibitemOpen
  \bibfield  {author} {\bibinfo {author} {\bibfnamefont {R.}~\bibnamefont
  {Shankar}},\ }\Doi {10.1103/RevModPhys.66.129} {\bibfield  {journal}
  {\bibinfo  {journal} {Rev. Mod. Phys.},\ }\textbf {\bibinfo {volume} {66}},\
  \bibinfo {pages} {129} (\bibinfo {year} {1994})}\BibitemShut {NoStop}%
\bibitem [{\citenamefont {Polchinski}(1984)}]{Polchinski:1984p7873}%
  \BibitemOpen
  \bibfield  {author} {\bibinfo {author} {\bibfnamefont {J.}~\bibnamefont
  {Polchinski}},\ }\href
  {http://www.sciencedirect.com/science/article/pii/0550321384902876}
  {\bibfield  {journal} {\bibinfo  {journal} {Nuclear Physics B},\ }\textbf
  {\bibinfo {volume} {231}},\ \bibinfo {pages} {269} (\bibinfo {year}
  {1984})}\BibitemShut {NoStop}%
\bibitem [{\citenamefont {Metzner}\ \emph {et~al.}(2012)\citenamefont
  {Metzner}, \citenamefont {Salmhofer}, \citenamefont {Honerkamp},
  \citenamefont {Meden},\ and\ \citenamefont
  {Sch{\"o}nhammer}}]{Metzner:2012p7861}%
  \BibitemOpen
  \bibfield  {author} {\bibinfo {author} {\bibfnamefont {W.}~\bibnamefont
  {Metzner}}, \bibinfo {author} {\bibfnamefont {M.}~\bibnamefont {Salmhofer}},
  \bibinfo {author} {\bibfnamefont {C.}~\bibnamefont {Honerkamp}}, \bibinfo
  {author} {\bibfnamefont {V.}~\bibnamefont {Meden}}, \ and\ \bibinfo {author}
  {\bibfnamefont {K.}~\bibnamefont {Sch{\"o}nhammer}},\ }\Doi
  {10.1103/RevModPhys.84.299} {\bibfield  {journal} {\bibinfo  {journal}
  {Reviews of Modern Physics},\ }\textbf {\bibinfo {volume} {84}},\ \bibinfo
  {pages} {299} (\bibinfo {year} {2012})}\BibitemShut {NoStop}%
\bibitem [{\citenamefont {Platt}\ \emph {et~al.}(2013)\citenamefont {Platt},
  \citenamefont {Hanke},\ and\ \citenamefont {Thomale}}]{Platt:2013p7896}%
  \BibitemOpen
  \bibfield  {author} {\bibinfo {author} {\bibfnamefont {C.}~\bibnamefont
  {Platt}}, \bibinfo {author} {\bibfnamefont {W.}~\bibnamefont {Hanke}}, \ and\
  \bibinfo {author} {\bibfnamefont {R.}~\bibnamefont {Thomale}},\ }\Doi
  {10.1080/00018732.2013.862020} {\bibfield  {journal} {\bibinfo  {journal}
  {Advances in Physics},\ }\textbf {\bibinfo {volume} {62}},\ \bibinfo {pages}
  {453} (\bibinfo {year} {2013})}\BibitemShut {NoStop}%
\bibitem [{\citenamefont {Foulkes}\ \emph {et~al.}(2001)\citenamefont
  {Foulkes}, \citenamefont {Mitas}, \citenamefont {Needs},\ and\ \citenamefont
  {Rajagopal}}]{Foulkes:2001p7937}%
  \BibitemOpen
  \bibfield  {author} {\bibinfo {author} {\bibfnamefont {W.~M.}\ \bibnamefont
  {Foulkes}}, \bibinfo {author} {\bibfnamefont {L.}~\bibnamefont {Mitas}},
  \bibinfo {author} {\bibfnamefont {R.~J.}\ \bibnamefont {Needs}}, \ and\
  \bibinfo {author} {\bibfnamefont {G.}~\bibnamefont {Rajagopal}},\ }\Doi
  {10.1103/RevModPhys.73.33} {\bibfield  {journal} {\bibinfo  {journal}
  {Reviews of Modern Physics},\ }\textbf {\bibinfo {volume} {73}},\ \bibinfo
  {pages} {33} (\bibinfo {year} {2001})}\BibitemShut {NoStop}%
\bibitem [{\citenamefont {Gros}(1989)}]{Gros:1989p7551}%
  \BibitemOpen
  \bibfield  {author} {\bibinfo {author} {\bibfnamefont {C.}~\bibnamefont
  {Gros}},\ }\Doi {10.1016/0003-4916(89)90077-8} {\bibfield  {journal}
  {\bibinfo  {journal} {Annals of Physics},\ }\textbf {\bibinfo {volume}
  {189}},\ \bibinfo {pages} {53} (\bibinfo {year} {1989})}\BibitemShut
  {NoStop}%
\bibitem [{\citenamefont {Schollw{\"o}ck}(2011)}]{Schollwock:2011p7856}%
  \BibitemOpen
  \bibfield  {author} {\bibinfo {author} {\bibfnamefont {U.}~\bibnamefont
  {Schollw{\"o}ck}},\ }\Doi {10.1016/j.aop.2010.09.012} {\bibfield  {journal}
  {\bibinfo  {journal} {Annals of Physics},\ }\textbf {\bibinfo {volume}
  {326}},\ \bibinfo {pages} {96} (\bibinfo {year} {2011})}\BibitemShut
  {NoStop}%
\bibitem [{\citenamefont {White}(1992)}]{White:1992p7946}%
  \BibitemOpen
  \bibfield  {author} {\bibinfo {author} {\bibfnamefont {S.~R.}\ \bibnamefont
  {White}},\ }\Doi {10.1103/PhysRevLett.69.2863} {\bibfield  {journal}
  {\bibinfo  {journal} {Physical Review Letters},\ }\textbf {\bibinfo {volume}
  {69}},\ \bibinfo {pages} {2863} (\bibinfo {year} {1992})}\BibitemShut
  {NoStop}%
\bibitem [{\citenamefont {Vidal}(2008)}]{Vidal:2008p7925}%
  \BibitemOpen
  \bibfield  {author} {\bibinfo {author} {\bibfnamefont {G.}~\bibnamefont
  {Vidal}},\ }\Doi {10.1103/PhysRevLett.101.110501} {\bibfield  {journal}
  {\bibinfo  {journal} {Physical Review Letters},\ }\textbf {\bibinfo {volume}
  {101}},\ \bibinfo {pages} {110501} (\bibinfo {year} {2008})}\BibitemShut
  {NoStop}%
\bibitem [{\citenamefont {Vidal}(2007)}]{Vidal:2007p7926}%
  \BibitemOpen
  \bibfield  {author} {\bibinfo {author} {\bibfnamefont {G.}~\bibnamefont
  {Vidal}},\ }\Doi {10.1103/PhysRevLett.99.220405} {\bibfield  {journal}
  {\bibinfo  {journal} {Physical Review Letters},\ }\textbf {\bibinfo {volume}
  {99}},\ \bibinfo {pages} {220405} (\bibinfo {year} {2007})}\BibitemShut
  {NoStop}%
\bibitem [{\citenamefont {Corboz}\ and\ \citenamefont
  {Vidal}(2009)}]{Corboz:2009p7934}%
  \BibitemOpen
  \bibfield  {author} {\bibinfo {author} {\bibfnamefont {P.}~\bibnamefont
  {Corboz}}\ and\ \bibinfo {author} {\bibfnamefont {G.}~\bibnamefont {Vidal}},\
  }\Doi {10.1103/PhysRevB.80.165129} {\bibfield  {journal} {\bibinfo  {journal}
  {Physical Review B},\ }\textbf {\bibinfo {volume} {80}},\ \bibinfo {pages}
  {165129} (\bibinfo {year} {2009})}\BibitemShut {NoStop}%
\bibitem [{\citenamefont {Corboz}\ \emph {et~al.}(2010)\citenamefont {Corboz},
  \citenamefont {Evenbly}, \citenamefont {Verstraete},\ and\ \citenamefont
  {Vidal}}]{Corboz:2010p7933}%
  \BibitemOpen
  \bibfield  {author} {\bibinfo {author} {\bibfnamefont {P.}~\bibnamefont
  {Corboz}}, \bibinfo {author} {\bibfnamefont {G.}~\bibnamefont {Evenbly}},
  \bibinfo {author} {\bibfnamefont {F.}~\bibnamefont {Verstraete}}, \ and\
  \bibinfo {author} {\bibfnamefont {G.}~\bibnamefont {Vidal}},\ }\Doi
  {10.1103/PhysRevA.81.010303} {\bibfield  {journal} {\bibinfo  {journal}
  {Physical Review A},\ }\textbf {\bibinfo {volume} {81}},\ \bibinfo {pages}
  {10303} (\bibinfo {year} {2010})}\BibitemShut {NoStop}%
\bibitem [{\citenamefont {Verstraete}\ and\ \citenamefont
  {Cirac}(2004)}]{Verstraete:2004p7928}%
  \BibitemOpen
  \bibfield  {author} {\bibinfo {author} {\bibfnamefont {F.}~\bibnamefont
  {Verstraete}}\ and\ \bibinfo {author} {\bibfnamefont {J.~I.}\ \bibnamefont
  {Cirac}},\ }\href {http://arxiv.org/abs/cond-mat/0407066v1} {\bibfield
  {journal} {\bibinfo  {journal} {arXiv},\ }\textbf {\bibinfo {volume}
  {cond-mat.str-el}} (\bibinfo {year} {2004})},\ \Eprint
  {http://arxiv.org/abs/cond-mat/0407066v1} {cond-mat/0407066v1} \BibitemShut
  {NoStop}%
\bibitem [{\citenamefont {Kraus}\ \emph {et~al.}(2010)\citenamefont {Kraus},
  \citenamefont {Schuch}, \citenamefont {Verstraete},\ and\ \citenamefont
  {Cirac}}]{Kraus:2010p7935}%
  \BibitemOpen
  \bibfield  {author} {\bibinfo {author} {\bibfnamefont {C.~V.}\ \bibnamefont
  {Kraus}}, \bibinfo {author} {\bibfnamefont {N.}~\bibnamefont {Schuch}},
  \bibinfo {author} {\bibfnamefont {F.}~\bibnamefont {Verstraete}}, \ and\
  \bibinfo {author} {\bibfnamefont {J.~I.}\ \bibnamefont {Cirac}},\ }\Doi
  {10.1103/PhysRevA.81.052338} {\bibfield  {journal} {\bibinfo  {journal}
  {Physical Review A},\ }\textbf {\bibinfo {volume} {81}},\ \bibinfo {pages}
  {52338} (\bibinfo {year} {2010})}\BibitemShut {NoStop}%
\bibitem [{\citenamefont {Yan}\ \emph {et~al.}(2011)\citenamefont {Yan},
  \citenamefont {Huse},\ and\ \citenamefont {White}}]{Yan:2011p7804}%
  \BibitemOpen
  \bibfield  {author} {\bibinfo {author} {\bibfnamefont {S.}~\bibnamefont
  {Yan}}, \bibinfo {author} {\bibfnamefont {D.~A.}\ \bibnamefont {Huse}}, \
  and\ \bibinfo {author} {\bibfnamefont {S.~R.}\ \bibnamefont {White}},\ }\Doi
  {10.1126/science.1201080} {\bibfield  {journal} {\bibinfo  {journal}
  {Science},\ }\textbf {\bibinfo {volume} {332}},\ \bibinfo {pages} {1173}
  (\bibinfo {year} {2011})}\BibitemShut {NoStop}%
\bibitem [{\citenamefont {Jiang}\ \emph {et~al.}(2012)\citenamefont {Jiang},
  \citenamefont {Yao},\ and\ \citenamefont {Balents}}]{Jiang:2012p7939}%
  \BibitemOpen
  \bibfield  {author} {\bibinfo {author} {\bibfnamefont {H.-C.}\ \bibnamefont
  {Jiang}}, \bibinfo {author} {\bibfnamefont {H.}~\bibnamefont {Yao}}, \ and\
  \bibinfo {author} {\bibfnamefont {L.}~\bibnamefont {Balents}},\ }\Doi
  {10.1103/PhysRevB.86.024424} {\bibfield  {journal} {\bibinfo  {journal}
  {Physical Review B},\ }\textbf {\bibinfo {volume} {86}},\ \bibinfo {pages}
  {24424} (\bibinfo {year} {2012})}\BibitemShut {NoStop}%
\bibitem [{\citenamefont {Gong}\ \emph
  {et~al.}(2013){\natexlab{a}}\citenamefont {Gong}, \citenamefont {Zhu},\ and\
  \citenamefont {Sheng}}]{Gong:2013p7942}%
  \BibitemOpen
  \bibfield  {author} {\bibinfo {author} {\bibfnamefont {S.-S.}\ \bibnamefont
  {Gong}}, \bibinfo {author} {\bibfnamefont {W.}~\bibnamefont {Zhu}}, \ and\
  \bibinfo {author} {\bibfnamefont {D.~N.}\ \bibnamefont {Sheng}},\ }\href
  {http://arxiv.org/abs/1312.4519v1} {\bibfield  {journal} {\bibinfo  {journal}
  {arXiv},\ }\textbf {\bibinfo {volume} {cond-mat.str-el}} (\bibinfo {year}
  {2013}{\natexlab{a}})},\ \Eprint {http://arxiv.org/abs/1312.4519v1}
  {1312.4519v1} \BibitemShut {NoStop}%
\bibitem [{\citenamefont {Gong}\ \emph
  {et~al.}(2013){\natexlab{b}}\citenamefont {Gong}, \citenamefont {Sheng},
  \citenamefont {Motrunich},\ and\ \citenamefont {Fisher}}]{Gong:2013p7940}%
  \BibitemOpen
  \bibfield  {author} {\bibinfo {author} {\bibfnamefont {S.-S.}\ \bibnamefont
  {Gong}}, \bibinfo {author} {\bibfnamefont {D.~N.}\ \bibnamefont {Sheng}},
  \bibinfo {author} {\bibfnamefont {O.~I.}\ \bibnamefont {Motrunich}}, \ and\
  \bibinfo {author} {\bibfnamefont {M.~P.~A.}\ \bibnamefont {Fisher}},\ }\Doi
  {10.1103/PhysRevB.88.165138} {\bibfield  {journal} {\bibinfo  {journal}
  {Physical Review B},\ }\textbf {\bibinfo {volume} {88}},\ \bibinfo {pages}
  {165138} (\bibinfo {year} {2013}{\natexlab{b}})}\BibitemShut {NoStop}%
\bibitem [{\citenamefont {Depenbrock}\ \emph {et~al.}(2012)\citenamefont
  {Depenbrock}, \citenamefont {McCulloch},\ and\ \citenamefont
  {Schollw{\"o}ck}}]{Depenbrock:2012p7944}%
  \BibitemOpen
  \bibfield  {author} {\bibinfo {author} {\bibfnamefont {S.}~\bibnamefont
  {Depenbrock}}, \bibinfo {author} {\bibfnamefont {I.~P.}\ \bibnamefont
  {McCulloch}}, \ and\ \bibinfo {author} {\bibfnamefont {U.}~\bibnamefont
  {Schollw{\"o}ck}},\ }\Doi {10.1103/PhysRevLett.109.067201} {\bibfield
  {journal} {\bibinfo  {journal} {Physical Review Letters},\ }\textbf {\bibinfo
  {volume} {109}},\ \bibinfo {pages} {67201} (\bibinfo {year}
  {2012})}\BibitemShut {NoStop}%
\bibitem [{Note1()}]{Note1}%
  \BibitemOpen
  \bibinfo {note} {The reason why we focus on commensurately doped systems is
  mainly due to technical considerations: at certain commensurate fillings,
  there can be very reasonable guesses for the candidate quantum phases, and
  explicitly constructing their wavefunctions is not too difficult within the
  currently available theoretical frameworks.}\BibitemShut {Stop}%
\bibitem [{\citenamefont {Li}(2012)}]{Li:2012p7905}%
  \BibitemOpen
  \bibfield  {author} {\bibinfo {author} {\bibfnamefont {T.}~\bibnamefont
  {Li}},\ }\Doi {10.1209/0295-5075/97/37001} {\bibfield  {journal} {\bibinfo
  {journal} {Europhysics Letters},\ }\textbf {\bibinfo {volume} {97}},\
  \bibinfo {pages} {37001} (\bibinfo {year} {2012})}\BibitemShut {NoStop}%
\bibitem [{\citenamefont {Raghu}\ \emph {et~al.}(2010)\citenamefont {Raghu},
  \citenamefont {Kivelson},\ and\ \citenamefont {Scalapino}}]{Raghu:2010p7897}%
  \BibitemOpen
  \bibfield  {author} {\bibinfo {author} {\bibfnamefont {S.}~\bibnamefont
  {Raghu}}, \bibinfo {author} {\bibfnamefont {S.~A.}\ \bibnamefont {Kivelson}},
  \ and\ \bibinfo {author} {\bibfnamefont {D.~J.}\ \bibnamefont {Scalapino}},\
  }\Doi {10.1103/PhysRevB.81.224505} {\bibfield  {journal} {\bibinfo  {journal}
  {Physical Review B},\ }\textbf {\bibinfo {volume} {81}},\ \bibinfo {pages}
  {224505} (\bibinfo {year} {2010})}\BibitemShut {NoStop}%
\bibitem [{\citenamefont {Nandkishore}\ \emph {et~al.}(2012)\citenamefont
  {Nandkishore}, \citenamefont {Levitov},\ and\ \citenamefont
  {Chubukov}}]{Nandkishore:2012p7877}%
  \BibitemOpen
  \bibfield  {author} {\bibinfo {author} {\bibfnamefont {R.}~\bibnamefont
  {Nandkishore}}, \bibinfo {author} {\bibfnamefont {L.~S.}\ \bibnamefont
  {Levitov}}, \ and\ \bibinfo {author} {\bibfnamefont {A.~V.}\ \bibnamefont
  {Chubukov}},\ }\Doi {doi:10.1038/nphys2208} {\bibfield  {journal} {\bibinfo
  {journal} {Nature Physics},\ }\textbf {\bibinfo {volume} {8}},\ \bibinfo
  {pages} {158} (\bibinfo {year} {2012})}\BibitemShut {NoStop}%
\bibitem [{\citenamefont {Wang}\ \emph
  {et~al.}(2012){\natexlab{a}}\citenamefont {Wang}, \citenamefont {Xiang},
  \citenamefont {Wang}, \citenamefont {Wang}, \citenamefont {Yang},\ and\
  \citenamefont {Lee}}]{Wang:2012p7894}%
  \BibitemOpen
  \bibfield  {author} {\bibinfo {author} {\bibfnamefont {W.-S.}\ \bibnamefont
  {Wang}}, \bibinfo {author} {\bibfnamefont {Y.-Y.}\ \bibnamefont {Xiang}},
  \bibinfo {author} {\bibfnamefont {Q.-H.}\ \bibnamefont {Wang}}, \bibinfo
  {author} {\bibfnamefont {F.}~\bibnamefont {Wang}}, \bibinfo {author}
  {\bibfnamefont {F.}~\bibnamefont {Yang}}, \ and\ \bibinfo {author}
  {\bibfnamefont {D.-H.}\ \bibnamefont {Lee}},\ }\Doi
  {10.1103/PhysRevB.85.035414} {\bibfield  {journal} {\bibinfo  {journal}
  {Physical Review B},\ }\textbf {\bibinfo {volume} {85}},\ \bibinfo {pages}
  {35414} (\bibinfo {year} {2012}{\natexlab{a}})}\BibitemShut {NoStop}%
\bibitem [{\citenamefont {Kiesel}\ \emph {et~al.}(2012)\citenamefont {Kiesel},
  \citenamefont {Platt}, \citenamefont {Hanke}, \citenamefont {Abanin},\ and\
  \citenamefont {Thomale}}]{Kiesel:2012p7895}%
  \BibitemOpen
  \bibfield  {author} {\bibinfo {author} {\bibfnamefont {M.~L.}\ \bibnamefont
  {Kiesel}}, \bibinfo {author} {\bibfnamefont {C.}~\bibnamefont {Platt}},
  \bibinfo {author} {\bibfnamefont {W.}~\bibnamefont {Hanke}}, \bibinfo
  {author} {\bibfnamefont {D.~A.}\ \bibnamefont {Abanin}}, \ and\ \bibinfo
  {author} {\bibfnamefont {R.}~\bibnamefont {Thomale}},\ }\Doi
  {10.1103/PhysRevB.86.020507} {\bibfield  {journal} {\bibinfo  {journal}
  {Physical Review B},\ }\textbf {\bibinfo {volume} {86}},\ \bibinfo {pages}
  {20507} (\bibinfo {year} {2012})}\BibitemShut {NoStop}%
\bibitem [{\citenamefont {Wu}\ \emph {et~al.}(2013)\citenamefont {Wu},
  \citenamefont {Scherer}, \citenamefont {Honerkamp},\ and\ \citenamefont
  {Hur}}]{Wu:2013p7952}%
  \BibitemOpen
  \bibfield  {author} {\bibinfo {author} {\bibfnamefont {W.}~\bibnamefont
  {Wu}}, \bibinfo {author} {\bibfnamefont {M.~M.}\ \bibnamefont {Scherer}},
  \bibinfo {author} {\bibfnamefont {C.}~\bibnamefont {Honerkamp}}, \ and\
  \bibinfo {author} {\bibfnamefont {K.~L.}\ \bibnamefont {Hur}},\ }\Doi
  {10.1103/PhysRevB.87.094521} {\bibfield  {journal} {\bibinfo  {journal}
  {Physical Review B},\ }\textbf {\bibinfo {volume} {87}},\ \bibinfo {pages}
  {94521} (\bibinfo {year} {2013})}\BibitemShut {NoStop}%
\bibitem [{\citenamefont {Pathak}\ \emph {et~al.}(2010)\citenamefont {Pathak},
  \citenamefont {Shenoy},\ and\ \citenamefont {Baskaran}}]{Pathak:2010p7901}%
  \BibitemOpen
  \bibfield  {author} {\bibinfo {author} {\bibfnamefont {S.}~\bibnamefont
  {Pathak}}, \bibinfo {author} {\bibfnamefont {V.~B.}\ \bibnamefont {Shenoy}},
  \ and\ \bibinfo {author} {\bibfnamefont {G.}~\bibnamefont {Baskaran}},\ }\Doi
  {10.1103/PhysRevB.81.085431} {\bibfield  {journal} {\bibinfo  {journal}
  {Physical Review B},\ }\textbf {\bibinfo {volume} {81}},\ \bibinfo {pages}
  {85431} (\bibinfo {year} {2010})}\BibitemShut {NoStop}%
\bibitem [{\citenamefont {Gu}\ \emph {et~al.}(2013)\citenamefont {Gu},
  \citenamefont {Jiang}, \citenamefont {Sheng}, \citenamefont {Yao},
  \citenamefont {Balents},\ and\ \citenamefont {Wen}}]{Gu:2013p7902}%
  \BibitemOpen
  \bibfield  {author} {\bibinfo {author} {\bibfnamefont {Z.-C.}\ \bibnamefont
  {Gu}}, \bibinfo {author} {\bibfnamefont {H.-C.}\ \bibnamefont {Jiang}},
  \bibinfo {author} {\bibfnamefont {D.~N.}\ \bibnamefont {Sheng}}, \bibinfo
  {author} {\bibfnamefont {H.}~\bibnamefont {Yao}}, \bibinfo {author}
  {\bibfnamefont {L.}~\bibnamefont {Balents}}, \ and\ \bibinfo {author}
  {\bibfnamefont {X.-G.}\ \bibnamefont {Wen}},\ }\Doi
  {10.1103/PhysRevB.88.155112} {\bibfield  {journal} {\bibinfo  {journal}
  {Physical Review B},\ }\textbf {\bibinfo {volume} {88}},\ \bibinfo {pages}
  {155112} (\bibinfo {year} {2013})}\BibitemShut {NoStop}%
\bibitem [{\citenamefont {Senthil}\ \emph {et~al.}(1999)\citenamefont
  {Senthil}, \citenamefont {Marston},\ and\ \citenamefont
  {Fisher}}]{Senthil:1999p7954}%
  \BibitemOpen
  \bibfield  {author} {\bibinfo {author} {\bibfnamefont {T.}~\bibnamefont
  {Senthil}}, \bibinfo {author} {\bibfnamefont {J.~B.}\ \bibnamefont
  {Marston}}, \ and\ \bibinfo {author} {\bibfnamefont {M.~P.~A.}\ \bibnamefont
  {Fisher}},\ }\Doi {10.1103/PhysRevB.60.4245} {\bibfield  {journal} {\bibinfo
  {journal} {Physical Review B (Condensed Matter and Materials Physics)},\
  }\textbf {\bibinfo {volume} {60}},\ \bibinfo {pages} {4245} (\bibinfo {year}
  {1999})}\BibitemShut {NoStop}%
\bibitem [{\citenamefont {Balents}(2010)}]{2010Natur.464..199B}%
  \BibitemOpen
  \bibfield  {author} {\bibinfo {author} {\bibfnamefont {L.}~\bibnamefont
  {Balents}},\ }\Doi {10.1038/nature08917} {\bibfield  {journal} {\bibinfo
  {journal} {Nature},\ }\textbf {\bibinfo {volume} {464}},\ \bibinfo {pages}
  {199} (\bibinfo {year} {2010})}\BibitemShut {NoStop}%
\bibitem [{\citenamefont {Han}\ \emph {et~al.}(2012)\citenamefont {Han},
  \citenamefont {Helton}, \citenamefont {Chu}, \citenamefont {Nocera},
  \citenamefont {Rodriguez-Rivera}, \citenamefont {Broholm},\ and\
  \citenamefont {Lee}}]{Han:2012p7973}%
  \BibitemOpen
  \bibfield  {author} {\bibinfo {author} {\bibfnamefont {T.-H.}\ \bibnamefont
  {Han}}, \bibinfo {author} {\bibfnamefont {J.~S.}\ \bibnamefont {Helton}},
  \bibinfo {author} {\bibfnamefont {S.}~\bibnamefont {Chu}}, \bibinfo {author}
  {\bibfnamefont {D.~G.}\ \bibnamefont {Nocera}}, \bibinfo {author}
  {\bibfnamefont {J.~A.}\ \bibnamefont {Rodriguez-Rivera}}, \bibinfo {author}
  {\bibfnamefont {C.}~\bibnamefont {Broholm}}, \ and\ \bibinfo {author}
  {\bibfnamefont {Y.~S.}\ \bibnamefont {Lee}},\ }\Doi {doi:10.1038/nature11659}
  {\bibfield  {journal} {\bibinfo  {journal} {Nature},\ }\textbf {\bibinfo
  {volume} {492}},\ \bibinfo {pages} {406} (\bibinfo {year}
  {2012})}\BibitemShut {NoStop}%
\bibitem [{\citenamefont {Watanabe}\ \emph {et~al.}(2012)\citenamefont
  {Watanabe}, \citenamefont {Yamashita}, \citenamefont {Tonegawa},
  \citenamefont {Oshima}, \citenamefont {Yamamoto}, \citenamefont {Kato},
  \citenamefont {Sheikin}, \citenamefont {Behnia}, \citenamefont {Terashima},
  \citenamefont {Uji}, \citenamefont {Shibauchi},\ and\ \citenamefont
  {Matsuda}}]{Watanabe:2012p7963}%
  \BibitemOpen
  \bibfield  {author} {\bibinfo {author} {\bibfnamefont {D.}~\bibnamefont
  {Watanabe}}, \bibinfo {author} {\bibfnamefont {M.}~\bibnamefont {Yamashita}},
  \bibinfo {author} {\bibfnamefont {S.}~\bibnamefont {Tonegawa}}, \bibinfo
  {author} {\bibfnamefont {Y.}~\bibnamefont {Oshima}}, \bibinfo {author}
  {\bibfnamefont {H.~M.}\ \bibnamefont {Yamamoto}}, \bibinfo {author}
  {\bibfnamefont {R.}~\bibnamefont {Kato}}, \bibinfo {author} {\bibfnamefont
  {I.}~\bibnamefont {Sheikin}}, \bibinfo {author} {\bibfnamefont
  {K.}~\bibnamefont {Behnia}}, \bibinfo {author} {\bibfnamefont
  {T.}~\bibnamefont {Terashima}}, \bibinfo {author} {\bibfnamefont
  {S.}~\bibnamefont {Uji}}, \bibinfo {author} {\bibfnamefont {T.}~\bibnamefont
  {Shibauchi}}, \ and\ \bibinfo {author} {\bibfnamefont {Y.}~\bibnamefont
  {Matsuda}},\ }\Doi {doi:10.1038/ncomms2082} {\bibfield  {journal} {\bibinfo
  {journal} {Nature Communications},\ }\textbf {\bibinfo {volume} {3}},\
  \bibinfo {pages} {1090} (\bibinfo {year} {2012})}\BibitemShut {NoStop}%
\bibitem [{\citenamefont {Cheng}\ \emph {et~al.}(2011)\citenamefont {Cheng},
  \citenamefont {Li}, \citenamefont {Balicas}, \citenamefont {Zhou},
  \citenamefont {Goodenough}, \citenamefont {Xu},\ and\ \citenamefont
  {Zhou}}]{Cheng:2011p7959}%
  \BibitemOpen
  \bibfield  {author} {\bibinfo {author} {\bibfnamefont {J.~G.}\ \bibnamefont
  {Cheng}}, \bibinfo {author} {\bibfnamefont {G.}~\bibnamefont {Li}}, \bibinfo
  {author} {\bibfnamefont {L.}~\bibnamefont {Balicas}}, \bibinfo {author}
  {\bibfnamefont {J.~S.}\ \bibnamefont {Zhou}}, \bibinfo {author}
  {\bibfnamefont {J.~B.}\ \bibnamefont {Goodenough}}, \bibinfo {author}
  {\bibfnamefont {C.}~\bibnamefont {Xu}}, \ and\ \bibinfo {author}
  {\bibfnamefont {H.~D.}\ \bibnamefont {Zhou}},\ }\Doi
  {10.1103/PhysRevLett.107.197204} {\bibfield  {journal} {\bibinfo  {journal}
  {Physical Review Letters},\ }\textbf {\bibinfo {volume} {107}},\ \bibinfo
  {pages} {197204} (\bibinfo {year} {2011})}\BibitemShut {NoStop}%
\bibitem [{\citenamefont {Pratt}\ \emph {et~al.}(2011)\citenamefont {Pratt},
  \citenamefont {Baker}, \citenamefont {Blundell}, \citenamefont {Lancaster},
  \citenamefont {Ohira-Kawamura}, \citenamefont {Baines}, \citenamefont
  {Shimizu}, \citenamefont {Kanoda}, \citenamefont {Watanabe},\ and\
  \citenamefont {Saito}}]{Pratt:2011p7961}%
  \BibitemOpen
  \bibfield  {author} {\bibinfo {author} {\bibfnamefont {F.~L.}\ \bibnamefont
  {Pratt}}, \bibinfo {author} {\bibfnamefont {P.~J.}\ \bibnamefont {Baker}},
  \bibinfo {author} {\bibfnamefont {S.~J.}\ \bibnamefont {Blundell}}, \bibinfo
  {author} {\bibfnamefont {T.}~\bibnamefont {Lancaster}}, \bibinfo {author}
  {\bibfnamefont {S.}~\bibnamefont {Ohira-Kawamura}}, \bibinfo {author}
  {\bibfnamefont {C.}~\bibnamefont {Baines}}, \bibinfo {author} {\bibfnamefont
  {Y.}~\bibnamefont {Shimizu}}, \bibinfo {author} {\bibfnamefont
  {K.}~\bibnamefont {Kanoda}}, \bibinfo {author} {\bibfnamefont
  {I.}~\bibnamefont {Watanabe}}, \ and\ \bibinfo {author} {\bibfnamefont
  {G.}~\bibnamefont {Saito}},\ }\Doi {doi:10.1038/nature09910} {\bibfield
  {journal} {\bibinfo  {journal} {Nature},\ }\textbf {\bibinfo {volume}
  {471}},\ \bibinfo {pages} {612} (\bibinfo {year} {2011})}\BibitemShut
  {NoStop}%
\bibitem [{\citenamefont {Yamashita}\ \emph {et~al.}(2010)\citenamefont
  {Yamashita}, \citenamefont {Nakata}, \citenamefont {Senshu}, \citenamefont
  {Nagata}, \citenamefont {Yamamoto}, \citenamefont {Kato}, \citenamefont
  {Shibauchi},\ and\ \citenamefont {Matsuda}}]{Yamashita:2010p7955}%
  \BibitemOpen
  \bibfield  {author} {\bibinfo {author} {\bibfnamefont {M.}~\bibnamefont
  {Yamashita}}, \bibinfo {author} {\bibfnamefont {N.}~\bibnamefont {Nakata}},
  \bibinfo {author} {\bibfnamefont {Y.}~\bibnamefont {Senshu}}, \bibinfo
  {author} {\bibfnamefont {M.}~\bibnamefont {Nagata}}, \bibinfo {author}
  {\bibfnamefont {H.~M.}\ \bibnamefont {Yamamoto}}, \bibinfo {author}
  {\bibfnamefont {R.}~\bibnamefont {Kato}}, \bibinfo {author} {\bibfnamefont
  {T.}~\bibnamefont {Shibauchi}}, \ and\ \bibinfo {author} {\bibfnamefont
  {Y.}~\bibnamefont {Matsuda}},\ }\Doi {10.1126/science.1188200} {\bibfield
  {journal} {\bibinfo  {journal} {Science},\ }\textbf {\bibinfo {volume}
  {328}},\ \bibinfo {pages} {1246} (\bibinfo {year} {2010})}\BibitemShut
  {NoStop}%
\bibitem [{\citenamefont {Neto}\ \emph {et~al.}(2009)\citenamefont {Neto},
  \citenamefont {Guinea}, \citenamefont {Peres}, \citenamefont {Novoselov},\
  and\ \citenamefont {Geim}}]{CastroNeto:2009p3917}%
  \BibitemOpen
  \bibfield  {author} {\bibinfo {author} {\bibfnamefont {A.~H.~C.}\
  \bibnamefont {Neto}}, \bibinfo {author} {\bibfnamefont {F.}~\bibnamefont
  {Guinea}}, \bibinfo {author} {\bibfnamefont {N.~M.~R.}\ \bibnamefont
  {Peres}}, \bibinfo {author} {\bibfnamefont {K.~S.}\ \bibnamefont
  {Novoselov}}, \ and\ \bibinfo {author} {\bibfnamefont {A.~K.}\ \bibnamefont
  {Geim}},\ }\Doi {10.1103/RevModPhys.81.109} {\bibfield  {journal} {\bibinfo
  {journal} {Reviews of Modern Physics},\ }\textbf {\bibinfo {volume} {81}},\
  \bibinfo {pages} {109} (\bibinfo {year} {2009})}\BibitemShut {NoStop}%
\bibitem [{\citenamefont {Nagaoka}(1966)}]{Nagaoka:1966p8011}%
  \BibitemOpen
  \bibfield  {author} {\bibinfo {author} {\bibfnamefont {Y.}~\bibnamefont
  {Nagaoka}},\ }\Doi {10.1103/PhysRev.147.392} {\bibfield  {journal} {\bibinfo
  {journal} {Phys. Rev.},\ }\textbf {\bibinfo {volume} {147}},\ \bibinfo
  {pages} {392} (\bibinfo {year} {1966})}\BibitemShut {NoStop}%
\bibitem [{\citenamefont {Kanamori}(1963)}]{Kanamori:1963p8010}%
  \BibitemOpen
  \bibfield  {author} {\bibinfo {author} {\bibfnamefont {J.}~\bibnamefont
  {Kanamori}},\ }\Doi {10.1143/PTP.30.275} {\bibfield  {journal} {\bibinfo
  {journal} {Prog. Theor. Phys.},\ }\textbf {\bibinfo {volume} {30}},\ \bibinfo
  {pages} {275} (\bibinfo {year} {1963})}\BibitemShut {NoStop}%
\bibitem [{\citenamefont {Zhang}\ \emph {et~al.}(2012)\citenamefont {Zhang},
  \citenamefont {Grover}, \citenamefont {Turner}, \citenamefont {Oshikawa},\
  and\ \citenamefont {Vishwanath}}]{Zhang:2012p7534}%
  \BibitemOpen
  \bibfield  {author} {\bibinfo {author} {\bibfnamefont {Y.}~\bibnamefont
  {Zhang}}, \bibinfo {author} {\bibfnamefont {T.}~\bibnamefont {Grover}},
  \bibinfo {author} {\bibfnamefont {A.}~\bibnamefont {Turner}}, \bibinfo
  {author} {\bibfnamefont {M.}~\bibnamefont {Oshikawa}}, \ and\ \bibinfo
  {author} {\bibfnamefont {A.}~\bibnamefont {Vishwanath}},\ }\Doi
  {10.1103/PhysRevB.85.235151} {\bibfield  {journal} {\bibinfo  {journal}
  {Physical Review B},\ }\textbf {\bibinfo {volume} {85}},\ \bibinfo {pages}
  {235151} (\bibinfo {year} {2012})}\BibitemShut {NoStop}%
\bibitem [{\citenamefont {Zhang}\ and\ \citenamefont
  {Vishwanath}(2013)}]{Zhang:2013p8017}%
  \BibitemOpen
  \bibfield  {author} {\bibinfo {author} {\bibfnamefont {Y.}~\bibnamefont
  {Zhang}}\ and\ \bibinfo {author} {\bibfnamefont {A.}~\bibnamefont
  {Vishwanath}},\ }\Doi {10.1103/PhysRevB.87.161113} {\bibfield  {journal}
  {\bibinfo  {journal} {Physical Review B},\ }\textbf {\bibinfo {volume}
  {87}},\ \bibinfo {pages} {161113} (\bibinfo {year} {2013})}\BibitemShut
  {NoStop}%
\bibitem [{\citenamefont {He}\ \emph {et~al.}(2013)\citenamefont {He},
  \citenamefont {Sheng},\ and\ \citenamefont {Chen}}]{He:2013p8022}%
  \BibitemOpen
  \bibfield  {author} {\bibinfo {author} {\bibfnamefont {Y.-C.}\ \bibnamefont
  {He}}, \bibinfo {author} {\bibfnamefont {D.~N.}\ \bibnamefont {Sheng}}, \
  and\ \bibinfo {author} {\bibfnamefont {Y.}~\bibnamefont {Chen}},\ }\href
  {http://arxiv.org/abs/1312.3461v2} {\bibfield  {journal} {\bibinfo  {journal}
  {arXiv},\ }\textbf {\bibinfo {volume} {cond-mat.str-el}} (\bibinfo {year}
  {2013})},\ \Eprint {http://arxiv.org/abs/1312.3461v2} {1312.3461v2}
  \BibitemShut {NoStop}%
\bibitem [{\citenamefont {Wen}(2012)}]{Wen:2012p8018}%
  \BibitemOpen
  \bibfield  {author} {\bibinfo {author} {\bibfnamefont {X.-G.}\ \bibnamefont
  {Wen}},\ }\href {http://arxiv.org/abs/1212.5121v2} {\bibfield  {journal}
  {\bibinfo  {journal} {arXiv},\ }\textbf {\bibinfo {volume} {cond-mat.str-el}}
  (\bibinfo {year} {2012})},\ \Eprint {http://arxiv.org/abs/1212.5121v2}
  {1212.5121v2} \BibitemShut {NoStop}%
\bibitem [{\citenamefont {Cincio}\ and\ \citenamefont
  {Vidal}(2013)}]{Cincio:2013p7854}%
  \BibitemOpen
  \bibfield  {author} {\bibinfo {author} {\bibfnamefont {L.}~\bibnamefont
  {Cincio}}\ and\ \bibinfo {author} {\bibfnamefont {G.}~\bibnamefont {Vidal}},\
  }\Doi {10.1103/PhysRevLett.110.067208} {\bibfield  {journal} {\bibinfo
  {journal} {Physical Review Letters},\ }\textbf {\bibinfo {volume} {110}},\
  \bibinfo {pages} {067208} (\bibinfo {year} {2013})}\BibitemShut {NoStop}%
\bibitem [{\citenamefont {Wang}\ and\ \citenamefont
  {Vishwanath}(2006)}]{Wang:2006p6704}%
  \BibitemOpen
  \bibfield  {author} {\bibinfo {author} {\bibfnamefont {F.}~\bibnamefont
  {Wang}}\ and\ \bibinfo {author} {\bibfnamefont {A.}~\bibnamefont
  {Vishwanath}},\ }\Doi {10.1103/PhysRevB.74.174423} {\bibfield  {journal}
  {\bibinfo  {journal} {Physical Review B},\ }\textbf {\bibinfo {volume}
  {74}},\ \bibinfo {pages} {174423} (\bibinfo {year} {2006})}\BibitemShut
  {NoStop}%
\bibitem [{\citenamefont {Sachdev}(1992)}]{Sachdev:1992p12377}%
  \BibitemOpen
  \bibfield  {author} {\bibinfo {author} {\bibfnamefont {S.}~\bibnamefont
  {Sachdev}},\ }\Doi {10.1103/PhysRevB.45.12377} {\bibfield  {journal}
  {\bibinfo  {journal} {Physical Review B},\ }\textbf {\bibinfo {volume}
  {45}},\ \bibinfo {pages} {12377} (\bibinfo {year} {1992})}\BibitemShut
  {NoStop}%
\bibitem [{\citenamefont {Sachdev}\ and\ \citenamefont
  {Read}(1991)}]{Sachdev:1991p8025}%
  \BibitemOpen
  \bibfield  {author} {\bibinfo {author} {\bibfnamefont {S.}~\bibnamefont
  {Sachdev}}\ and\ \bibinfo {author} {\bibfnamefont {N.}~\bibnamefont {Read}},\
  }\href {http://www.worldscientific.com/doi/abs/10.1142/S0217979291000158}
  {\bibfield  {journal} {\bibinfo  {journal} {Int J Mod Phys B},\ }\textbf
  {\bibinfo {volume} {5}},\ \bibinfo {pages} {219} (\bibinfo {year}
  {1991})}\BibitemShut {NoStop}%
\bibitem [{\citenamefont {Read}\ and\ \citenamefont
  {Sachdev}(1991)}]{Read:1991p5262}%
  \BibitemOpen
  \bibfield  {author} {\bibinfo {author} {\bibfnamefont {N.}~\bibnamefont
  {Read}}\ and\ \bibinfo {author} {\bibfnamefont {S.}~\bibnamefont {Sachdev}},\
  }\Doi {10.1103/PhysRevLett.66.1773} {\bibfield  {journal} {\bibinfo
  {journal} {Physical Review Letters},\ }\textbf {\bibinfo {volume} {66}},\
  \bibinfo {pages} {1773} (\bibinfo {year} {1991})}\BibitemShut {NoStop}%
\bibitem [{\citenamefont {Arovas}\ and\ \citenamefont
  {Auerbach}(1988)}]{Arovas:1988p7948}%
  \BibitemOpen
  \bibfield  {author} {\bibinfo {author} {\bibfnamefont {D.~P.}\ \bibnamefont
  {Arovas}}\ and\ \bibinfo {author} {\bibfnamefont {A.}~\bibnamefont
  {Auerbach}},\ }\Doi {10.1103/PhysRevB.38.316} {\bibfield  {journal} {\bibinfo
   {journal} {Physical Review B (Condensed Matter)},\ }\textbf {\bibinfo
  {volume} {38}},\ \bibinfo {pages} {316} (\bibinfo {year} {1988})}\BibitemShut
  {NoStop}%
\bibitem [{\citenamefont {Wen}(2002){\natexlab{a}}}]{Wen:2002p6309}%
  \BibitemOpen
  \bibfield  {author} {\bibinfo {author} {\bibfnamefont {X.-G.}\ \bibnamefont
  {Wen}},\ }\Doi {10.1103/PhysRevB.65.165113} {\bibfield  {journal} {\bibinfo
  {journal} {Physical Review B},\ }\textbf {\bibinfo {volume} {65}},\ \bibinfo
  {pages} {165113} (\bibinfo {year} {2002}{\natexlab{a}})}\BibitemShut
  {NoStop}%
\bibitem [{\citenamefont {Wen}(2002){\natexlab{b}}}]{Wen:2002p8024}%
  \BibitemOpen
  \bibfield  {author} {\bibinfo {author} {\bibfnamefont {X.-G.}\ \bibnamefont
  {Wen}},\ }\Doi {10.1016/S0375-9601(02)00808-3} {\bibfield  {journal}
  {\bibinfo  {journal} {Physics Letters A},\ }\textbf {\bibinfo {volume}
  {300}},\ \bibinfo {pages} {175} (\bibinfo {year}
  {2002}{\natexlab{b}})}\BibitemShut {NoStop}%
\bibitem [{\citenamefont {Haldane}(1988)}]{Haldane:1988p5265}%
  \BibitemOpen
  \bibfield  {author} {\bibinfo {author} {\bibfnamefont {F.~D.~M.}\
  \bibnamefont {Haldane}},\ }\Doi {10.1103/PhysRevLett.61.2015} {\bibfield
  {journal} {\bibinfo  {journal} {Physical Review Letters},\ }\textbf {\bibinfo
  {volume} {61}},\ \bibinfo {pages} {2015} (\bibinfo {year}
  {1988})}\BibitemShut {NoStop}%
\bibitem [{\citenamefont {Kitaev}(2003)}]{Kitaev:2003p6185}%
  \BibitemOpen
  \bibfield  {author} {\bibinfo {author} {\bibfnamefont {A.~Y.}\ \bibnamefont
  {Kitaev}},\ }\Doi {10.1016/S0003-4916(02)00018-0} {\bibfield  {journal}
  {\bibinfo  {journal} {Annals of Physics},\ }\textbf {\bibinfo {volume}
  {303}},\ \bibinfo {pages} {2} (\bibinfo {year} {2003})}\BibitemShut {NoStop}%
\bibitem [{\citenamefont {Morita}\ \emph {et~al.}(2002)\citenamefont {Morita},
  \citenamefont {Watanabe},\ and\ \citenamefont {Imada}}]{Morita:2002p8014}%
  \BibitemOpen
  \bibfield  {author} {\bibinfo {author} {\bibfnamefont {H.}~\bibnamefont
  {Morita}}, \bibinfo {author} {\bibfnamefont {S.}~\bibnamefont {Watanabe}}, \
  and\ \bibinfo {author} {\bibfnamefont {M.}~\bibnamefont {Imada}},\ }\Doi
  {10.1143/JPSJ.71.2109} {\bibfield  {journal} {\bibinfo  {journal} {Journal of
  the Physical Society of Japan},\ }\textbf {\bibinfo {volume} {71}},\ \bibinfo
  {pages} {2109} (\bibinfo {year} {2002})}\BibitemShut {NoStop}%
\bibitem [{\citenamefont {Koretsune}\ \emph {et~al.}(2007)\citenamefont
  {Koretsune}, \citenamefont {Motome},\ and\ \citenamefont
  {Furusaki}}]{Koretsune:2007p8016}%
  \BibitemOpen
  \bibfield  {author} {\bibinfo {author} {\bibfnamefont {T.}~\bibnamefont
  {Koretsune}}, \bibinfo {author} {\bibfnamefont {Y.}~\bibnamefont {Motome}}, \
  and\ \bibinfo {author} {\bibfnamefont {A.}~\bibnamefont {Furusaki}},\ }\Doi
  {10.1143/JPSJ.76.074719} {\bibfield  {journal} {\bibinfo  {journal} {Journal
  of the Physical Society of Japan},\ }\textbf {\bibinfo {volume} {76}},\
  \bibinfo {pages} {4719} (\bibinfo {year} {2007})}\BibitemShut {NoStop}%
\bibitem [{\citenamefont {Kyung}\ and\ \citenamefont
  {Tremblay}(2006)}]{Kyung:2006p8015}%
  \BibitemOpen
  \bibfield  {author} {\bibinfo {author} {\bibfnamefont {B.}~\bibnamefont
  {Kyung}}\ and\ \bibinfo {author} {\bibfnamefont {A.-M.~S.}\ \bibnamefont
  {Tremblay}},\ }\Doi {10.1103/PhysRevLett.97.046402} {\bibfield  {journal}
  {\bibinfo  {journal} {Physical Review Letters},\ }\textbf {\bibinfo {volume}
  {97}},\ \bibinfo {pages} {46402} (\bibinfo {year} {2006})}\BibitemShut
  {NoStop}%
\bibitem [{\citenamefont {Tay}\ and\ \citenamefont
  {Motrunich}(2011)}]{Tay:2011p7822}%
  \BibitemOpen
  \bibfield  {author} {\bibinfo {author} {\bibfnamefont {T.}~\bibnamefont
  {Tay}}\ and\ \bibinfo {author} {\bibfnamefont {O.~I.}\ \bibnamefont
  {Motrunich}},\ }\Doi {10.1103/PhysRevB.84.020404} {\bibfield  {journal}
  {\bibinfo  {journal} {Physical Review B},\ }\textbf {\bibinfo {volume}
  {84}},\ \bibinfo {pages} {020404} (\bibinfo {year} {2011})}\BibitemShut
  {NoStop}%
\bibitem [{\citenamefont {Wang}(2010)}]{Wang:2010p6724}%
  \BibitemOpen
  \bibfield  {author} {\bibinfo {author} {\bibfnamefont {F.}~\bibnamefont
  {Wang}},\ }\Doi {10.1103/PhysRevB.82.024419} {\bibfield  {journal} {\bibinfo
  {journal} {Physical Review B},\ }\textbf {\bibinfo {volume} {82}},\ \bibinfo
  {pages} {24419} (\bibinfo {year} {2010})}\BibitemShut {NoStop}%
\bibitem [{Note2()}]{Note2}%
  \BibitemOpen
  \bibinfo {note} {For $X\times Y\times 2$ lattices, $X$=$Y$ is required to
  respect point-group symmetry of the honeycomb lattice, and $X$ needs to be an
  even integer so that $1/4$ doping can be accommodated.}\BibitemShut {Stop}%
\bibitem [{Note3()}]{Note3}%
  \BibitemOpen
  \bibinfo {note} {On the honeycomb lattice, apart from the $2N\times 2N\times
  2$ samples, there exists a second sequence of finite-size samples respecting
  the full point group symmetry, and accommodating the 1/4 doping. By tripling
  the unit cell, namely treating each hexagon in the honeycomb lattice as one
  unit cell, one can obtain this second sequence as $2N\times 2N\times 6$
  lattices. Among this sequence, although the 6-site sample is very small, the
  24-site sample considered here has a reasonable size to investigate the bulk
  physics.}\BibitemShut {Stop}%
\bibitem [{Note4()}]{Note4}%
  \BibitemOpen
  \bibinfo {note} {This energy is between values for two larger samples, inset
  of Fig.~\ref {fig:did_relative_energy}.}\BibitemShut {Stop}%
\bibitem [{\citenamefont {Horsch}\ and\ \citenamefont
  {Kaplan}()}]{Horsch:2014p7574}%
  \BibitemOpen
  \bibfield  {author} {\bibinfo {author} {\bibfnamefont {P.}~\bibnamefont
  {Horsch}}\ and\ \bibinfo {author} {\bibfnamefont {T.~A.}\ \bibnamefont
  {Kaplan}},\ }\Doi {10.1088/0022-3719/16/35/002} {\bibfield  {journal}
  {\bibinfo  {journal} {J. Phys. C: Solid State Phys.},\ }\textbf {\bibinfo
  {volume} {16}},\ \bibinfo {pages} {L1203}}\BibitemShut {NoStop}%
\bibitem [{Note5()}]{Note5}%
  \BibitemOpen
  \bibinfo {note} {For instance, on the 32-site sample the ratio of correlation
  between farthest sites and nearest neighbor sites is typically around 5 times
  smaller than in the DMRG state in c-SDW/SCCL state (Fig.~\ref
  {fig:szszfar}).}\BibitemShut {Stop}%
\bibitem [{ite(2014)}]{itensor}%
  \BibitemOpen
  \href@noop {} {\bibfield  {journal} {\bibinfo  {journal} {itensor.org}}
  (\bibinfo {year} {2014})}\BibitemShut {NoStop}%
\bibitem [{Note6()}]{Note6}%
  \BibitemOpen
  \bibinfo {note} {We also check that projecting to opposite rotation sector
  (e.g., $\protect \qopname \relax o{exp}(+i\pi /3)$ in low-$J$ phase) reverses
  the sign of chirality.}\BibitemShut {Stop}%
\bibitem [{\citenamefont {Wen}\ and\ \citenamefont
  {Zee}(1992)}]{Wen:1992p6753}%
  \BibitemOpen
  \bibfield  {author} {\bibinfo {author} {\bibfnamefont {X.-G.}\ \bibnamefont
  {Wen}}\ and\ \bibinfo {author} {\bibfnamefont {A.}~\bibnamefont {Zee}},\
  }\Doi {10.1103/PhysRevB.46.2290} {\bibfield  {journal} {\bibinfo  {journal}
  {Physical Review B (Condensed Matter)},\ }\textbf {\bibinfo {volume} {46}},\
  \bibinfo {pages} {2290} (\bibinfo {year} {1992})}\BibitemShut {NoStop}%
\bibitem [{\citenamefont {Girvin}\ and\ \citenamefont
  {MacDonald}(1987)}]{Girvin:1987p8027}%
  \BibitemOpen
  \bibfield  {author} {\bibinfo {author} {\bibfnamefont {S.~M.}\ \bibnamefont
  {Girvin}}\ and\ \bibinfo {author} {\bibfnamefont {A.~H.}\ \bibnamefont
  {MacDonald}},\ }\Doi {10.1103/PhysRevLett.58.1252} {\bibfield  {journal}
  {\bibinfo  {journal} {Physical Review Letters},\ }\textbf {\bibinfo {volume}
  {58}},\ \bibinfo {pages} {1252} (\bibinfo {year} {1987})}\BibitemShut
  {NoStop}%
\bibitem [{\citenamefont {Zhang}\ \emph {et~al.}(1989)\citenamefont {Zhang},
  \citenamefont {Hansson},\ and\ \citenamefont {Kivelson}}]{Zhang:1989p4231}%
  \BibitemOpen
  \bibfield  {author} {\bibinfo {author} {\bibfnamefont {S.-C.}\ \bibnamefont
  {Zhang}}, \bibinfo {author} {\bibfnamefont {T.~H.}\ \bibnamefont {Hansson}},
  \ and\ \bibinfo {author} {\bibfnamefont {S.}~\bibnamefont {Kivelson}},\ }\Doi
  {10.1103/PhysRevLett.62.82} {\bibfield  {journal} {\bibinfo  {journal}
  {Physical Review Letters (ISSN 0031-9007)},\ }\textbf {\bibinfo {volume}
  {62}},\ \bibinfo {pages} {82} (\bibinfo {year} {1989})}\BibitemShut {NoStop}%
\bibitem [{\citenamefont {Wen}(1990)}]{Wen:1990p7458}%
  \BibitemOpen
  \bibfield  {author} {\bibinfo {author} {\bibfnamefont {X.~G.}\ \bibnamefont
  {Wen}},\ }\Doi {10.1142/S0217979290000139} {\bibfield  {journal} {\bibinfo
  {journal} {Int J Mod Phys B},\ }\textbf {\bibinfo {volume} {4}},\ \bibinfo
  {pages} {239} (\bibinfo {year} {1990})}\BibitemShut {NoStop}%
\bibitem [{\citenamefont {Wen}\ and\ \citenamefont
  {Niu}(1990)}]{Wen:1990p5870}%
  \BibitemOpen
  \bibfield  {author} {\bibinfo {author} {\bibfnamefont {X.-G.}\ \bibnamefont
  {Wen}}\ and\ \bibinfo {author} {\bibfnamefont {Q.}~\bibnamefont {Niu}},\
  }\Doi {10.1103/PhysRevB.41.9377} {\bibfield  {journal} {\bibinfo  {journal}
  {Physical Review B (Condensed Matter)},\ }\textbf {\bibinfo {volume} {41}},\
  \bibinfo {pages} {9377} (\bibinfo {year} {1990})}\BibitemShut {NoStop}%
\bibitem [{\citenamefont {Essin}\ and\ \citenamefont
  {Hermele}(2013)}]{Essin:2013p7585}%
  \BibitemOpen
  \bibfield  {author} {\bibinfo {author} {\bibfnamefont {A.~M.}\ \bibnamefont
  {Essin}}\ and\ \bibinfo {author} {\bibfnamefont {M.}~\bibnamefont
  {Hermele}},\ }\Doi {10.1103/PhysRevB.87.104406} {\bibfield  {journal}
  {\bibinfo  {journal} {Physical Review B},\ }\textbf {\bibinfo {volume}
  {87}},\ \bibinfo {pages} {104406} (\bibinfo {year} {2013})}\BibitemShut
  {NoStop}%
\bibitem [{\citenamefont {Chen}\ \emph {et~al.}(2014)\citenamefont {Chen},
  \citenamefont {Burnell}, \citenamefont {Vishwanath},\ and\ \citenamefont
  {Fidkowski}}]{Chen:2014p8001}%
  \BibitemOpen
  \bibfield  {author} {\bibinfo {author} {\bibfnamefont {X.}~\bibnamefont
  {Chen}}, \bibinfo {author} {\bibfnamefont {F.~J.}\ \bibnamefont {Burnell}},
  \bibinfo {author} {\bibfnamefont {A.}~\bibnamefont {Vishwanath}}, \ and\
  \bibinfo {author} {\bibfnamefont {L.}~\bibnamefont {Fidkowski}},\ }\href
  {http://arxiv.org/abs/1403.6491v1} {\bibfield  {journal} {\bibinfo  {journal}
  {arXiv},\ }\textbf {\bibinfo {volume} {cond-mat.str-el}} (\bibinfo {year}
  {2014})},\ \Eprint {http://arxiv.org/abs/1403.6491v1} {1403.6491v1}
  \BibitemShut {NoStop}%
\bibitem [{\citenamefont {Hung}\ and\ \citenamefont
  {Wen}(2012)}]{Hung:2012p7537}%
  \BibitemOpen
  \bibfield  {author} {\bibinfo {author} {\bibfnamefont {L.-Y.}\ \bibnamefont
  {Hung}}\ and\ \bibinfo {author} {\bibfnamefont {X.-G.}\ \bibnamefont {Wen}},\
  }\href {http://arxiv.org/abs/1212.1827v1} {\bibfield  {journal} {\bibinfo
  {journal} {arXiv},\ }\textbf {\bibinfo {volume} {cond-mat.str-el}} (\bibinfo
  {year} {2012})},\ \Eprint {http://arxiv.org/abs/1212.1827v1} {1212.1827v1}
  \BibitemShut {NoStop}%
\bibitem [{\citenamefont {Mesaros}\ and\ \citenamefont
  {Ran}(2012)}]{Mesaros:2012p7535}%
  \BibitemOpen
  \bibfield  {author} {\bibinfo {author} {\bibfnamefont {A.}~\bibnamefont
  {Mesaros}}\ and\ \bibinfo {author} {\bibfnamefont {Y.}~\bibnamefont {Ran}},\
  }\href {http://arxiv.org/abs/1212.0835v3} {\bibfield  {journal} {\bibinfo
  {journal} {arXiv},\ }\textbf {\bibinfo {volume} {cond-mat.str-el}} (\bibinfo
  {year} {2012})},\ \Eprint {http://arxiv.org/abs/1212.0835v3} {1212.0835v3}
  \BibitemShut {NoStop}%
\bibitem [{\citenamefont {Lu}\ and\ \citenamefont
  {Vishwanath}(2013)}]{Lu:2013p7858}%
  \BibitemOpen
  \bibfield  {author} {\bibinfo {author} {\bibfnamefont {Y.-M.}\ \bibnamefont
  {Lu}}\ and\ \bibinfo {author} {\bibfnamefont {A.}~\bibnamefont
  {Vishwanath}},\ }\href {http://arxiv.org/abs/1302.2634v2} {\bibfield
  {journal} {\bibinfo  {journal} {arXiv},\ }\textbf {\bibinfo {volume}
  {cond-mat.str-el}} (\bibinfo {year} {2013})},\ \Eprint
  {http://arxiv.org/abs/1302.2634v2} {1302.2634v2} \BibitemShut {NoStop}%
\bibitem [{\citenamefont {Wen}(1995)}]{Wen:1995p8031}%
  \BibitemOpen
  \bibfield  {author} {\bibinfo {author} {\bibfnamefont {X.-G.}\ \bibnamefont
  {Wen}},\ }\href
  {http://www.tandfonline.com/doi/abs/10.1080/00018739500101566%23.Uz8L_17YLl4}
  {\bibfield  {journal} {\bibinfo  {journal} {Advances in Physics},\ }\textbf
  {\bibinfo {volume} {44}},\ \bibinfo {pages} {405} (\bibinfo {year}
  {1995})}\BibitemShut {NoStop}%
\bibitem [{\citenamefont {Lu}\ and\ \citenamefont
  {Vishwanath}(2012)}]{Lu:2012p7468}%
  \BibitemOpen
  \bibfield  {author} {\bibinfo {author} {\bibfnamefont {Y.-M.}\ \bibnamefont
  {Lu}}\ and\ \bibinfo {author} {\bibfnamefont {A.}~\bibnamefont
  {Vishwanath}},\ }\Doi {10.1103/PhysRevB.86.125119} {\bibfield  {journal}
  {\bibinfo  {journal} {Physical Review B},\ }\textbf {\bibinfo {volume}
  {86}},\ \bibinfo {pages} {125119} (\bibinfo {year} {2012})}\BibitemShut
  {NoStop}%
\bibitem [{\citenamefont {Kane}\ and\ \citenamefont
  {Fisher}(1992)}]{Kane:1992p8028}%
  \BibitemOpen
  \bibfield  {author} {\bibinfo {author} {\bibfnamefont {C.}~\bibnamefont
  {Kane}}\ and\ \bibinfo {author} {\bibfnamefont {M.~P.~A.}\ \bibnamefont
  {Fisher}},\ }\href
  {http://journals.aps.org/prb/abstract/10.1103/PhysRevB.46.15233} {\bibfield
  {journal} {\bibinfo  {journal} {Physical Review B},\ }\textbf {\bibinfo
  {volume} {46}},\ \bibinfo {pages} {15233} (\bibinfo {year}
  {1992})}\BibitemShut {NoStop}%
\bibitem [{\citenamefont {Fisher}(1994)}]{Fisher:1994p8066}%
  \BibitemOpen
  \bibfield  {author} {\bibinfo {author} {\bibfnamefont {M.~P.~A.}\
  \bibnamefont {Fisher}},\ }\Doi {10.1103/PhysRevB.49.14550} {\bibfield
  {journal} {\bibinfo  {journal} {Physical Review B (Condensed Matter)},\
  }\textbf {\bibinfo {volume} {49}},\ \bibinfo {pages} {1332495} (\bibinfo
  {year} {1994})}\BibitemShut {NoStop}%
\bibitem [{\citenamefont {Stone}\ and\ \citenamefont
  {Fisher}(1994)}]{Stone:1994p8029}%
  \BibitemOpen
  \bibfield  {author} {\bibinfo {author} {\bibfnamefont {M.}~\bibnamefont
  {Stone}}\ and\ \bibinfo {author} {\bibfnamefont {M.~P.~A.}\ \bibnamefont
  {Fisher}},\ }\href
  {http://www.worldscientific.com/doi/abs/10.1142/S0217979294001020} {\bibfield
   {journal} {\bibinfo  {journal} {Int J Mod Phys B},\ }\textbf {\bibinfo
  {volume} {8}},\ \bibinfo {pages} {2539} (\bibinfo {year} {1994})}\BibitemShut
  {NoStop}%
\bibitem [{\citenamefont {de~C~Chamon}\ and\ \citenamefont
  {Fradkin}(1997)}]{Chamon:1997p7909}%
  \BibitemOpen
  \bibfield  {author} {\bibinfo {author} {\bibfnamefont {C.}~\bibnamefont
  {de~C~Chamon}}\ and\ \bibinfo {author} {\bibfnamefont {E.}~\bibnamefont
  {Fradkin}},\ }\href {http://prb.aps.org/abstract/PRB/v56/i4/p2012_1}
  {\bibfield  {journal} {\bibinfo  {journal} {Physical Review B},\ }\textbf
  {\bibinfo {volume} {56}},\ \bibinfo {pages} {2012} (\bibinfo {year}
  {1997})}\BibitemShut {NoStop}%
\bibitem [{\citenamefont {White}\ and\ \citenamefont
  {Scalapino}(2000)}]{White:2000p8048}%
  \BibitemOpen
  \bibfield  {author} {\bibinfo {author} {\bibfnamefont {S.~R.}\ \bibnamefont
  {White}}\ and\ \bibinfo {author} {\bibfnamefont {D.~J.}\ \bibnamefont
  {Scalapino}},\ }\Doi {10.1103/PhysRevB.61.6320} {\bibfield  {journal}
  {\bibinfo  {journal} {Physical Review B (Condensed Matter and Materials
  Physics)},\ }\textbf {\bibinfo {volume} {61}},\ \bibinfo {pages} {6320}
  (\bibinfo {year} {2000})}\BibitemShut {NoStop}%
\bibitem [{Note7()}]{Note7}%
  \BibitemOpen
  \bibinfo {note} {The Fermi surface is only a mean-field level description.
  For example, one could imagine the situation that the Fermi surface is
  strongly coupled with dynamical gauge fields.\cite
  {Kaul:2008p8050}}\BibitemShut {NoStop}%
\bibitem [{\citenamefont {Luttinger}(1960)}]{Luttinger:1960p8033}%
  \BibitemOpen
  \bibfield  {author} {\bibinfo {author} {\bibfnamefont {J.~M.}\ \bibnamefont
  {Luttinger}},\ }\Doi {10.1103/PhysRev.119.1153} {\bibfield  {journal}
  {\bibinfo  {journal} {Phys. Rev.},\ }\textbf {\bibinfo {volume} {119}},\
  \bibinfo {pages} {1153} (\bibinfo {year} {1960})}\BibitemShut {NoStop}%
\bibitem [{\citenamefont {Sheng}\ \emph {et~al.}(2011)\citenamefont {Sheng},
  \citenamefont {Gu}, \citenamefont {Sun},\ and\ \citenamefont
  {Sheng}}]{Sheng:2011p8046}%
  \BibitemOpen
  \bibfield  {author} {\bibinfo {author} {\bibfnamefont {D.~N.}\ \bibnamefont
  {Sheng}}, \bibinfo {author} {\bibfnamefont {Z.-C.}\ \bibnamefont {Gu}},
  \bibinfo {author} {\bibfnamefont {K.}~\bibnamefont {Sun}}, \ and\ \bibinfo
  {author} {\bibfnamefont {L.}~\bibnamefont {Sheng}},\ }\Doi
  {doi:10.1038/ncomms1380} {\bibfield  {journal} {\bibinfo  {journal} {Nature
  Communications},\ }\textbf {\bibinfo {volume} {2}},\ \bibinfo {pages} {389}
  (\bibinfo {year} {2011})}\BibitemShut {NoStop}%
\bibitem [{\citenamefont {Regnault}\ and\ \citenamefont
  {Bernevig}(2011)}]{Regnault:2011p8044}%
  \BibitemOpen
  \bibfield  {author} {\bibinfo {author} {\bibfnamefont {N.}~\bibnamefont
  {Regnault}}\ and\ \bibinfo {author} {\bibfnamefont {B.~A.}\ \bibnamefont
  {Bernevig}},\ }\Doi {10.1103/PhysRevX.1.021014} {\bibfield  {journal}
  {\bibinfo  {journal} {Physical Review X},\ }\textbf {\bibinfo {volume} {1}},\
  \bibinfo {pages} {21014} (\bibinfo {year} {2011})}\BibitemShut {NoStop}%
\bibitem [{\citenamefont {Wang}\ \emph
  {et~al.}(2012){\natexlab{b}}\citenamefont {Wang}, \citenamefont {Yao},
  \citenamefont {Gu}, \citenamefont {Gong},\ and\ \citenamefont
  {Sheng}}]{Wang:2012p5933}%
  \BibitemOpen
  \bibfield  {author} {\bibinfo {author} {\bibfnamefont {Y.-F.}\ \bibnamefont
  {Wang}}, \bibinfo {author} {\bibfnamefont {H.}~\bibnamefont {Yao}}, \bibinfo
  {author} {\bibfnamefont {Z.-C.}\ \bibnamefont {Gu}}, \bibinfo {author}
  {\bibfnamefont {C.-D.}\ \bibnamefont {Gong}}, \ and\ \bibinfo {author}
  {\bibfnamefont {D.~N.}\ \bibnamefont {Sheng}},\ }\Doi
  {10.1103/PhysRevLett.108.126805} {\bibfield  {journal} {\bibinfo  {journal}
  {Physical Review Letters},\ }\textbf {\bibinfo {volume} {108}},\ \bibinfo
  {pages} {126805} (\bibinfo {year} {2012}{\natexlab{b}})}\BibitemShut
  {NoStop}%
\bibitem [{\citenamefont {Lu}\ and\ \citenamefont {Ran}(2012)}]{Lu:2012p6109}%
  \BibitemOpen
  \bibfield  {author} {\bibinfo {author} {\bibfnamefont {Y.-M.}\ \bibnamefont
  {Lu}}\ and\ \bibinfo {author} {\bibfnamefont {Y.}~\bibnamefont {Ran}},\ }\Doi
  {10.1103/PhysRevB.85.165134} {\bibfield  {journal} {\bibinfo  {journal}
  {Physical Review B},\ }\textbf {\bibinfo {volume} {85}},\ \bibinfo {pages}
  {165134} (\bibinfo {year} {2012})}\BibitemShut {NoStop}%
\bibitem [{\citenamefont {Liu}\ \emph {et~al.}(2012)\citenamefont {Liu},
  \citenamefont {Bergholtz}, \citenamefont {Fan},\ and\ \citenamefont
  {L{\"a}uchli}}]{Liu:2012p8034}%
  \BibitemOpen
  \bibfield  {author} {\bibinfo {author} {\bibfnamefont {Z.}~\bibnamefont
  {Liu}}, \bibinfo {author} {\bibfnamefont {E.~J.}\ \bibnamefont {Bergholtz}},
  \bibinfo {author} {\bibfnamefont {H.}~\bibnamefont {Fan}}, \ and\ \bibinfo
  {author} {\bibfnamefont {A.~M.}\ \bibnamefont {L{\"a}uchli}},\ }\Doi
  {10.1103/PhysRevLett.109.186805} {\bibfield  {journal} {\bibinfo  {journal}
  {Physical Review Letters},\ }\textbf {\bibinfo {volume} {109}},\ \bibinfo
  {pages} {186805} (\bibinfo {year} {2012})}\BibitemShut {NoStop}%
\bibitem [{\citenamefont {Kourtis}\ \emph {et~al.}(2012)\citenamefont
  {Kourtis}, \citenamefont {Venderbos},\ and\ \citenamefont
  {Daghofer}}]{Kourtis:2012p8039}%
  \BibitemOpen
  \bibfield  {author} {\bibinfo {author} {\bibfnamefont {S.}~\bibnamefont
  {Kourtis}}, \bibinfo {author} {\bibfnamefont {J.~W.~F.}\ \bibnamefont
  {Venderbos}}, \ and\ \bibinfo {author} {\bibfnamefont {M.}~\bibnamefont
  {Daghofer}},\ }\Doi {10.1103/PhysRevB.86.235118} {\bibfield  {journal}
  {\bibinfo  {journal} {Physical Review B},\ }\textbf {\bibinfo {volume}
  {86}},\ \bibinfo {pages} {235118} (\bibinfo {year} {2012})}\BibitemShut
  {NoStop}%
\bibitem [{\citenamefont {Jian}\ and\ \citenamefont
  {Qi}(2013)}]{Jian:2013p8035}%
  \BibitemOpen
  \bibfield  {author} {\bibinfo {author} {\bibfnamefont {C.-M.}\ \bibnamefont
  {Jian}}\ and\ \bibinfo {author} {\bibfnamefont {X.-L.}\ \bibnamefont {Qi}},\
  }\Doi {10.1103/PhysRevB.88.165134} {\bibfield  {journal} {\bibinfo  {journal}
  {Physical Review B},\ }\textbf {\bibinfo {volume} {88}},\ \bibinfo {pages}
  {165134} (\bibinfo {year} {2013})}\BibitemShut {NoStop}%
\bibitem [{\citenamefont {Grushin}\ \emph {et~al.}(2012)\citenamefont
  {Grushin}, \citenamefont {Neupert}, \citenamefont {Chamon},\ and\
  \citenamefont {Mudry}}]{Grushin:2012p8036}%
  \BibitemOpen
  \bibfield  {author} {\bibinfo {author} {\bibfnamefont {A.~G.}\ \bibnamefont
  {Grushin}}, \bibinfo {author} {\bibfnamefont {T.}~\bibnamefont {Neupert}},
  \bibinfo {author} {\bibfnamefont {C.}~\bibnamefont {Chamon}}, \ and\ \bibinfo
  {author} {\bibfnamefont {C.}~\bibnamefont {Mudry}},\ }\Doi
  {10.1103/PhysRevB.86.205125} {\bibfield  {journal} {\bibinfo  {journal}
  {Physical Review B},\ }\textbf {\bibinfo {volume} {86}},\ \bibinfo {pages}
  {205125} (\bibinfo {year} {2012})}\BibitemShut {NoStop}%
\bibitem [{\citenamefont {Maciejko}\ \emph {et~al.}(2010)\citenamefont
  {Maciejko}, \citenamefont {Qi}, \citenamefont {Karch},\ and\ \citenamefont
  {Zhang}}]{Maciejko:2010p7431}%
  \BibitemOpen
  \bibfield  {author} {\bibinfo {author} {\bibfnamefont {J.}~\bibnamefont
  {Maciejko}}, \bibinfo {author} {\bibfnamefont {X.-L.}\ \bibnamefont {Qi}},
  \bibinfo {author} {\bibfnamefont {A.}~\bibnamefont {Karch}}, \ and\ \bibinfo
  {author} {\bibfnamefont {S.-C.}\ \bibnamefont {Zhang}},\ }\Doi
  {10.1103/PhysRevLett.105.246809} {\bibfield  {journal} {\bibinfo  {journal}
  {Physical Review Letters},\ }\textbf {\bibinfo {volume} {105}},\ \bibinfo
  {pages} {246809} (\bibinfo {year} {2010})}\BibitemShut {NoStop}%
\bibitem [{\citenamefont {Levin}\ and\ \citenamefont
  {Stern}(2012)}]{Levin:2012p7429}%
  \BibitemOpen
  \bibfield  {author} {\bibinfo {author} {\bibfnamefont {M.}~\bibnamefont
  {Levin}}\ and\ \bibinfo {author} {\bibfnamefont {A.}~\bibnamefont {Stern}},\
  }\Doi {10.1103/PhysRevB.86.115131} {\bibfield  {journal} {\bibinfo  {journal}
  {Physical Review B},\ }\textbf {\bibinfo {volume} {86}},\ \bibinfo {pages}
  {115131} (\bibinfo {year} {2012})}\BibitemShut {NoStop}%
\bibitem [{\citenamefont {Cho}\ \emph {et~al.}(2012)\citenamefont {Cho},
  \citenamefont {Lu},\ and\ \citenamefont {Moore}}]{Cho:2012p7426}%
  \BibitemOpen
  \bibfield  {author} {\bibinfo {author} {\bibfnamefont {G.}~\bibnamefont
  {Cho}}, \bibinfo {author} {\bibfnamefont {Y.-M.}\ \bibnamefont {Lu}}, \ and\
  \bibinfo {author} {\bibfnamefont {J.}~\bibnamefont {Moore}},\ }\Doi
  {10.1103/PhysRevB.86.125101} {\bibfield  {journal} {\bibinfo  {journal}
  {Physical Review B},\ }\textbf {\bibinfo {volume} {86}},\ \bibinfo {pages}
  {125101} (\bibinfo {year} {2012})}\BibitemShut {NoStop}%
\bibitem [{\citenamefont {Swingle}\ \emph {et~al.}(2011)\citenamefont
  {Swingle}, \citenamefont {Barkeshli}, \citenamefont {Mcgreevy},\ and\
  \citenamefont {Senthil}}]{Swingle:2011p7432}%
  \BibitemOpen
  \bibfield  {author} {\bibinfo {author} {\bibfnamefont {B.}~\bibnamefont
  {Swingle}}, \bibinfo {author} {\bibfnamefont {M.}~\bibnamefont {Barkeshli}},
  \bibinfo {author} {\bibfnamefont {J.}~\bibnamefont {Mcgreevy}}, \ and\
  \bibinfo {author} {\bibfnamefont {T.}~\bibnamefont {Senthil}},\ }\Doi
  {10.1103/PhysRevB.83.195139} {\bibfield  {journal} {\bibinfo  {journal}
  {Physical Review B},\ }\textbf {\bibinfo {volume} {83}},\ \bibinfo {pages}
  {195139} (\bibinfo {year} {2011})}\BibitemShut {NoStop}%
\bibitem [{\citenamefont {Kataev}\ \emph {et~al.}(2005)\citenamefont {Kataev},
  \citenamefont {M{\"o}ller}, \citenamefont {L{\"o}w}, \citenamefont {Jung},
  \citenamefont {Schittner}, \citenamefont {Kriener},\ and\ \citenamefont
  {Freimuth}}]{Kataev:2005p8065}%
  \BibitemOpen
  \bibfield  {author} {\bibinfo {author} {\bibfnamefont {V.}~\bibnamefont
  {Kataev}}, \bibinfo {author} {\bibfnamefont {A.}~\bibnamefont {M{\"o}ller}},
  \bibinfo {author} {\bibfnamefont {U.}~\bibnamefont {L{\"o}w}}, \bibinfo
  {author} {\bibfnamefont {W.}~\bibnamefont {Jung}}, \bibinfo {author}
  {\bibfnamefont {N.}~\bibnamefont {Schittner}}, \bibinfo {author}
  {\bibfnamefont {M.}~\bibnamefont {Kriener}}, \ and\ \bibinfo {author}
  {\bibfnamefont {A.}~\bibnamefont {Freimuth}},\ }\href
  {http://www.sciencedirect.com.ezp-prod1.hul.harvard.edu/science/article/pii/%
S0304885304014623} {\bibfield  {journal} {\bibinfo  {journal} {Journal of
  Magnetism and Magnetic Materials},\ }\textbf {\bibinfo {volume} {290}},\
  \bibinfo {pages} {310} (\bibinfo {year} {2005})}\BibitemShut {NoStop}%
\bibitem [{\citenamefont {Das}\ \emph {et~al.}(2008)\citenamefont {Das},
  \citenamefont {Pisana}, \citenamefont {Chakraborty}, \citenamefont
  {Piscanec}, \citenamefont {Saha}, \citenamefont {Waghmare}, \citenamefont
  {Novoselov}, \citenamefont {Krishnamurthy}, \citenamefont {Geim},
  \citenamefont {Ferrari},\ and\ \citenamefont {Sood}}]{Das:2008p8054}%
  \BibitemOpen
  \bibfield  {author} {\bibinfo {author} {\bibfnamefont {A.}~\bibnamefont
  {Das}}, \bibinfo {author} {\bibfnamefont {S.}~\bibnamefont {Pisana}},
  \bibinfo {author} {\bibfnamefont {B.}~\bibnamefont {Chakraborty}}, \bibinfo
  {author} {\bibfnamefont {S.}~\bibnamefont {Piscanec}}, \bibinfo {author}
  {\bibfnamefont {S.~K.}\ \bibnamefont {Saha}}, \bibinfo {author}
  {\bibfnamefont {U.~V.}\ \bibnamefont {Waghmare}}, \bibinfo {author}
  {\bibfnamefont {K.~S.}\ \bibnamefont {Novoselov}}, \bibinfo {author}
  {\bibfnamefont {H.~R.}\ \bibnamefont {Krishnamurthy}}, \bibinfo {author}
  {\bibfnamefont {A.~K.}\ \bibnamefont {Geim}}, \bibinfo {author}
  {\bibfnamefont {A.~C.}\ \bibnamefont {Ferrari}}, \ and\ \bibinfo {author}
  {\bibfnamefont {A.~K.}\ \bibnamefont {Sood}},\ }\Doi
  {doi:10.1038/nnano.2008.67} {\bibfield  {journal} {\bibinfo  {journal}
  {Nature Nanotechnology},\ }\textbf {\bibinfo {volume} {3}},\ \bibinfo {pages}
  {210} (\bibinfo {year} {2008})}\BibitemShut {NoStop}%
\bibitem [{\citenamefont {McChesney}\ \emph {et~al.}(2010)\citenamefont
  {McChesney}, \citenamefont {Bostwick}, \citenamefont {Ohta}, \citenamefont
  {Seyller}, \citenamefont {Horn}, \citenamefont {Gonz{\'a}lez},\ and\
  \citenamefont {Rotenberg}}]{McChesney:2010p7899}%
  \BibitemOpen
  \bibfield  {author} {\bibinfo {author} {\bibfnamefont {J.~L.}\ \bibnamefont
  {McChesney}}, \bibinfo {author} {\bibfnamefont {A.}~\bibnamefont {Bostwick}},
  \bibinfo {author} {\bibfnamefont {T.}~\bibnamefont {Ohta}}, \bibinfo {author}
  {\bibfnamefont {T.}~\bibnamefont {Seyller}}, \bibinfo {author} {\bibfnamefont
  {K.}~\bibnamefont {Horn}}, \bibinfo {author} {\bibfnamefont {J.}~\bibnamefont
  {Gonz{\'a}lez}}, \ and\ \bibinfo {author} {\bibfnamefont {E.}~\bibnamefont
  {Rotenberg}},\ }\Doi {10.1103/PhysRevLett.104.136803} {\bibfield  {journal}
  {\bibinfo  {journal} {Physical Review Letters},\ }\textbf {\bibinfo {volume}
  {104}},\ \bibinfo {pages} {136803} (\bibinfo {year} {2010})}\BibitemShut
  {NoStop}%
\bibitem [{\citenamefont {Xiao}\ \emph {et~al.}(2011)\citenamefont {Xiao},
  \citenamefont {Zhu}, \citenamefont {Ran}, \citenamefont {Nagaosa},\ and\
  \citenamefont {Okamoto}}]{1Xiao:2011p7761}%
  \BibitemOpen
  \bibfield  {author} {\bibinfo {author} {\bibfnamefont {D.}~\bibnamefont
  {Xiao}}, \bibinfo {author} {\bibfnamefont {W.}~\bibnamefont {Zhu}}, \bibinfo
  {author} {\bibfnamefont {Y.}~\bibnamefont {Ran}}, \bibinfo {author}
  {\bibfnamefont {N.}~\bibnamefont {Nagaosa}}, \ and\ \bibinfo {author}
  {\bibfnamefont {S.}~\bibnamefont {Okamoto}},\ }\Doi {10.1038/ncomms1602}
  {\bibfield  {journal} {\bibinfo  {journal} {Nature Communications},\ }\textbf
  {\bibinfo {volume} {2}},\ \bibinfo {pages} {596} (\bibinfo {year}
  {2011})}\BibitemShut {NoStop}%
\bibitem [{\citenamefont {Middey}\ \emph {et~al.}(2012)\citenamefont {Middey},
  \citenamefont {Meyers}, \citenamefont {Kareev}, \citenamefont {Moon},
  \citenamefont {Gray}, \citenamefont {Liu}, \citenamefont {Freeland},\ and\
  \citenamefont {Chakhalian}}]{Middey:2012p7893}%
  \BibitemOpen
  \bibfield  {author} {\bibinfo {author} {\bibfnamefont {S.}~\bibnamefont
  {Middey}}, \bibinfo {author} {\bibfnamefont {D.}~\bibnamefont {Meyers}},
  \bibinfo {author} {\bibfnamefont {M.}~\bibnamefont {Kareev}}, \bibinfo
  {author} {\bibfnamefont {E.~J.}\ \bibnamefont {Moon}}, \bibinfo {author}
  {\bibfnamefont {B.~A.}\ \bibnamefont {Gray}}, \bibinfo {author}
  {\bibfnamefont {X.}~\bibnamefont {Liu}}, \bibinfo {author} {\bibfnamefont
  {J.~W.}\ \bibnamefont {Freeland}}, \ and\ \bibinfo {author} {\bibfnamefont
  {J.}~\bibnamefont {Chakhalian}},\ }\Doi {10.1063/1.4773375} {\bibfield
  {journal} {\bibinfo  {journal} {Applied Physics Letters},\ }\textbf {\bibinfo
  {volume} {101}},\ \bibinfo {pages} {1602} (\bibinfo {year}
  {2012})}\BibitemShut {NoStop}%
\bibitem [{\citenamefont {Okamoto}(2013)}]{Okamoto:2013p8067}%
  \BibitemOpen
  \bibfield  {author} {\bibinfo {author} {\bibfnamefont {S.}~\bibnamefont
  {Okamoto}},\ }\Doi {10.1103/PhysRevLett.110.066403} {\bibfield  {journal}
  {\bibinfo  {journal} {Physical Review Letters},\ }\textbf {\bibinfo {volume}
  {110}},\ \bibinfo {pages} {66403} (\bibinfo {year} {2013})}\BibitemShut
  {NoStop}%
\bibitem [{\citenamefont {Doennig}\ \emph {et~al.}(2013)\citenamefont
  {Doennig}, \citenamefont {Pickett},\ and\ \citenamefont
  {Pentcheva}}]{Doennig:2013p8051}%
  \BibitemOpen
  \bibfield  {author} {\bibinfo {author} {\bibfnamefont {D.}~\bibnamefont
  {Doennig}}, \bibinfo {author} {\bibfnamefont {W.~E.}\ \bibnamefont
  {Pickett}}, \ and\ \bibinfo {author} {\bibfnamefont {R.}~\bibnamefont
  {Pentcheva}},\ }\Doi {10.1103/PhysRevLett.111.126804} {\bibfield  {journal}
  {\bibinfo  {journal} {Physical Review Letters},\ }\textbf {\bibinfo {volume}
  {111}},\ \bibinfo {pages} {126804} (\bibinfo {year} {2013})}\BibitemShut
  {NoStop}%
\bibitem [{\citenamefont {Polini}\ \emph {et~al.}(2013)\citenamefont {Polini},
  \citenamefont {Guinea}, \citenamefont {Lewenstein}, \citenamefont
  {Manoharan},\ and\ \citenamefont {Pellegrini}}]{Polini:2013p8068}%
  \BibitemOpen
  \bibfield  {author} {\bibinfo {author} {\bibfnamefont {M.}~\bibnamefont
  {Polini}}, \bibinfo {author} {\bibfnamefont {F.}~\bibnamefont {Guinea}},
  \bibinfo {author} {\bibfnamefont {M.}~\bibnamefont {Lewenstein}}, \bibinfo
  {author} {\bibfnamefont {H.~C.}\ \bibnamefont {Manoharan}}, \ and\ \bibinfo
  {author} {\bibfnamefont {V.}~\bibnamefont {Pellegrini}},\ }\Doi
  {doi:10.1038/nnano.2013.161} {\bibfield  {journal} {\bibinfo  {journal}
  {Nature Nanotechnology},\ }\textbf {\bibinfo {volume} {8}},\ \bibinfo {pages}
  {625} (\bibinfo {year} {2013})}\BibitemShut {NoStop}%
\bibitem [{\citenamefont {Uehlinger}\ \emph {et~al.}(2013)\citenamefont
  {Uehlinger}, \citenamefont {Jotzu}, \citenamefont {Messer}, \citenamefont
  {Greif}, \citenamefont {Hofstetter}, \citenamefont {Bissbort},\ and\
  \citenamefont {Esslinger}}]{Uehlinger:2013p8070}%
  \BibitemOpen
  \bibfield  {author} {\bibinfo {author} {\bibfnamefont {T.}~\bibnamefont
  {Uehlinger}}, \bibinfo {author} {\bibfnamefont {G.}~\bibnamefont {Jotzu}},
  \bibinfo {author} {\bibfnamefont {M.}~\bibnamefont {Messer}}, \bibinfo
  {author} {\bibfnamefont {D.}~\bibnamefont {Greif}}, \bibinfo {author}
  {\bibfnamefont {W.}~\bibnamefont {Hofstetter}}, \bibinfo {author}
  {\bibfnamefont {U.}~\bibnamefont {Bissbort}}, \ and\ \bibinfo {author}
  {\bibfnamefont {T.}~\bibnamefont {Esslinger}},\ }\Doi
  {10.1103/PhysRevLett.111.185307} {\bibfield  {journal} {\bibinfo  {journal}
  {Physical Review Letters},\ }\textbf {\bibinfo {volume} {111}},\ \bibinfo
  {pages} {185307} (\bibinfo {year} {2013})}\BibitemShut {NoStop}%
\bibitem [{\citenamefont {Kaul}\ \emph {et~al.}(2008)\citenamefont {Kaul},
  \citenamefont {Kim}, \citenamefont {Sachdev},\ and\ \citenamefont
  {Senthil}}]{Kaul:2008p8050}%
  \BibitemOpen
  \bibfield  {author} {\bibinfo {author} {\bibfnamefont {R.~K.}\ \bibnamefont
  {Kaul}}, \bibinfo {author} {\bibfnamefont {Y.~B.}\ \bibnamefont {Kim}},
  \bibinfo {author} {\bibfnamefont {S.}~\bibnamefont {Sachdev}}, \ and\
  \bibinfo {author} {\bibfnamefont {T.}~\bibnamefont {Senthil}},\ }\Doi
  {10.1038/nphys790} {\bibfield  {journal} {\bibinfo  {journal} {Nature
  Physics},\ }\textbf {\bibinfo {volume} {4}},\ \bibinfo {pages} {28} (\bibinfo
  {year} {2008})}\BibitemShut {NoStop}%
\end{thebibliography}%

\end{document}